\documentclass[graybox, nosecnum]{svmult}

\usepackage{amsmath}
\usepackage{amssymb}

\usepackage{mathptmx}       
\usepackage{helvet}         
\usepackage{courier}        
\usepackage{type1cm}        
\usepackage{makeidx}         
\usepackage{graphicx}        
\usepackage{multicol}        
\usepackage[bottom]{footmisc}
\usepackage{hyperref}        
\usepackage{soul}            
\usepackage{comment}
\hypersetup{colorlinks=true,urlcolor=blue}
\usepackage[square,numbers]{natbib}
\makeindex             
                       
\def\msun{$M_{\odot}$}    
\newcommand{\angs}{\textup{\AA}}

\begin{document}
\title*{$r$-process nucleosynthesis from compact binary mergers}
\author{A. Perego\thanks{corresponding author}, F.-K. Thielemann and G. Cescutti}
\institute{A. Perego \at Department of Physics, University of Trento, Via Sommarive 14, IT-38123 - Trento, Italy.
\email{albino.perego@unitn.it}
\and F.-K. Thielemann \at Department of Physics, University of Basel, Klingelbergstrasse 82, CH-4056, Basel, Switzerland;
\at GSI Helmholtz Center for Heavy Ion Research, Planckstrasse 1, D-64291
- Darmstadt, Germany. 
\email{f-k.thielemann@unibas.ch}
\and G. Cescutti \at 
INAF, Trieste Astronomical Observatory, Via Giambattista Tiepolo, 11, IT-34131
- Trieste, Italy, 
 \email{gabriele.cescutti@inaf.it}
 }
%
%
\maketitle
\abstract{The merger of two neutron stars or of a neutron star and a black hole often result in the ejection of a few percents of a solar mass of matter expanding at high speed in space. Being matter coming from the violent disruption of a neutron star, these ejecta are initially very dense, hot and extremely rich in neutrons. The few available protons form heavy nuclei (``seeds") that absorb the more abundant free neutrons, increasing their size. The neutron density is so high that a substantial number of neutron captures occur before the resulting unstable nuclei can decay toward more stable configurations, converting neutrons into protons. Depending mostly on the initial neutron richness, this mechanism leads to the formation of up to half of the heavy elements that we observe in nature and it is called rapid neutron capture process (``$r$-process"). The prediction of the precise composition of the ejecta requires a detailed knowledge of the properties of very exotic nuclei, that have never been produced in a laboratory. Despite having long been a speculative scenario, nowadays several observational evidences point to compact binary mergers as one of the major sites where heavy elements are formed in the Universe. The most striking one was the detection of a kilonova following the merger of a neutron star binary: the light emitted by this astronomical transient is indeed powered by the radioactive decay of freshly synthesized neutron-rich nuclei and testifies the actual nature of compact binary mergers as cosmic forges.}
\section{Keywords} 
$r$-process nucleosynthesis;  
neutron captures;  
beta decays;       
nuclear fission;   
dynamical ejecta;  
disk outflows;     
kilonovae;         
AT2017gfo;         
heavy elements;    
metal-poor stars.  

\section{Introduction}

After the discovery of the first stellar system formed by two neutron stars (NSs) \cite{Hulse.Taylor1975}, it was immediately realized that the decompression of NS matter following the merger of a binary neutron star (BNS) or of a black hole (BH)-neutron star (BHNS) system produces an ideal environment where the rapid neutron capture process ($r$-process) nucleosynthesis can take place.
The very first calculations were carried out for BHNS binaries \cite{Lattimer:1974a,Lattimer:1976}: in this scenario, during the last orbits of the gravitational wave (GW)-driven inspiral, the NS is tidally disrupted by the gravity of the more massive BH and a fraction of it is ejected into space.
The coalescence of two NSs could also eject neutron-rich matter into the interstellar medium (ISM) through an even richer dynamics \cite{Eichler:1989}.
The first modeling of BNS merger nucleosynthesis was indeed accomplished a few years later \cite{Symbalisty:1982,Freiburghaus.etal:1999}. Since then, major progress has been made in understanding the mechanisms behind the ejection of matter and the properties of these ejecta.

$r$-process nucleosynthesis is one of the fundamental processes responsible for the production of the heaviest elements in the Universe (e.g. \cite{BBFH:1957,Cameron:1957}, see also \cite{Cowan.etal:2019} for a recent review).
The binding energy per nucleon in nuclei increases almost steadily from lithium up to ${}^{56}{\rm Fe}$ and ${}^{56}{\rm Ni}$. This allows the production of nuclei starting from H and He (mostly produced during the big bang) up to iron inside massive stars for increasing plasma temperatures in hydrostatic conditions (see e.g. \cite{Clayton:1983}).
The production of heavier elements through reactions involving charged nuclei would require even larger temperatures to overcome Coulomb repulsion. The kinetic energy necessary to synthesise very heavy elements through fusion reactions 
becomes soon prohibitive for all plausible astrophysical scenarios. Moreover, at such high temperatures disintegration reactions would even dominate due to highly energetic photons. This strongly limits the nucleosynthesis yields produced through this path.
The capture of a free neutron on a nucleus has instead the clear advantage of not having any Coulomb barrier to overcome and it is indeed the key process to produce the heaviest elements.
However, free neutrons in non-degenerate conditions are unstable against $\beta^-$ decay. Thus, neutron capture nucleosynthesis requires a source of neutrons lasting for the relevant timescale over which the nucleosynthesis takes place.
Moreover, the neutron-rich nuclei produced by neutron captures are unstable against $\beta^-$ decay. While decay rates are constants, neutron capture rates crucially depend on the neutron density: 
if the neutron density is high enough ($n_n \gtrsim 10^{20}~{\rm cm^{-3}}$), at least for a short timescale ($\sim 1$s), neutrons are rapidly captured (increasing the mass number) before $\beta$ decays increase the charge number and therefore produce isotopes of the next heavier element. 
Due to its rapidity, this process can happen in explosive environments. This is the basic idea behind $r$-process nucleosynthesis and matter ejected from compact binary mergers provides precisely the conditions necessary for the $r$-process nucleosynthesis to occur.

Over the past few years, several observational evidences have accumulated pointing to compact binary mergers as one of the main sites in the Universe where $r$-process nucleosynthesis takes place, including the observation of elements synthesised through the $r$-process in the atmosphere of very old metal poor stars (see e.g. \cite{Sneden.etal:2008}). The first unambiguous detection of a kilonova (also called macronova), AT2017gfo, as one of the electromagnetic counterparts of the GW signal GW170817 (compatible with a BNS merger) represented the strongest evidence, so far, of this picture (e.g. \cite{Abbott:2017b,Smartt.etal:2017,Kasen.etal:2017}.

In this chapter, we present the most relevant aspects of the $r$-process nucleosynthesis in compact binary mergers. We start by reviewing the conditions of matter expelled during a BNS or a BHNS merger. After that, we present how $r$-process nucleosynthesis proceeds in these ejecta. Finally, we overview the main observational evidences supporting compact binary mergers as major astrophysical sites for $r$-process nucleosynthesis.
In most of our calculations we use cgs units. 
The physical constants employed through the text are the speed of light $c$, the reduced Planck constant $\hbar$, the Boltzmann constant $k_B$, the gravitational constant $G$, the Stefan-Boltzmann constant $\sigma_{\rm SB}$, the solar mass \msun, the masses of the electron, proton and neutron $m_e$, $m_p$, $m_n$, and a generic baryon mass $m_{b}$, which for our purposes can be assumed $m_{b} \approx m_{n}$.
Temperatures are expressed both in Kelvin and in MeV (i.e. as $k_B T$), depending on the context. We recall here that the conversion factor between GK and MeV is roughly one tenth, i.e. $ k_{\rm B} \times 1 {\rm GK} \approx 0.086~{\rm MeV}$.
The distribution of a generic nuclear species $i$ can be expressed either in terms of its mass fraction $X_i$ or number abundance $Y_i$. The former is defined as the ratio between the mass (density) of the species $i$ over the total mass (density): $X_i = m_i/m_{\rm tot} = \rho_i / \rho$, while the latter as the ratio between the number (density) of the $i$ species over the total baryon number (density), $Y_i = N_i/N_{b} = n_i / n_b$. Clearly, $n_{b} \approx \rho/m_{b}$ and $Y_i = X_i/A_i $, where $A_i$ is the atomic number of the species $i$. The electron abundance is defined as $Y_e = n_e/n_{b}$, where $n_e$ is the net density of electrons (i.e. the density of electrons minus the one of positrons). Due to charge neutrality, $Y_e = Y_{p,{\rm free}} + Y_{p,{\rm nuclei}} $, where $Y_{p,\rm free}$ and $Y_{p,\rm bound}$ are the abundance of free protons and of protons bound in nuclei, respectively.


\section{\textit{Matter ejection from compact binary mergers}}
\label{sec: ejecta}



Matter ejection in compact binary mergers happens through different channels. These channels are characterized by specific ejection mechanisms, which operate on different timescales and leave an imprint on the ejecta properties and, ultimately, on the nucleosynthesis. 
In the following, we review the properties of the ejecta from BNS and BHNS mergers by directly relating them to their merger dynamics.
Before doing that, we briefly present some of the most relevant features that characterize the modeling of compact binary mergers and some of the more fundamental processes that influence the ejecta properties. For more extended and complete information about these topics, we refer to the dedicated Chapters and to a few recent reviews, e.g. \cite{Shibata.Hotokezaka:2019,Radice.etal:2020}, where detailed references to the original works can be found.

The dynamics of the merger and of the ejecta expulsion depends on several intrinsic parameters of the binary: first of all, on the nature of the coalescing objects (i.e. if it is a BNS or a BHNS system), but also on their masses and spins. Another relevant ingredient is the still uncertain nuclear equation of state (EOS) for matter at supranuclear density \cite{Oertel.etal:2017}.
Any quantitative statement (and even a robust qualitative understanding) about the merger dynamics relies on detailed numerical simulations. The latter solve the equations of relativistic neutrino-radiation ($\nu$-radiation) hydrodynamics coupled with dynamical space-time evolution.
The hydrodynamics equations are closed by a finite temperature, composition dependent EOS describing the microphysical properties of matter for a rest mass density that varies between stellar densities (a few ${\rm g~cm^{-3}}$) up to several times $10^{15}{\rm g~cm^{-3}}$, corresponding to more than 10 times nuclear saturation density.

Both in BNS and BHNS mergers an accretion disk around a central remnant is expected to form after the merger. In the case of BHNS systems, the remnant is always represented by a BH, while in the case of BNS systems a massive NS, possibly collapsing to a BH on a variable timescale, usually forms. 
While the inspiraling NSs are in cold neutrino-less weak equilibrium, hot matter inside the merger remnant is out of equilibrium and its neutron-to-proton content changes due to neutrino-matter interactions.
Several hydrodynamics processes increase matter temperature during the merger and the relevant temperatures range between 0 and $\sim$150~MeV.
Neutrino production is strongly boosted in hot and dense matter. Due to their low opacity, neutrinos become the dominant cooling source and their luminosity is of the order of a few $10^{53}{\rm erg~s^{-1}}$, at least at merger and during the early aftermath.
The decompression and heating of neutron-rich matter favors initially $\bar{\nu}_e$ luminosity through the reaction $n + e^{+} \rightarrow p + \bar{\nu}_e$ on the thermally produced positrons and $L_{\bar{\nu}_e} \gtrsim 2 L_{\nu_e} $. The resulting net effect is to increase the electron fraction (leptonization), balancing the proton-to-neutron ratio such that at later times $L_{\nu_e} \approx L_{\bar{\nu}_e}$.
In the densest part of the remnant, where matter density is above $\rho \sim 10^{12}{\rm g~cm^{-3}}$, and the temperature $T_{\rm rem}$ is of the order of 10 MeV, the neutrino mean free path $\ell_\nu = 1/(n_{\rm B} \sigma_{\nu})$ is smaller than the size of the system ($\sim 10$-$100~{\rm km}$) for thermal neutrinos of energy $E_{\nu} \sim 3.15 T_{\rm rem}$ :
$
\ell_{\nu} \sim 
250~{\rm m} 
\left( \rho /10^{12}{\rm g~cm^{-3}} \right)^{-1}
\left( T_{\rm rem}/10~{\rm MeV} \right)^{-2} \, .
$
For this estimate we have used an approximated expression for the cross-section of neutrino scattering off free nucleons, $\sigma_{\nu} \approx \sigma_0 (E_\nu/m_e c^2)^2$ with $\sigma_0 = 1.76\times 10^{-44}{\rm cm^{2}}$, to evaluate the neutrino-matter cross-section. Absorption cross-section on free baryons have similar magnitudes and dependences.
Then deep inside the remnant neutrinos equilibrate with matter and diffuse out on the diffusion timescale ($\sim$ seconds).
Due to the presence of an accretion disk, neutrinos are emitted preferentially along the polar direction rather than along the equator, such that the polar flux can be a few times the equatorial one.   
The most relevant neutrino decoupling surfaces are located in the density interval $10^{11-12}{\rm g~cm^{-3}}$, with $\nu_e$'s decoupling at lower densities and temperatures than $\bar{\nu}_e$'s and $\nu_{\mu,\tau}$'s, since the neutron richness favors $n+\nu_e \rightarrow p + e^-$ as absorption process over other inelastic neutrino-matter processes. Typical neutrino mean energies in the decoupling region are $E_{\nu_e} \sim 10~{\rm MeV}$, $E_{\bar{\nu}_e} \sim 15~{\rm MeV}$ and $E_{\nu_{\mu,\tau}} \sim 20~{\rm MeV}$. 
Matter at even lower density, usually located at larger distances from the center, is irradiated by the neutrinos emitted at the inner decoupling surfaces and this irradiation can change the neutron-to-proton content (i.e. $Y_e$) through neutrino absorption on neutrons, protons, and nuclei.

Finally, we recall that NSs are magnetized objects. During the merger and the subsequent remnant evolution, several mechanisms (e.g. dynamo amplification, magneto-rotational instabilities, Kelvin-Helmholtz instabilities) amplify the field strength. Even for initially low magnetic field ($B \sim 10^{9-10} \rm{G}$), the field can be amplified up to several $\sim 10^{15} \rm{G}$ and it becomes dynamically relevant during the merger aftermath, also for matter ejection.

\subsection{\textit{Ejecta from binary neutron stars mergers}}

\begin{figure}[t!]
    \centering
    \includegraphics[width=0.75 \linewidth]{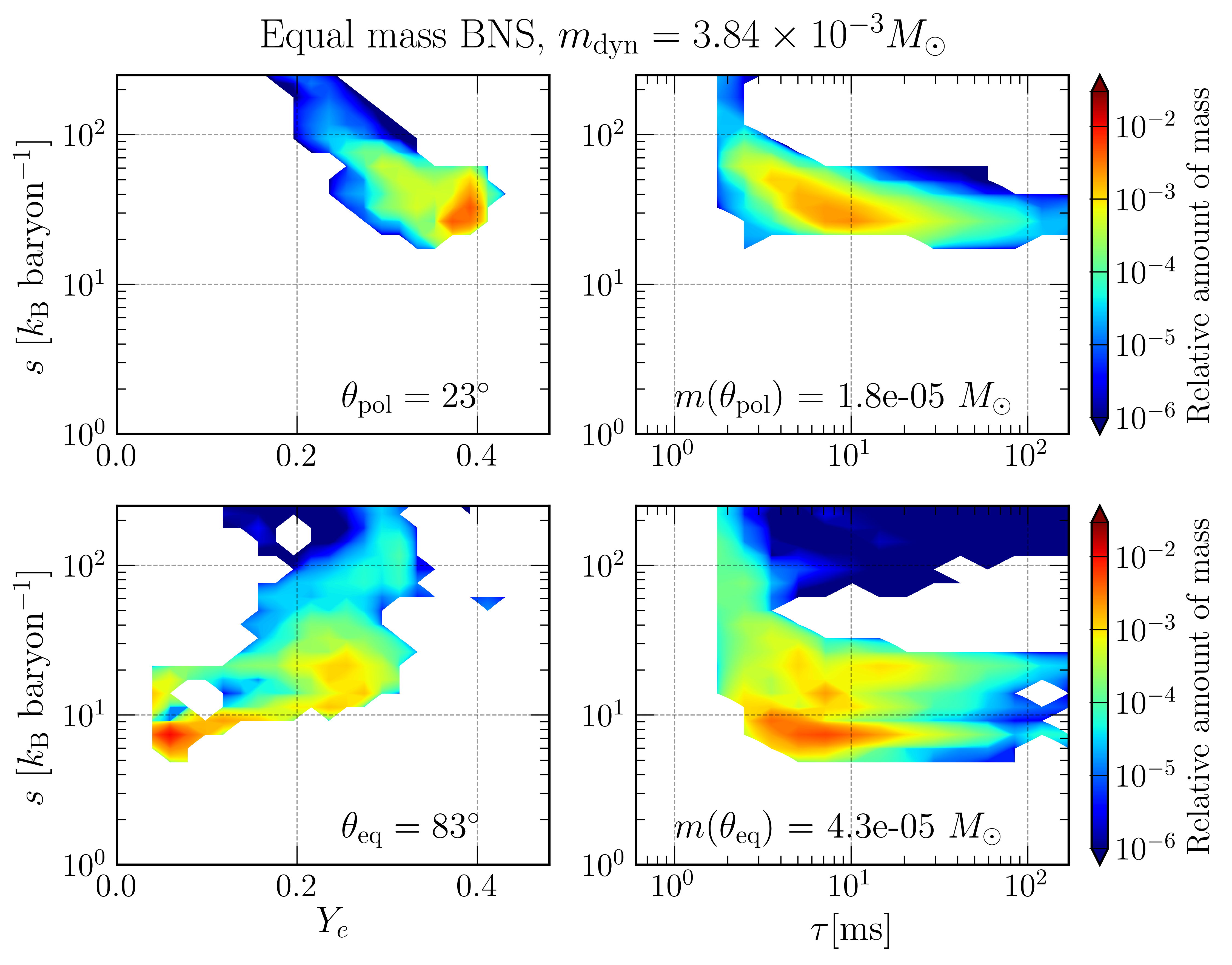}
    \caption{Color coded two-dimensional histograms of the conditions of the dynamical ejecta, as obtained by an equal mass BNS merger simulation (BLh $q=1$ run presented in \protect\cite{Bernuzzi.etal:2020}). On the left panel, the ejecta are characterized in terms of their specific entropy and electron fraction, while on the right panels of their specific entropy and expansion timescale. The top panels refer to an angular slice close to the rotational axis of the binary, while the bottom one to a slice close to the equator.}
    \label{fig: NR histograms}
\end{figure}

\begin{figure}[t!]
    \centering
    \includegraphics[width=0.75 \linewidth]{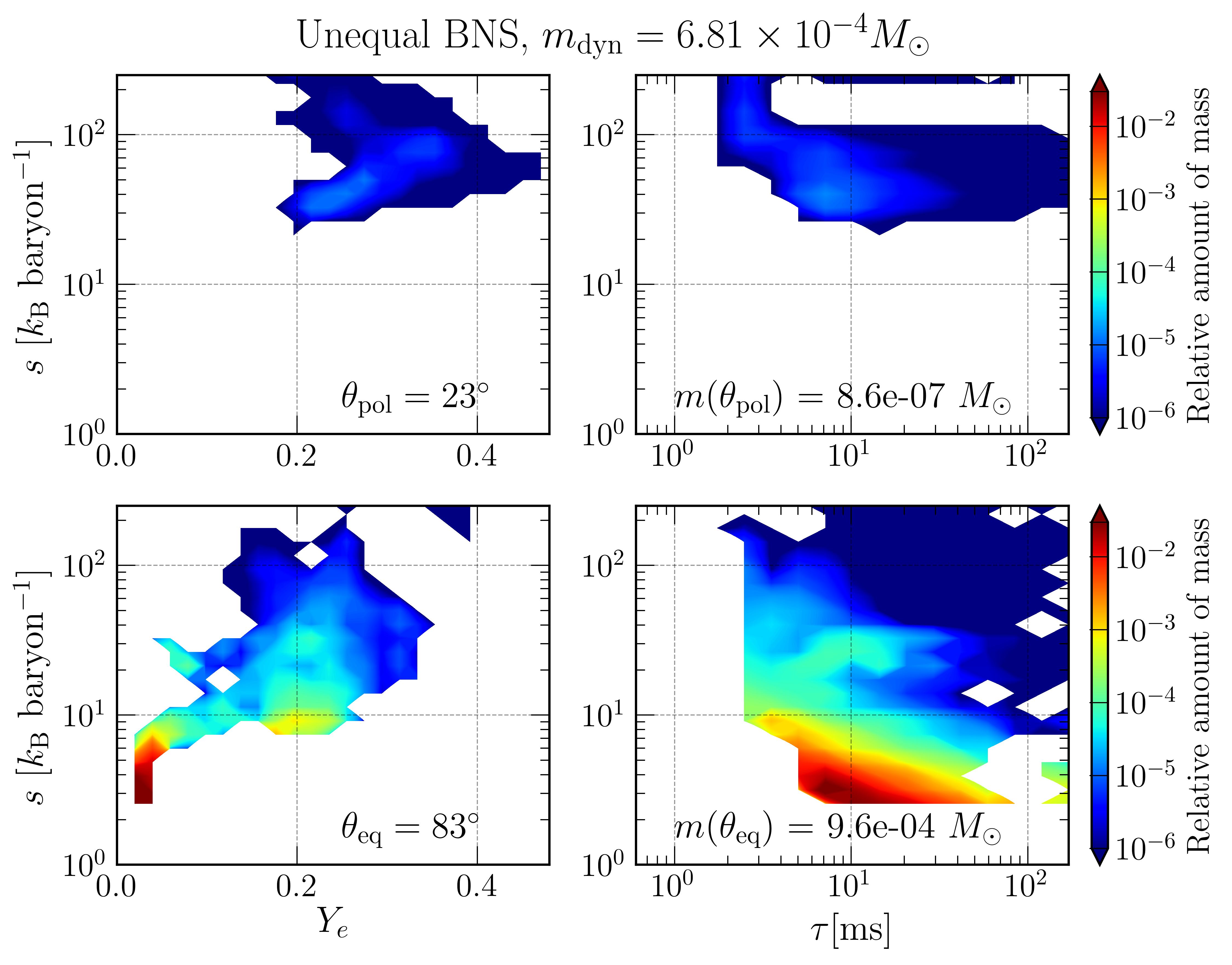}
    \caption{Same as in Figure~\ref{fig: NR histograms}, but for a very unequal mass BNS merger (BLh $q=1.8$ run presented in \protect\cite{Bernuzzi.etal:2020}). These conditions are qualitatively similar to the onse observed also in BHNS mergers.}
    \label{fig: NR histograms 2}
\end{figure}

The first kind of ejecta emerging from a BNS merger are the \textit{dynamical ejecta}. The dynamical timescale of the merger is set by the orbital (angular) velocity at the last orbit before merger, $v_{\rm dyn}$ ($\Omega_{\rm dyn}$), as $t_{\rm dyn} \sim 2 \pi /\Omega_{\rm dyn}  \sim \left( 4 \pi~R_{\rm NS} \right)/v_{\rm dyn} $ where $R_{\rm NS}$ is the NS radius. Assuming a Keplerian behavior, $v_{\rm dyn} \sim \sqrt{GM/(2R_{\rm NS})}$, we obtain:
\begin{equation}
v_{\rm dyn} \sim 0.4~c \left(\frac{M}{2.7 M_{\odot}} \right)^{1/2} \left(\frac{R_{\rm NS}}{12~{\rm km}} \right)^{-1/2} \, ,
\label{eq: orbital speed}
\end{equation}
and 
\begin{equation}
t_{\rm dyn} \sim 1.23~{\rm ms} \left(\frac{M}{2.7 M_{\odot}} \right)^{-1/2} \left(\frac{R_{\rm NS}}{12~{\rm km}} \right)^{3/2} \, ,
\label{eq: dynamical timescale}
\end{equation}
for a typical total binary mass $M$ of 2.7 \msun and a NS radius of 12km \cite{Abbott:2017a,Abbott.etal:2018}.
During the last orbits, when the two NSs approach each other, each of them gets deformed by the tidal field of the companion. As soon as the NSs touch, a large fraction of their kinetic orbital energy is converted into internal energy, while the tidal tails 
retain their orbital speed, Eq.~(\ref{eq: orbital speed}). Since this is larger than the radial escape velocity, matter in the tails is ballistically expelled with velocity $v_{\rm ej} \approx 0.1$-$0.3 c \lesssim v_{\rm dyn}$. This is the \textit{tidal component} of the dynamical ejecta and it mainly develops across the equatorial plane of the binary. 
The temperature in these ejecta is only marginally increased by the tidal compression that precedes the merger ($T \lesssim 1~{\rm MeV}$) and the emission of neutrinos is too weak to change $Y_e$ significantly. Thus, these ejecta mostly retain their original $Y_e$, $0.05 \lesssim Y_e \lesssim 0.15$, and they increase their specific entropy only up to a few $k_B$ per baryon, $s \lesssim 5 k_{\rm B}~{\rm baryon^{-1}}$.
If the total mass of the binary is too large for the matter pressure and for the rotational support to sustain the forming remnant, the latter collapses immediately to a BH.
Otherwise, a (possibly metastable) massive NS forms in the center. This object is far from equilibrium and it bounces as the result of the gravitational pull and matter pressure response. As the sound waves generated by these oscillations travel through the remnant and reach lower density regions, they convert into shock waves, triggering the ejection of matter heated by compression and shocks from the outer edge of the remnant.
These ejecta are called \textit{shock-heated ejecta} and present a broad distribution of expansion velocity, usually peaked around 0.2-0.3$c$, but with a possible high velocity tail extending up to 0.6-0.8$c$.
Due to the action of shocks and compression the ejecta entropy increases, typically reaching values around 10-20 $k_{\rm B}~{\rm baryon}^{-1}$, but with a low-mass, high-entropy, high speed tail extending up to $\sim$100 $k_{\rm B}~{\rm baryon}^{-1}$. The corresponding increase in temperature (initially, up to several tens of MeV before dropping due to matter expansion) produces a large density of electron-positron pairs and determines an increase in $Y_e$ in the ejecta due to positron captures on neutrons.
Moreover, neutrino irradiation coming from the forming remnant can further increase $Y_e$ through $\nu_e$ absorption on neutrons \cite{Wetal:2014}.
Since neutrino emission is more efficient along the polar direction, 
the effect of irradiation is more evident at high latitudes.
The combined effect of tidal tail interactions, hydrodynamics shocks and weak processes is the expansion of the dynamical ejecta over the entire solid angle, still with a preference along the equatorial plane (the mass distribution retains a $\sin^2{\theta}$ dependence on the polar angle $\theta$), with a clear gradient in the $Y_e$ distribution moving from the equator ($Y_e \approx 0.1$) to the poles ($Y_e \lesssim 0.4$). In particular, matter above $\theta \sim 45^\circ$ is expected to have $Y_e \gtrsim 0.25$.
The ejection of dynamical ejecta lasts for a few ms after merger, see Eq.~(\ref{eq: dynamical timescale}), and its amount ranges between $\sim 10^{-4}$ and $\sim 10^{-2}$ \msun, depending on the binary properties and on the nuclear EOS (see e.g. \cite{Radice.etal:2018}). The tidal component is more relevant if the high density part of the EOS is rather stiff or if the two NSs in the binary have very different masses. In this case, at least one of the two NSs is not very compact (i.e. $R_{\rm NS}$ is larger) and the tidal disruption is very effective.
The presence of shock-heated ejecta is instead more relevant in the case of equal mass mergers and for a soft nuclear EOS. In these cases, the NSs are more compact (i.e. $R_{\rm NS}$ is smaller), the collision velocity at merger is larger (see Eq.~\ref{eq: orbital speed}) and the shocks are more violent.
Typical conditions of the dynamical ejecta are presented in Figures \ref{fig: NR histograms} and \ref{fig: NR histograms 2} for both an equal and a very unequal mass BNS merger, as obtained by detailed merger simulations in Numerical Relativity, at two different polar angles .

While the massive NS forms in the center, matter compressed and heated up at the contact interface between the two NS cores is expelled outward. Conservation of angular momentum drives the formation of a rotationally supported, thick accretion disk of radial and vertical extension $R_{\rm disk}$ and $H_{\rm disk}$, such that its aspect ratio is $(H/R)_{\rm disk} \sim 1/3$. The typical disk mass $M_{\rm disk}$ (where the disk is usually defined as the part of the remnant with density below $10^{13}{\rm g~cm^{-3}}$ or, in the presence of a BH, outside of the horizon) ranges between $10^{-3}$ and 0.3$M_{\odot}$. A prompt collapse to BH stops the disk formation, leading to lighter disks. An interesting exception is represented by very asymmetric binaries: in this case, the sudden BH formation is accompanied by a very efficient tidal disruption of the secondary NS, such that a significant fraction of it settles in Keplerian orbital motion outside the BH horizon \cite{Bernuzzi.etal:2020}. 

The evolution of the rotating disk is governed by hydrodynamics, magnetic and weak processes. Several mechanisms are responsible for the local amplification of the magnetic field.
The resulting viscosity of turbulent origin drives matter accretion onto the central object on the accretion timescale:
\begin{equation}
t_{\rm acc} \sim \frac{1}{\alpha} \left( \frac{H}{R} \right)^{-2} \Omega_{\rm disk}^{-1} \approx
0.76~{\rm s} 
\left( \frac{\alpha}{0.02} \right)^{-1}
\left( \frac{H/R}{1/3} \right)^{-2}
\left( \frac{M_{\rm rem}}{2.6 M_{\odot}} \right)^{-1/2}
\left( \frac{R_{\rm disk}}{100~{\rm km}} \right)^{3/2}
\label{eq: disk timescale}
\end{equation}
where $\alpha$ is an effective viscosity parameter, $\Omega_{\rm disk}$ the Keplerian angular velocity, and for the remnant mass $M_{\rm rem}$ and disk radial scale $R_{\rm disk}$ we set as characteristic values 2.6\msun and 100~km, respectively.

During its secular evolution, the processes that determine the disk evolution can produce mass outflows, known as \textit{disk wind ejecta}.
Neutrino absorption, for example, redistributes energy and momentum inside the remnant, from hot and dense regions ($\rho > 10^{12}{\rm g~cm^{-3}}$) to regions where the density decreases to $\rho \sim 10^{10-11}{\rm g~cm^{-3}}$. This process inflates the disk, mainly in the vertical direction (\textit{neutrino-driven winds}). 
At the same time, the accretion process implies an angular momentum redistribution inside the remnant: while mass has a net inflow and the bulk of the disk is accreted, a fraction of it expands radially (\textit{viscosity-driven winds}). 
In both cases, the disk expansion determines a drop in temperature and when $T \lesssim 5-6{\rm GK}$ free neutrons and protons recombine first in $\alpha$ particles and then in heavier nuclei. The energy released is on average $\epsilon_{\rm nucl} \approx 8.6~{\rm MeV~baryon^{-1}}$ and the corresponding expansion velocity can be estimated by equating the kinetic energy at infinity with the sum of the gravitational and nuclear energy released by recombination:
$ v/c \lesssim \sqrt{2\left( 
\epsilon_{\rm nucl}/m_{b} c^2 -
GM_{\rm rem}/(Rc^2) \right)}
$,
where $R_{\rm rec}$ is the radial scale where recombination occurs. Assuming $M_{\rm rem} = 2.6 M_{\odot}$, and $R_{\rm rec}\approx 450-600~{\rm km}$ for disk winds, one obtains $v \approx 0.034-0.074~c$.
Neutrino-driven winds can emerge on a few tens of millisecond timescale, while viscosity-driven winds on the longer accretion timescale. The earlier the wind develops, the faster the ejecta travel. Indeed, at earlier time the disk is hotter and the recombination radius is larger.

Other processes happening inside the disk can also drive disk winds. For example, if the central remnant does not collapse to a BH, non-axisymmetric bars extending inside the disk (mainly $m=1$ and $m=2$ spiral modes) acts continuously on its innermost part, producing a net outflow of angular momentum that crosses the disk and expels matter from its edge (\textit{spiral-wave winds}). Spiral-wave winds develops immediately after disk formation and possibly last up to the point where the central massive NS collapses or the bars are dissipated by GW emission.
Moreover, if in addition to disordered local fields, large scale magnetic fields develop inside the remnant, magnetic pressure and the Lorentz force can further accelerate matter producing \textit{magnetically-driven wind disks}. Both in the magnetic and in the spiral-wave wind cases, expansion velocities are intermediate between the fast dynamical ejecta and the slower recombination disk winds, $v \sim 0.1-0.2c$.

\begin{figure}[t!]
    \centering
    \includegraphics[width=\linewidth]{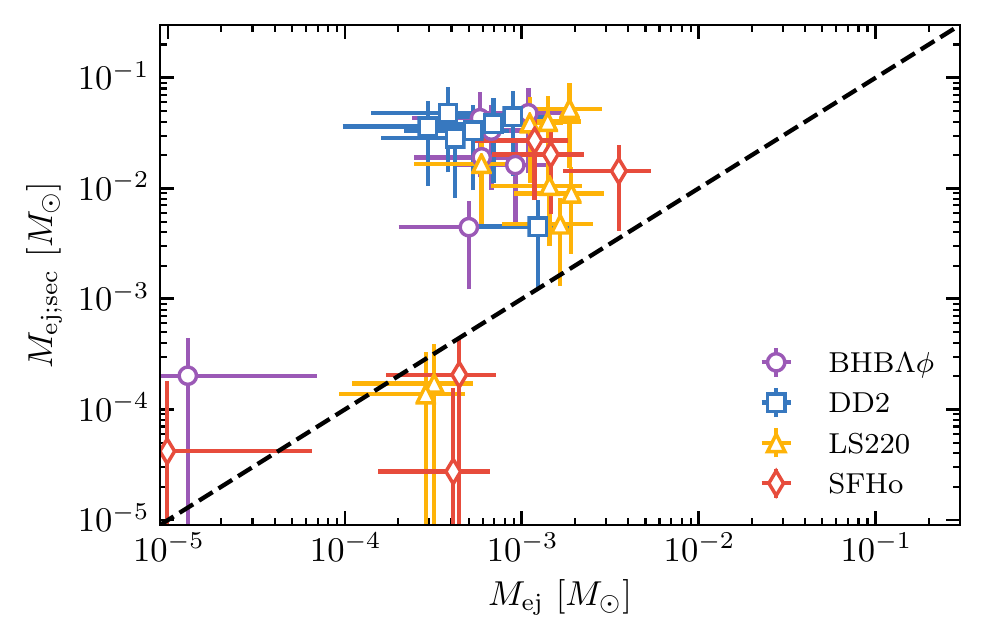}
    \caption{Comparison between the dynamical and the disk-wind ejecta, as obtained by a large set of BNS merger models, employing several nuclear EOSs. While the amount of dynamical ejecta are computed within the simulations, the mass of the wind-disk is assumed to be 20\% of the mass of the accretion disk at the end of the simulations. Figure taken from \protect\cite{Radice.etal:2018}.}
    \label{fig: disk VS dyn ejecta}
\end{figure}

The many processes taking place inside the disk are very effective in unbinding mass from it. The scale that sets the ejecta amount is the mass of the disk itself. While neutrinos alone are able to unbind only a few percents of the disk (and even less in absence of the very luminous massive NS), the other mechanisms unbind between 0.1 and 0.4 $M_{\rm disk}$. It is worth mentioning that these mechanisms can work at the same time, with disk consumption being the only really competing factor. Disk-wind ejecta are thus likely the most relevant source of ejecta for BNS mergers, as visible in Figure~\ref{fig: disk VS dyn ejecta}.

The ejection timescale of disk winds is comparable to the weak reaction timescale. As for the dynamical ejecta, while the initial $Y_e$ in the disk is set by the cold weak equilibrium of the merging NSs, the hot temperature increases $Y_e$ inside the expanding winds due to positron and neutrino absorption on neutrons. Assuming that neutrino irradiation is effective and long enough to reach equilibrium, $Y_e$ tends progressively toward \cite{Qian.Woosley:1996,Martinez-Pinedo.etal:2012}:
\begin{equation}
Y_{e,{\rm eq}} \approx \left( 1+ 
\frac{L_{\bar{\nu}_e}}{L_{\nu_e}}
\frac{W_{\bar{\nu}_e}}{W_{\nu_e}}
\frac{\epsilon_{\bar{\nu}_e} - 2 \Delta + 1.2 \Delta^2/\epsilon_{\bar{\nu}_e}}
{\epsilon_{{\nu}_e} + 2 \Delta + 1.2 \Delta^2/\epsilon_{{\nu}_e}}
\right)^{-1} \, ,
\end{equation}
where $\Delta = (m_n - m_p) c^2 \approx 1.29~{\rm MeV}$, $\epsilon_{\nu}$ is the ratio between the average squared neutrino energy and the average neutrino energy (which for relevant spectral distribution gives $\epsilon_{\nu} \approx 1.2~E_{\nu}$), and $W_i \approx 1 + \eta \langle E_i \rangle/m_b c^2$ with $\eta_{\nu_e} = 1.01$ and $\eta_{\bar{\nu}_e} = -7.22$ is the weak magnetism correction factor.
For typical neutrino luminosities and mean energies, $Y_{e,{\rm eq}} \approx 0.45 < 0.5 $. Thus, depending on the ejection time and on the strength of the neutrino irradiation, the electron fraction in the wind ejecta shows a broad distribution between 0.1 and 0.45. In the case of short-lived massive NS the bulk of the disk wind ejecta have $Y_e$ between $0.2$ and $0.3$, while in the case of long-lived remnant, the higher neutrino luminosity drives the $Y_e$ distribution toward $Y_e \gtrsim 0.3$.  
Matter in the disk is shocked by waves produced by the central remnant and it is heated by viscous dissation. Then its entropy increases to an average value of 15-25 $k_{\rm B}~{\rm baryon}^{-1}$, with possibly high entropy tails extending also in this case up to $\sim$100 $k_{\rm B}~{\rm baryon}^{-1}$.

\subsection{\textit{Ejecta from neutron star-black hole mergers}}

The ejection of matter from BHNS systems shares many similarities with the one from BNS mergers, but it also present crucial differences.
Astrophysical BHs are characterized by their mass, $M_{\rm BH}$, and spin, $\mathbf{J}_{\rm BH}$.
In the following we assume that the orbital angular momentum of the binary points toward the positive $z$ direction and that the BH is significantly more massive than the NS ($M_{\rm BH} \gtrsim 5$ \msun, while $M_{\rm NS} \lesssim 2$ \msun). Assuming $\mathbf{J}_{\rm BH}$ to be also along the $z$ axis, we characterize it through the dimensionless spin parameter $a_{\rm BH} = \pm J_{\rm BH} c/(GM_{\rm BH}^2)$, where $0 \leq a < 1$ if the spin points toward the positive $z$ direction, $-1 < a < 0$ otherwise.
When at the end of the GW-driven inspiral phase the two compact objects approach each other, the merger fate depends on the location of the BH last stable circular orbit, $R_{\rm ISCO}$ (i.e. the radius inside of which circular time-like test-mass orbits in the equatorial plane become unstable to small perturbations), with respect to the tidal distance, $d_{\rm tidal}$ (i.e. the BH-NS distance at which the gravitational force on a test mass at the NS surface equals the tidal force pulling the mass toward the BH). For Kerr BH $R_{\rm ISCO} = M_{\rm BH} f(a)$, where $f(a)$ is a monotonically decreasing function such that $1 \leq f(a) \leq 9$ and $f(0)=6$, while $d_{\rm tidal} \approx (2 M_{\rm BH}/M_{\rm NS})^{1/3} R_{\rm NS}$.
If $d_{\rm tidal} \lesssim R_{\rm ISCO}$ the NS is swallowed by the BH before a significant tidal disruption can occur. No mass is practically left outside the horizon to form a disk or to become unbound. 
Otherwise the NS experiences a partial tidal disruption before most of its mass gets inside the BH horizon. Mass coming from the farther NS edge can become unbound in the form of tidal dynamical ejecta inside a crescent. The rest of the unswallowed mass sets into a Keplerian accretion torus around a spinning BH. 
From the above relations, it is clear that the probability of leaving mass outside the BH horizon increases for stiffer nuclear EOS, smaller mass ratios, and larger $a_{\rm BH}$ .
Numerical simulations show that $ 0 \leq M_{\rm dyn} \lesssim 0.15$\msun for the dynamical ejecta while the mass in the torus is such that $ 0 \leq M_{\rm disk} \lesssim 0.5$\msun. While the dynamical ejecta mass decreases for larger NS masses, the most massive torii are observed for very massive NSs (and for smaller mass ratios and larger spins).
The NS gets compressed during the inspiral. However, the tidal nature of the merger keeps the entropy in the dynamical ejecta low ($s \sim {\rm a~few~}k_{\rm B}$) such that $k_{\rm B}T \lesssim 1~{\rm MeV}$ always. The electron fraction of the ejecta stays also very close to its cold neutrino-less weak equilibrium value, i.e. $Y_e \sim 0.05$.
Inside the torus, even in absence of shocks produced by the central remnant, accretion and disk dynamics heat up matter up to $k_{\rm B}T \sim10~{\rm MeV}$, producing $\sim 10^{52}{\rm erg~s^{-1}}$ of accretion-powered neutrino luminosities. The subsequent production of disk wind ejecta is similar to the one observed in BNS mergers.
Due to the lower neutrino luminosity, the electron fraction of the disk wind ejecta shows a broad distribution, $0.1 \lesssim Y_e \lesssim 0.35 $, with a lower average value, compared with the wind disk ejecta produced by long-lived NS remnants.

\subsection{\textit{Ejecta expansion and thermodynamics}}

As we have seen both for BNS and for BHNS systems, during the merger cold and neutron-rich nuclear matter is heated up and leptonized by several processes, including matter compression, hydrodynamics shocks and neutrino irradiation. 
The bulk of the dynamical ejecta and of the remnant disk originate from inside the outer NS core, where $\rho \gtrsim 10^{14}{\rm g~cm^{-3}}$ and $Y_e \lesssim 0.1$.
In these conditions, nuclei are fully dissociated in free neutrons and protons (homogeneous nuclear matter), and due to charge neutrality their initial abundances are related to the electron fraction by $Y_p = Y_e \lesssim 0.1$ and $Y_n = (1-Y_e) \gtrsim 0.9$.
After having reached its peak temperature, matter is expelled and the density and temperature drop, while neutrons and protons start to form nuclei. In the following we will always assume that the peak temperature of the ejecta exceeds $4~{\rm GK}$.

Matter and radiation in the expanding ejecta can be considered as a neutrally charged plasma consisting of nuclei (often distinguished in free neutrons $n$'s, free protons $p$'s, $\alpha$ particles and a distribution of all other nuclei), electrons, photons and neutrinos.
During most of the relevant nucleosynthesis timescale, photons, electrons and nuclei are in thermodynamics equilibrium. It is thus useful to consider an expanding fluid element as a Lagrangian particle characterized by an evolving density $\rho=\rho(t)$ and temperature $T=T(t)$.
Nuclei and nucleons can be treated as an ideal, non relativistic classical gas, obeying Maxwell-Boltzmann statistics. Electrons are degenerate in the early phase of the expansion before entering a classical gas phase, and they can be described by an ideal Fermi gas of arbitrary degeneracy. Photons are always characterized by a black body spectrum, typical of massless bosons in thermal equilibrium with matter.
Once the density, temperature and composition are given, the entropy of the system can also be computed through the resulting equation of state, $s=s(t)$.


In the first phase of the ejecta expansion, when hydrodynamics processes are still active, the fluid density decreases approximately in an exponential way: 
$\rho(t) \approx \rho_0~\exp{\left( -(t-t_0)/\tau \right)}$,
where $\rho_0$ is the density at the onset of the expansion, $t=t_0$. The expansion timescale, $\tau$, quantifies how fast the density drops during the first phase. Faster dynamical ejecta have smaller $\tau$'s ($1 \lesssim \tau[{\rm ms}] \lesssim 10 $) than slower disk winds ($10 \lesssim \tau[{\rm ms}] \lesssim 100 $). The expansion timescale can be related to the expansion velocity $v_{\rm ej}$ through:
\begin{equation}
\tau \approx \frac{\tilde{R}}{v_{\rm ej}} \approx 16.7~{\rm ms} 
\left( \frac{\tilde{R}}{500~{\rm km}} \right)
\left( \frac{v_{\rm ej}}{0.1 c} \right)^{-1}
\label{eq: expansion timescale VS velocity}
\end{equation}
where $\tilde{R}$ is the lengthscale where matter becomes unbound, depending on the kind of ejecta and ejection mechanism. For the dynamical ejecta, the more impulsive expulsion set a lower $\tilde{R} \sim 300~{\rm km}$ than for the disk wind ejecta, $\tilde{R} \sim 600~{\rm km}$. After a timescale $t_{\rm hom} \sim 3\tau $, internal dynamics and momentum redistribution cease and fluid elements start to expand with an approximately constant velocity: the fastest fluid elements at the front, the slowest at the bottom. 
Such expansion profile is said homologous since $v \propto r$ and
in this phase the density evolves as:
\begin{equation}
\rho(t) = \rho_{\rm hom} \left( 
\frac{t_{\rm hom}}{t} \right)^{3} \approx 100~{\rm g~cm^{-3}}
\left( \frac{\rho_{\rm hom}}{3 \times 10^6{\rm g~cm^{-3}}} \right)
\left( \frac{\tau}{10~{\rm ms}} \right)^{3}
\left( \frac{t}{1~{\rm s}} \right)^{-3}
\label{eq: density evolution}
\end{equation}
with $\rho_{\rm hom}$  varying inside the range $10^{4}-10^{8}{\rm g~cm^{-3}}$, and the larger values are for low entropy, fast expanding ejecta. This density evolution is a very relevant input for the $r$-process \cite{Rosswog.etal:2014}. 
After the nucleosynthesis, the internal energy of the ejecta is subdominant with respect to the kinetic energy and the expansion proceeds adiabatically. Moreover, for the relevant density and temperature conditions, the EOS is dominated by the relativistic photons and electrons. For such a gas, an adiabatic expansion satisfies $\rho T^3 \approx {\rm const}$ and thus $T(t) \propto t^{-1}$.
While this temperature evolution profile is not accurate during the nucleosynthesis epoch, it provides a good approximation after it.


\section{\textit{$r$-process nucleosynthesis in compact mergers}}


We now move to the study of the nucleosynthesis happening inside the ejecta. Before speaking about the $r$-process, we introduce a few basics, although necessary, concepts in nuclear reaction theory. We address the reader to Refs.~\cite{Clayton:1983,Iliadis:2007,Martinez-Pinedo:2008} for general introductions and to \cite{Thielemann.etal:2017,Cowan.etal:2019} for more specific reviews on the $r$-process.

Nuclear abundances inside the ejecta evolve in time
as a result of nuclear (both strong and weak) and electromagnetic reactions.
The usage of abundances in nucleosynthesis calculations is very sensible because, being the ratio of densities, it allows to decouple the effects of reactions on the composition from the effects due to matter expansion.
Strong and electromagnetic reactions conserve separately the number of protons and neutrons. 
Strong nuclear reactions include fusion reactions among nuclei. In the case of transfer reactions, they are often indicated as $B(i,o)C$ to emphasise the transfer particles $i$ and $o$. For example, 
in a $(n,\alpha)$ reaction a free neutron is absorbed by a nucleus $B=(A,Z)$, and a nucleus $C=(A-3,Z-2)$ and an $\alpha$ particle are produced in the final state.
Strong nuclear interactions include also $\alpha$ decays, $(A,Z) \rightarrow (A-4,Z-2) + \alpha$, and spontaneous or induced fission processes.
The most relevant electromagnetic reactions involving nuclei are the photodisintegration reactions, $B(\gamma,o)C$, and their inverse absorption processes, $B(i,\gamma)C$. For example, in the case of a $(\gamma,n)$ reaction, the absorption of a photon on a nucleus $B=(A,Z)$ produces a nucleus $C=(A-1,Z)$ together with a free neutron. 
The opposite reaction $(n,\gamma)$, called neutron capture, consists in the absorption of a free neutron on a nucleus $B=(A,Z)$, producing a nucleus $C=(A+1,Z)$ and a photon.

Weak nuclear reactions include $\beta^{\pm}$ decays, electron and neutrino captures, and they convert neutrons into protons and vice-versa, changing the electron fraction of matter. 
In the expanding neutron-rich ejecta, $\beta^-$ decays, $(A,Z) \rightarrow (A,Z+1) + e^{-} + \bar{\nu}_e$, are the most important weak reactions while in the initial expansion phase neutrino irradiation acts through neutrino absorption, as previously discussed.

Any of the transfer and absorption reactions $B(i,o)C$ is characterized by its cross-section, $\sigma_{B(i,o)C}$, which in general depends on the energy of the colliding particles. In an astrophysical plasma, the distribution of the colliding energy depends on the local thermodynamics properties and the reaction rate, $r_{B(i,o)C}$ (defined as the number of reactions occurring per unit time, per units volume and per reactant pair), is given by $ r_{B(i,o),C} = \langle \sigma v \rangle_{B(i,o)C}~n_B n_i $
where $\langle \sigma v \rangle_{B(i,o)C}$ is the product of the reaction cross section times the relative velocity between $B$ and $i$, averaged over their thermal distributions.

In the case of photon or neutrino absorptions on a nucleus $B$, the reaction rate can be expressed in terms of an effective destruction/decay rate, $\lambda_{B(\gamma/\nu,o)C}$, as $r_{B(\gamma/\nu,o)C} = \lambda_{B(\gamma/\nu,o)C}~n_B$,
where $\lambda_{B,\gamma}$ is a function of $T$ only (since the photon gas properties depend only on temperature), while $\lambda_{B,\nu}$ depends on the neutrino spectrum, which is in general not in equilibrium with the plasma. Then, strong and electromagnetic reactions depend only on the local plasma properties.
Alpha and beta decays are instead characterized by a constant decay rate $\lambda_{\alpha / \beta}$, related to the nucleus half-life $t_{1/2}$ by $\lambda_{\alpha / \beta} = \ln{2}/t_{1/2}$. The lifetime of a nucleus against a certain reaction can be defined as the inverse of the corresponding rate,  $\tau_i = 1/ \lambda_i$.

For large temperatures, strong and photodisintegration reactions are characterized by large reaction rates.
For $T \gtrsim 4-5~{\rm GK}$ the resulting fusion and photodisintegration timescales become much shorter than the weak and dynamical timescales in the ejecta. Thus, these reactions can be considered in equilibrium among them and with their inverse reactions.
This condition is called Nuclear Statistical Equilibrium (NSE) and the nuclear abundance in NSE are fully determined by the local thermodynamical conditions, i.e. by $\rho$, $T$ and $Y_e$. While $\rho$ and $T$ vary due to the expansion, $Y_e$ changes due to weak processes, but in both cases on much longer timescales than the nuclear NSE timescale. 
When the temperature decreases below $\sim 4-5~{\rm GK}$ some reactions characterized by small $Q$-values become slow enough that NSE is no more guaranteed across the entire nuclear distribution, and especially for nuclei characterized by magic nuclear numbers, i.e. close to shell closure conditions. This transition is called \textit{NSE freeze-out}. 
Since a large fraction of direct and inverse nuclear reactions are still very fast, the nuclear distribution splits into areas of Quasi Statistical Equilibrium (QSE), where equilibrium conditions still apply on sub-sets of nuclei.
When the temperature decreases even further, no equilibrium arguments apply and fully out-of-equilibrium nucleosynthesis occurs. 
While in NSE conditions accurate abundances can be computed even without precise information on the reaction rates, a detailed knowledge of the properties of the ejecta and of nuclei all over the nuclear chart are requested to predict accurate abundances after NSE freeze-out.
Nevertheless, equilibrium arguments still provide a useful tool to understand the basic feature of nucleosynthesis in QSE conditions.

\subsection{\textit{Compact binary mergers as $r$-process site}}

Before entering the details of the $r$-process, we first motivate why the ejecta of compact binary mergers represent a suitable environment for $r$-process nucleosyntheshis.
The ejecta come from high density conditions and usually experience high enough temperatures such that matter is mostly dissociated into free neutrons and protons under NSE conditions at their peak temperature.
As temperature drops, neutrons and protons recombine first into $\alpha$ particles. The subsequent building of the most tightly bound iron group nuclei depends on the three body reactions responsible for the assembly of heavier nuclei, namely $2 \alpha + {\rm n} \rightarrow {}^{9}{\rm Be} + \gamma $ and  $3 \alpha \rightarrow {}^{12}{\rm C}+ \gamma $.
The first one is the most relevant in neutron-rich conditions. 
Triple reactions are in competition with their inverse photodestruction reactions. 
Due to their triple nature, the former are favored by larger densities, while the latter are strongly enhanced by higher temperatures since for the photon density and mean energy one has $n_{\gamma} \propto T^3$ and $\langle E_{\gamma} \rangle \propto T$, respectively.
If the plasma is radiation-dominated, the value of the specific entropy ultimately determine whether iron group nuclei can form. Indeed, if $s \gtrsim s_{\gamma} + s_{e^{\pm}} \approx 2 s_{\gamma}$, the density can be computed as:
\begin{equation} 
\rho \lesssim 2
\left( \frac{4 \pi^2 m_{b}}{45 \left( \hbar c \right)^3} \right) \frac{k_{\rm B}^3 T^3}{s/k_{\rm B}}
\approx 3.02 \times 10^{6}~{\rm g~cm^{-3}} 
\left( \frac{T}{5~{\rm GK}} \right)^3
\left( \frac{s}{10~k_{\rm B}} \right)^{-1}
\, .
\label{eq: density photon-dominated gas}
\end{equation}
In very neutron rich conditions ($Y_e \sim 0.1$), when temperature decreases between 5 and 2.5 GK, $\alpha \alpha n$ reactions occur more efficiently than their inverse photodestruction reactions only for $\rho \gtrsim 3 \times 10^5~{\rm g~cm^{-3}}$,
i.e. for $s \lesssim 100 k_{\rm B}~{\rm baryon}^{-1}$.
This implies that in low and moderate entropy conditions (typical of the merger ejecta) iron group nuclei are formed in NSE conditions, while in the high entropy tail $\alpha$ particles are mostly produced when NSE equilibrium is no more guaranteed ($\alpha$-rich freeze out).

Almost all protons are bound inside nuclei and a distribution of heavy nuclei around the iron group (called \textit{seed nuclei}) is present at the NSE freeze-out.
For the seed nuclei, one can introduce a representative average nucleus $(\langle A \rangle_{\rm seed},\langle Z \rangle_{\rm seed})$ defined such that:
\begin{equation}
\langle A \rangle_{\rm seed} \equiv \left( \sum\limits_{(A,Z),A > 4}~A ~ Y_{(A,Z)} \right) /Y_{\rm seed} \, ,
\quad
\langle Z \rangle_{\rm seed} \equiv \left( \sum\limits_{(A,Z),A > 4}~Z ~ Y_{(A,Z)} \right) /Y_{\rm seed} \, ,
\end{equation}
where we have defined $Y_{\rm seed} \equiv \sum_{(A,Z),A > 4}~Y_{(A,Z)}$.
While for $Y_e \approx 0.5$ the nuclear abundance distribution has its peak around $^{56}{\rm Fe}$ and $^{56}{\rm Ni}$ (the most bound, symmetric nuclei, characterized by the presence of a $Z=28$ proton shell closure), for $Y_e < 0.5$ the distribution moves toward more exotic, neutron-rich iron group nuclei.
However, there is a limit for the amount of neutrons that can be bound inside a nucleus:
for $20 \lesssim Z \lesssim 40$, nuclei located at the neutron drip line have $(Z/A)_{\rm min} \sim 0.3$. 
Moreover, due to the neutron richness it becomes energetically favorable to have a certain fraction of free neutrons even for $Y_e > 0.3$.
Starting from baryon number conservation written as $X_n + X_{n,{\rm seed}} + X_{p,{\rm seed}} \approx 1$ (where $X_{n/p,{\rm seed}}$ indicate the fractions of neutrons or protons bound in seed nuclei), from the charge neutrality condition $Y_e \approx \sum_{(A,Z),A> 4} ~\left( Z~Y_{(A,Z)} \right)$, and using the above definitions, the free neutron fraction can be evaluated as:
\begin{equation}
Y_n \approx 1 - \left( \langle A \rangle_{\rm seed} / \langle Z \rangle_{\rm seed} \right) Y_e \, .
\label{eq:Yn formula}
\end{equation}
Assuming for simplicity that $\langle Z \rangle_{\rm seed} \approx 28 $, and that the dependence of $\langle A \rangle_{\rm seed}$ on $Y_e$ is approximately linear between $A/Z = 0.5$ and $A/Z = \left(A/Z\right)_{\rm min}$:
\begin{equation}
\langle A \rangle_{\rm seed} \approx 56 + 70 \left( 0.5 - Y_e \right) \, ,
\label{eq:Aseed estimate}
\end{equation}
we obtain an approximation for $Y_n$:
\begin{equation}
Y_n \approx 1 - 3.25 Y_e + 2.5 Y_e^2  \, ,
\label{eq:Yn estimate}
\end{equation}
that varies between 0 for $Y_e = 0.5$ and 1 for $Y_e = 0$.
Since almost all protons are contained inside the seed nuclei, we can also estimate the seed abundance as:
\begin{equation}
Y_{\rm seed} \approx Y_e/\langle Z \rangle_{\rm seed} \approx Y_e/28 \, .
\label{eq:Yseed formula}
\end{equation}

Neutron capture is the key reaction to produce heavy nuclei beyond the iron group, once NSE freeze-out has occurred.
In accordance with the definition of reaction rate and in analogy to the effective rate definition, the lifetime of a generic seed nucleus against a neutron capture reaction can be estimated as:
\begin{equation}
\tau_{(n,\gamma)} = 1/ \left( n_n \langle \sigma v \rangle_{(n,\gamma)} \right) \, ,
\label{eq: n-capture lifetime}
\end{equation}
where $n_n$ is the free neutron density. 
The $Q$-value of an $(n,\gamma)$ reaction is the energy gained by the nucleus $(A,Z)$ by acquiring a neutron and it is computed as $Q = \left( m_{(A,Z)}  + m_n c^2 - m_{(A+1,Z)} \right) c^2$. It is also equal to the energy required to remove a neutron from a $(A+1,Z)$ nucleus, called \textit{neutron separation energy} of the $(A+1,Z)$ nucleus, $S_{n,(A+1,Z)}$.
Neutron captures have $Q$-values that range from 0 for exotic neutron-rich nuclei at the neutron drip line up to $\sim 15~{\rm MeV}$ close to the valley of stability.
The leading contribution to $\sigma_{(n,\gamma)}$ is provided by the $s$-wave term of the partial wave expansion:
$ \sigma_{n} \approx \left( \pi/k^2 \right) T_s$,
where $T_s \approx 4k/k'$ is the transmission coefficient obtained by considering a neutron moving against the potential barrier of the nucleus, and $k,k'$ are the wave numbers of the particles in the initial and final state, respectively.
If $\mu \approx m_n$ is the reduced mass of the parent state and $E \sim k_{\rm B}T$ the thermal energy scale, for non-relativistic energies ($k_{\rm B}T \sim 0.1~{\rm MeV}\ll Q \ll m_n c^2 $) the relative speed 
is $v = \sqrt{2E/\mu}$,
and the wave numbers are $k \approx \sqrt{2 \mu E}/\hbar$ and $ k' \approx  \sqrt{2 \mu (S_{n}+E)}/ \hbar \approx \sqrt{2 \mu S_{n}}/\hbar $, so that the typical cross-section is:
\begin{equation}
\sigma_{(n,\gamma)} \sim 3.70~{\rm barn} \left( \frac{k_{\rm B}T}{0.1~{\rm MeV}} \right)^{-1/2} \left( \frac{S_{n}}{5~{\rm MeV}} \right)^{-1/2} \, .
\label{eq: cross section estimate}
\end{equation}
Since $\sigma_n \propto 1/\sqrt{T}$ and $v \propto \sqrt{T}$, $\langle \sigma v \rangle_{n,\gamma} $ is approximately constant for thermal, non-relativistic neutrons. Thus, the lifetime against neutron capture, Eq.~(\ref{eq: n-capture lifetime}), depends mainly on $n_n$, which for a plasma of density $\rho$ and free neutron fraction $Y_n$ is simply
$n_n = \rho Y_n/m_{\rm B} \approx 2.99 \times 10^{24} {\rm cm^{-3}}~(\rho/10~{\rm g~cm^{-3}})~(Y_n/0.5)$,
and $\tau_{(n,\gamma)}$ becomes:
\begin{equation}
\tau_{(n,\gamma)} \sim 0.21~{\rm ns}
\left( \frac{\rho}{10~{\rm g~cm^{-3}}} \right)^{-1}
\left( \frac{Y_n}{ 0.5} \right)^{-1}
\left( \frac{S_{n}}{5~{\rm MeV}} \right)^{1/2} \, .
\label{eq:tau_n_gamma estimate}
\end{equation}
The density in the ejecta changes considerably with time as a result of the homologous expansion, Eq~(\ref{eq: density evolution}). If we further consider that $\rho_{\rm hom} \sim 10^6{\rm g~cm^{-3}} $ at NSE freeze-out for $s \sim 10~k_{\rm B}~{\rm baryon^{-1}} $, see Eq~(\ref{eq: density photon-dominated gas}), then the free neutron density at time $t$ is:
\begin{equation}
n_n \approx 8.08 \times 10^{24}~{\rm cm^{-3}} 
\left( \frac{\rho_{\rm hom}}{10^6~{\rm g~cm^{-3}}} \right)
\left( \frac{Y_n}{0.5} \right)
\left( \frac{\tau}{10~{\rm ms}} \right)^3
\left( \frac{t}{1~{\rm s}} \right)^{-3} \, ,
\label{eq:tau_n_gamma estimate 2}
\end{equation}
and $\tau_{(n,\gamma)}$ can be also expressed as a function of time after merger as:
\begin{eqnarray}
\tau_{(n,\gamma)} & \sim & 1.18~{\rm \mu s}
\left( \frac{Y_n~\rho_{\rm hom}}{0.5 \times 10^6~{\rm g~cm^{-3}}} \right)^{-1}
\left( \frac{\tau}{10~{\rm ms}} \right)^{-3} 
\left( \frac{S_{n}}{5~{\rm MeV}} \right)^{1/2}
\left( \frac{t}{1~{\rm s}} \right)^{3} \, .
\label{eq:tau_n_gamma estimate 2}
\end{eqnarray}

The capture of one or more neutrons increases the mass number by one or more units without increasing the atomic number. Then, neutron captures move the nuclear abundances toward the neutron-rich side of the nuclear chart.
Two kinds of reactions compete with neutron capture in producing heavier and heavier neutron-rich nuclei: $(\gamma,n)$ reactions and the $\beta^{-}$ decays.

For the $(\gamma,n)$ reactions, while the high neutron density guarantees high $(n,\gamma)$ rates, high energy photons are required to knock a neutron off a nucleus, overcoming the neutron separation energy $S_n$ and boosting the photodestruction rates.
For $T \gtrsim 4{\rm GK}$ the two set of reactions are in NSE, meaning that the temperature is large enough to bring $(n,\gamma)$-$(n,\gamma)$ at equilibrium everywhere, also close to the valley of stability where $S_n \sim 8-10~{\rm MeV}$. In neutron-rich conditions, when the nuclear distribution is shifted toward the neutron-drip line, the relevant $S_n$ can be as low as $\sim 1-2~{\rm MeV}$. Thus, temperatures lower than 4 GK 
are enough to preserve the $(n,\gamma)$-$(n,\gamma)$ equilibrium after NSE-freeze-out on the neutron rich side of the nuclear chart. This is indeed the QSE typical of neutron-rich conditions.
Since $(\gamma,n)$ reactions are the inverse of $(n,\gamma)$ reactions, their rates are related by detailed balance conditions: $ n_{(A,Z)} n_n ~ \langle \sigma v \rangle_{(n,\gamma)} = n_{(A+1,Z)}\lambda_{(\gamma,n)} $,
where $\lambda_{(\gamma,n)}$ is the photodisintegration rate of $(\gamma,n)$ reactions.
Assuming these reactions to be in equilibrium, i.e. $(A,Z) + n \leftrightarrow (A+1,Z) + \gamma$, the chemical potential of the different nuclear species involved are related by $\mu_{(A,Z)} + \mu_n = \mu_{(A+1,Z)}$ (we recall that $\mu_{\gamma}=0$).
Using the expression of the chemical potentials for an ideal Maxwell-Boltzmann gas in its relativistic version (i.e., including the rest mass contribution), one obtains:
\begin{equation}
\frac{n_{(A+1,Z)}}{n_{(A,Z)}~n_{n}} =  
\left( \frac{2 \pi \hbar^2}{m_{b} k_{\rm B} T} \right)^{3/2} \frac{G_{(A+1,Z)}(T)}{2~G_{(A,Z)}(T)} 
\left( \frac{A+1}{A} \right)^{3/2} \exp\left(\frac{S_{n,(A+1,Z)}}{k_{\rm B}T}\right) \, ,
\label{eq: chain abundaces}
\end{equation}
where $G_{(A,Z)}(T)$ is the nuclear partition function, dependent on the matter temperature.
Assuming $G_{(A+1,Z)}/G_{(A,Z)} \sim 1 $ and $(A+1)/A \sim 1 $, we finally obtain an expression for $\tau_{(\gamma,n)} \equiv 1/\lambda_{(\gamma,n)}$, the lifetime of a seed nucleus against photodestruction:
\begin{equation}
\tau_{(\gamma,n)} 
\approx 0.19~{\rm \mu s}
\left( \frac{k_{\rm B}T}{0.167 {\rm MeV}} \right)^{-3/2}
\left( \frac{S_n}{5 {\rm MeV}} \right)^{1/2}
\frac{\exp{\left( S_n / k_{\rm B}T \right) }}{\exp{(30)}} \, ,
\label{eq: tau_g_n estimate}
\end{equation}
where the reference temperature (corresponding to $T \approx 2~{\rm GK}$) and neutron separation energy are chosen such that $S_n/k_{\rm B}T \approx 30$.
This timescale depends heavily on the temperature due to the presence of the exponential factor that rises by orders of magnitudes as the temperature decreases.
Additionally, it depends also on the nuclear masses through $S_n$ and the latter sets the scale in the exponential argument, meaning that for a given temperature $\tau_{(\gamma,n)}$ changes dramatically between the valley of stability and the drip line.

Initially, at NSE freeze out ($T \sim 4~{\rm GK}$), the temperature and the density are large enough such that both $\tau_{\rm n,\gamma}$ and $\tau_{\rm \gamma,n}$ are much smaller than the dynamical timescale. The exponential term in  Eq.~(\ref{eq: tau_g_n estimate}) set the typical $S_{n}$ necessary to guarantee the equilibrium between the two reaction sets. As $T$ and $\rho$ decrease, $\tau_{(\gamma,n)}$ changes much more sensibly and $(n,\gamma)$-$(\gamma,n)$ equilibrium can establish only down to $T \approx 1~{\rm GK}$, assuming typical $S_n \approx 1$-$3~{\rm MeV}$.

For the $\beta^-$ decays, using the low-energy limit of the weak interaction theory at leading order, the transition matrix element $\mathcal{M}$ is proportional to the Fermi coupling constant, $G_{\rm F}$. Remembering that in natural units $G_{\rm F}$ is the reciprocal of an energy squared, that the time is the reciprocal of an energy, and that the only relevant energy scale in the process is the $Q$-value of the decay, the lifetime of a nucleus against $\beta$ decay,  $\tau_{\beta}$, must be proportional to $Q^{-5}$.
This dependence can be seen also as a consequence of the three-body nature of the final state.
The $Q$-value of $\beta^-$ decays involving neutron-rich nuclei is roughly proportional to the neutron excess, $D = N-Z$, and varies between a fraction of MeV close to the valley of stability and $\sim$5-15 MeV at the neutron drip line (with larger values at lower mass numbers, where the neutron excess can also be much larger).
Using the decay of the free $n$ as representative $\beta^-$ reaction (for which $Q=\Delta$ and $D=1$), we estimate the typical $\beta$-decay lifetime as:
\begin{eqnarray}
\tau_{\beta} \sim \tau_{n} \left(\frac{\Delta}{Q} \right)^5 
&  \approx & 3.19~{\rm ms} \left( \frac{Q}{10~{\rm MeV}} \right)^{-5} \approx
8.82 {\rm ms} \left( \frac{D}{10} \right)^{-5}
\, ,
\label{eq: tau_beta estimate}
\end{eqnarray}
where in the last step we have further assumed that $Q \sim D \Delta$. 
Then, $\beta^{-}$ decays act on much longer timescale than neutron captures at NSE freeze-out and during the $(n,\gamma)-(\gamma,n)$ equilibrium, and become competitive only when the temperature and density have significantly dropped.
This difference in the neutron capture and in the $\beta^-$ decay timescale qualifies the the ejecta from compact binary mergers as one of the astrophysical sites for $r$-process nucleosynthesis in the Universe.

Starting from a certain seed nuclei distribution, the fast neutron and photon captures move abundances within the same isotopic chain (i.e., the sequence of nuclei characterized by the same $Z$ and by an increasing $N$). However, nuclear decay is a stochastic process and a fraction of nuclei decay from one isotopic chain to the next one even if $\tau_{\beta}$ is significantly larger than $\tau_{(n,\gamma)}$, especially if the $(n,\gamma)$-$(\gamma,n)$ equilibrium is maintained on a timescale comparable to or longer than $\tau_{\beta}$. 
Moreover, $S_n$ is not a smooth, monotonic function inside the nuclear chart. The closure of neutron shells inside the nucleus at the magic numbers $N=28,50,82,126$ determines an increase of $S_n$ around those values. The corresponding nuclei become \textit{waiting points}, where $\tau_{(n,\gamma)}$ increases and matter tends to accumulate. For these nuclei the $\beta^-$ decay starts earlier to be competitive and matter flows through them from an isotopic chain to the next one.

The net results of the combination of neutron captures, photodestructions and $\beta^-$-decays on the whole distribution of nuclei emerging from NSE during the $(n,\gamma)$-$(\gamma,n)$ equilibrium is a characteristic nuclear distribution that proceeds as a river inside the neutron-rich side of the nuclear chart. This is called the \textit{r-process path}.

The final point of the $r$-process nucleosynthesis depends on how many free neutrons are available to be captured by the seed nuclei. This number is called the \textit{neutron-to-seed ratio}, $Y_n/Y_{\rm seed}$, and the end point of the $r$-process can be estimated as:
\begin{equation}
\langle A \rangle_{\rm final} \sim \langle A \rangle_{\rm seed} + Y_n/Y_{\rm seed} \, ,
\end{equation}
where $\langle A \rangle_{\rm seed}$ is the average mass number of the seed nuclei. For example, assuming $\langle A \rangle_{\rm seed} \sim 80-90$ (see Eq.~\ref{eq:Aseed estimate}), to produce element with $A \approx 130$ it is necessary to have $Y_n/Y_{\rm seed} \approx 40-50$; for $A \approx 195$, $Y_n/Y_{\rm seed} \approx 95-115$ (we will see later than these are the mass numbers of the so-called second and third $r$-process peaks); while for uranium and thorium ($A \approx 235$) $Y_n/Y_{\rm seed} \approx 145-155$.
For low or moderate entropy ejecta, the value of $Y_e$ primary determines the neutron-to-seed ratio and, from that, how far the $r$-process nucleosynthesis proceeds in producing heavy elements starting from iron group seed nuclei.
Using Eqs.~(\ref{eq:Yn estimate}) and (\ref{eq:Yseed formula}) we can obtain a simple estimate for the neutron-to-seed ratio as a function of $Y_e$ in low entropy conditions:
\begin{equation}
Y_n/Y_{\rm seed} \sim \langle Z \rangle_{\rm seed}/Y_e - \langle A \rangle_{\rm seed} \approx 28/Y_e - 70~Y_e - 21 \, .
\end{equation}
This expression diverges for $Y_e \rightarrow 0$, as there are no seeds, while it goes to 0 for $Y_e = 0.5$. For intermediate values $Y_e = 0.1,0.2, 0.3, 0.4$ we obtain  $Y_n/Y_{\rm seed} \approx 250, 105, 51, 21$, respectively.
These formulae, relying on the simple assumption of a linear dependence of $\langle A \rangle_{\rm seed}$ on $Y_e$, Eq~(\ref{eq:Aseed estimate}), must be understood as very rough estimates that nevertheless catch the most relevant trends for typical entropy ($s \sim 10~k_{\rm B}{\rm baryon^{-1}}$) and expansion timescale ($\tau \sim 10~{\rm ms}$) in the ejecta. More detailed and physically motivated calculations (e.g. \cite{Hoffman.etal:1997,Lippuner.Roberts:2015}) extending to broader ranges of possible conditions, show that the neutron-to-seed ratio has a more complex dependence on the (thermo)dynamics conditions at NSE freeze-out, in particular, $Y_n/Y_{\rm seed}$ is larger for lower $Y_e$, smaller $\tau$ and larger $s$. For example, for $Y_e \gtrsim 0.4$ the neutron-to-seed ratio is such that $Y_n/Y_{\rm seed} \propto s^3/ Y_e^3 \tau$.

\subsection{\textit{The working of the $r$-process in compact binary mergers}}

\begin{figure}[t!]
    \centering
    \includegraphics[width=\linewidth]{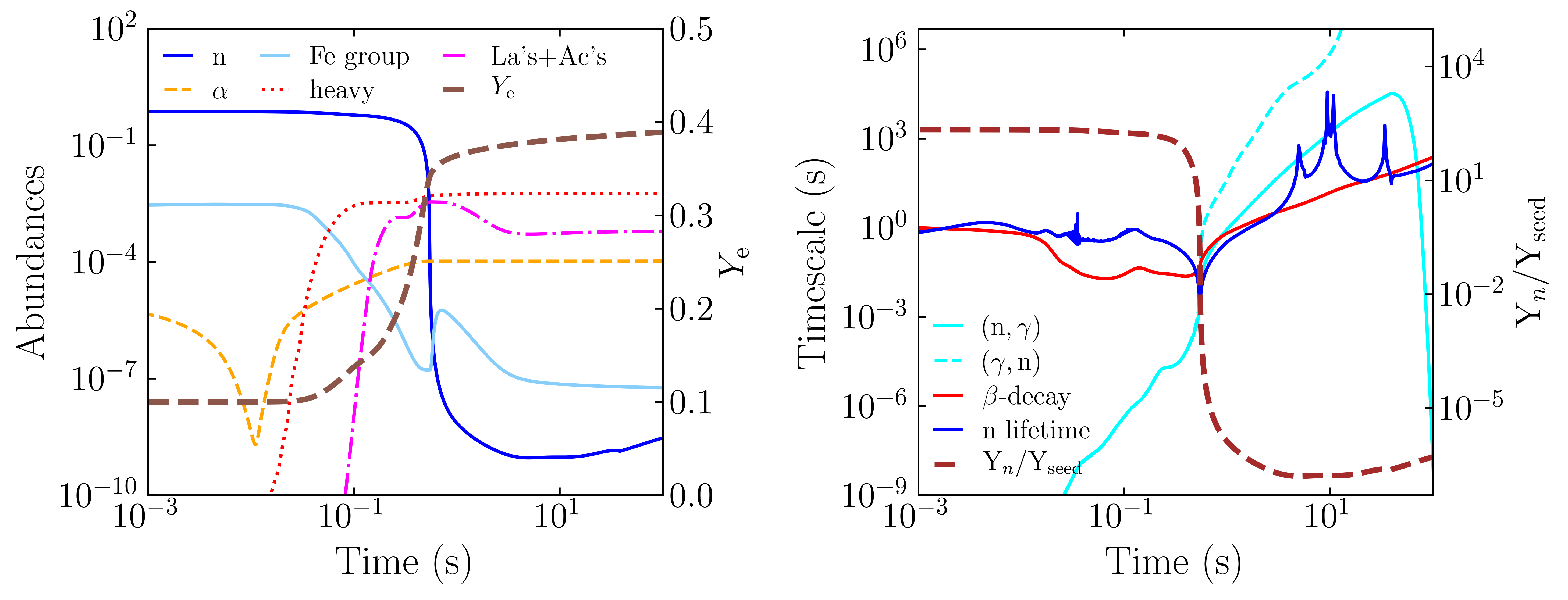}
    \caption{Evolution of a few selected abundances computed for a fluid elements expanding in space with homologous expansion, see Eq.~({\protect\ref{eq: density evolution}}), and initially characterized by a specific entropy $s = 10~k_{\rm B}{\rm baryon}^{-1}$, an initial $Y_e = 0.10$ and expansion timescale $\tau = 10{\rm ms}$, corresponding to typical compact binary merger ejecta conditions (left panel). Heavy nuclei are defined as nuclei for which $A \geq 120$. In addition, relevant timescales as computed according to Eqs.~(\protect\ref{eq: neutron timescale})-(\protect\ref{eq: tau beta network}) are also shown (right panel).
    Calculations were performed using the SkyNet nuclear network \protect\cite{Lippuner.Roberts:2017} (Courtesy of D. Vescovi).}
    \label{fig: skynet trajectory}
\end{figure}

\begin{figure}[t!]
    \centering
    \includegraphics[width=\linewidth]{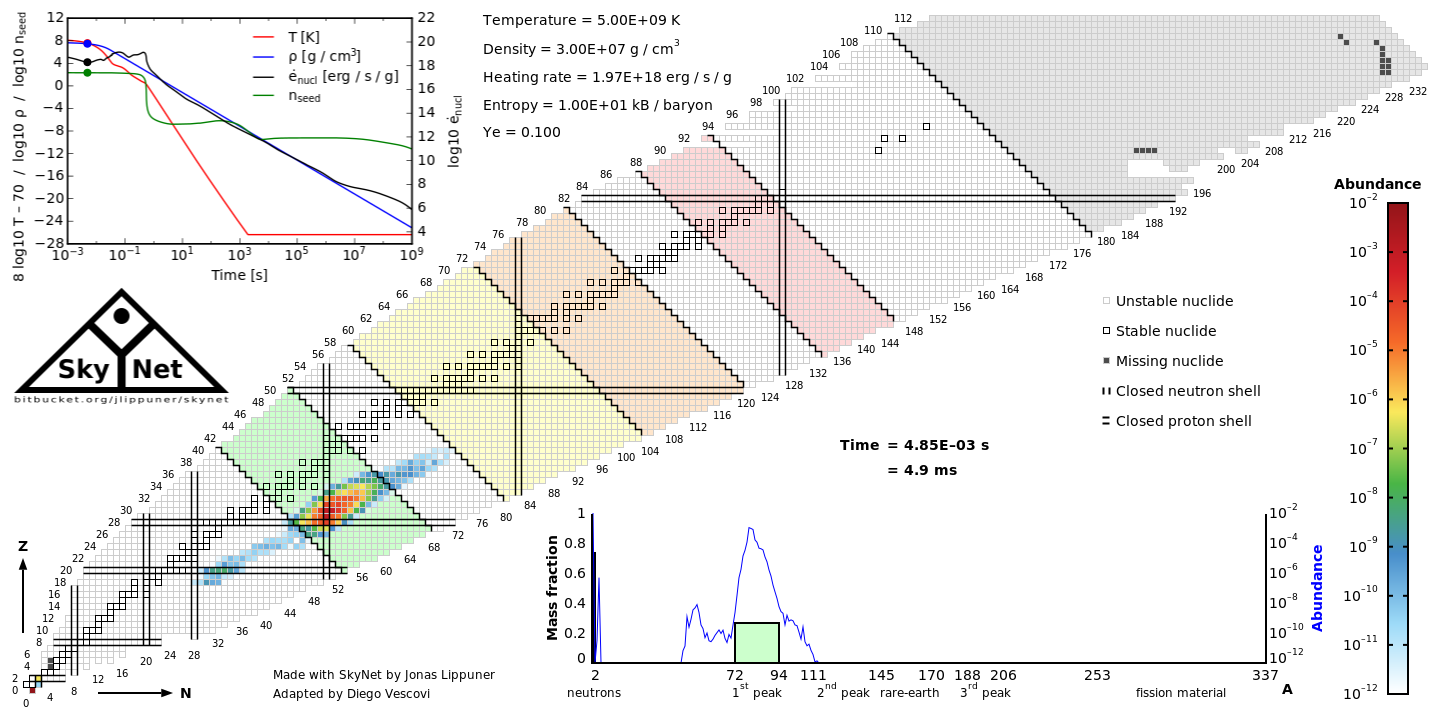}
    \vspace{0.5cm}\\
    \includegraphics[width=\linewidth]{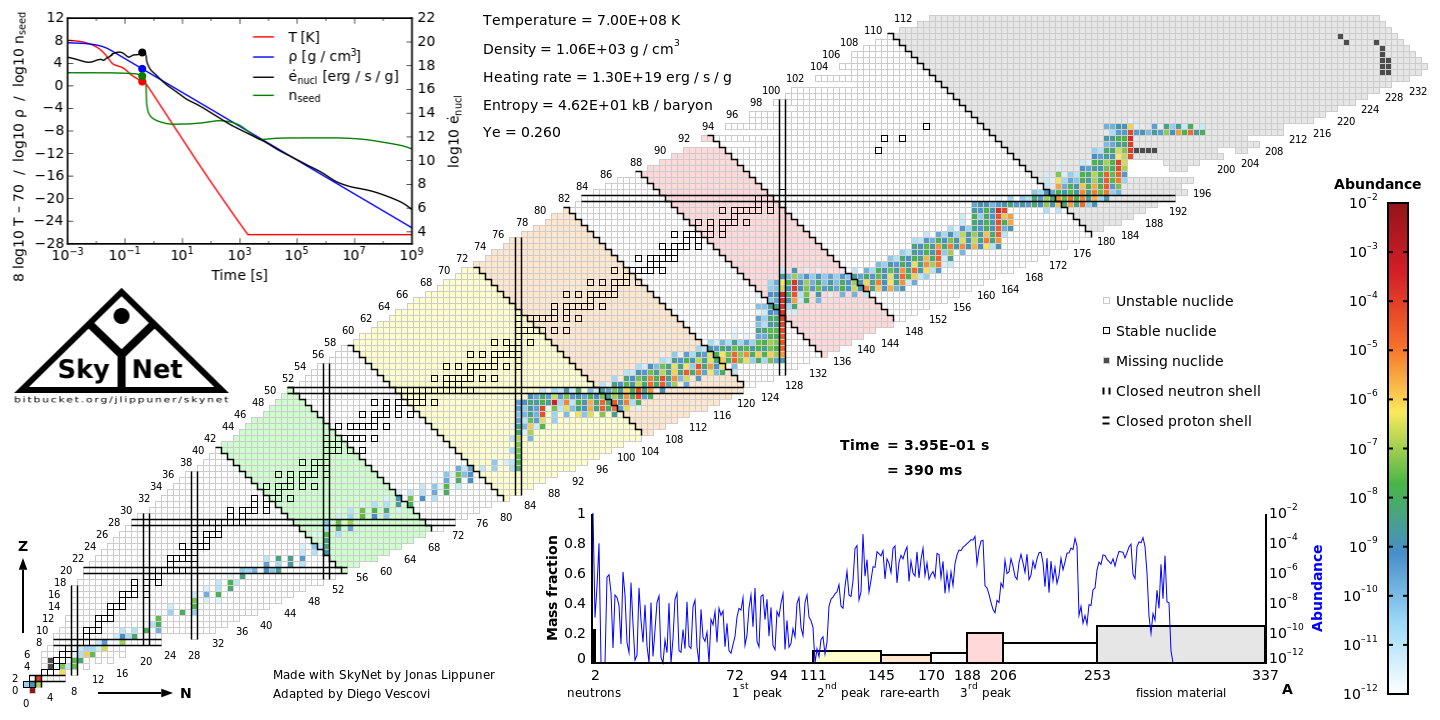}
    \caption{For the same trajectory used for Figure~\ref{fig: skynet trajectory}, we also present detailed nuclear compositions on the nuclear chart. In this two panels we show the abundances at the end of the NSE phase (top panel), and the end of the $(n,\gamma)$-$(\gamma,n)$) equilibrium (bottom panel). In the latter case, the $r$-process path is clearly visible. These pictures were produced using the SkyNet nuclear network \protect\cite{Lippuner.Roberts:2017} and the dedicated visualization software (Courtesy of D. Vescovi).}
    \label{fig: skynet chart 1}
\end{figure}

\begin{figure}[t!]
    \centering
    \includegraphics[width=\linewidth]{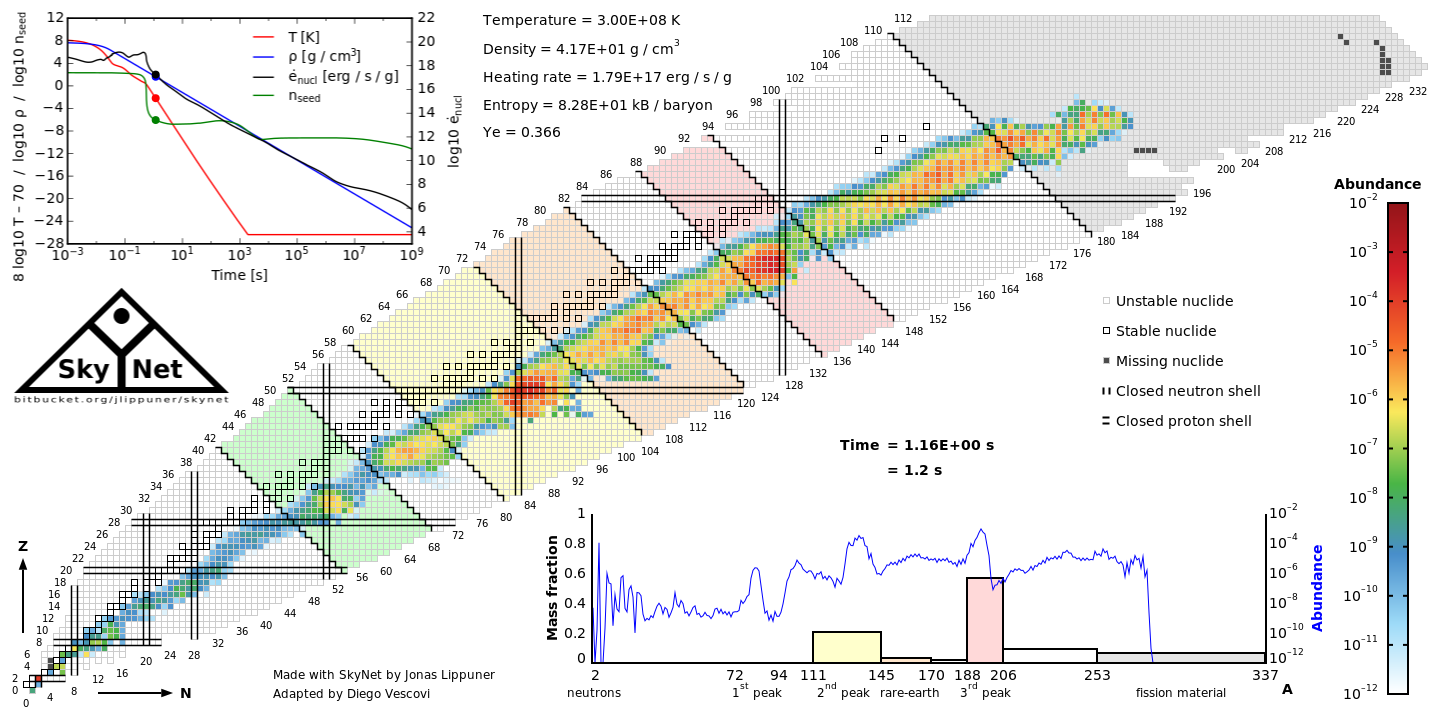}
    \vspace{0.5cm}\\
    \includegraphics[width=\linewidth]{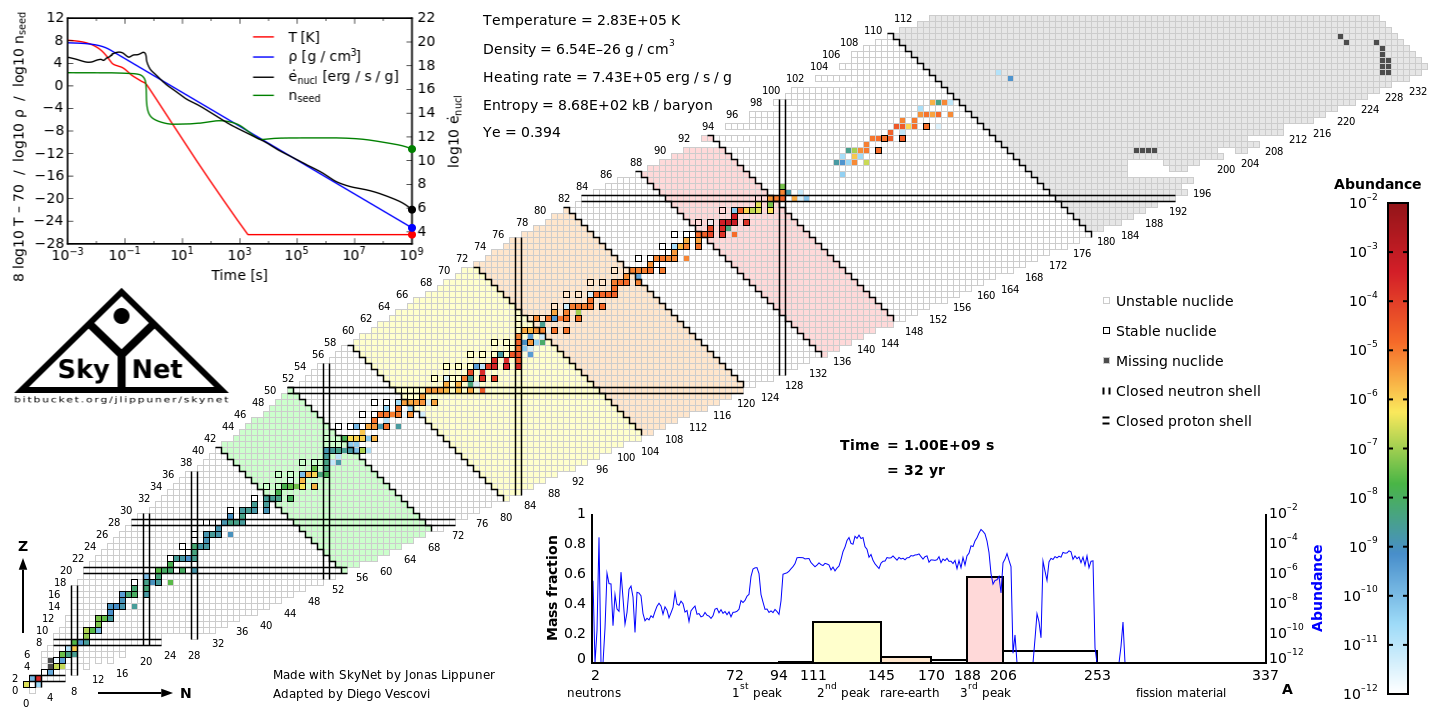}
    \caption{Same as in Figure \ref{fig: skynet chart 1}, but during the decay phase, just after neutron freeze-out (top panel) and at $10^9{\rm s}$ after merger (bottom panel). 
    These pictures were produced using the SkyNet nuclear network \protect\cite{Lippuner.Roberts:2017} and the dedicated visualization software (Courtesy of D. Vescovi).}\label{fig: skynet chart 2}
\end{figure}

The evolution of the nuclear abundances and the calculation of the nuclear energy released during the nucleosynthesis is computed by nuclear reaction networks. A nuclear network is a large system of coupled ordinary differential equations. For each nucleus $i\equiv (A_i,Z_i)$, its abundance $Y_i \equiv Y_(A_i,Z_i)$ evolves according to:
\begin{equation}
    \frac{{\rm d}Y_i}{{\rm d}t} = 
    \sum_{j} \mathcal{N}^{i}_{j} \lambda_j Y_j +
    \sum_{j,k} \mathcal{N}^{i}_{j,k} \frac{\rho}{m_b} \langle \sigma v \rangle_{j,k} Y_j Y_k +    
    \sum_{j,k,l} \mathcal{N}^{i}_{j,k,l} \frac{\rho^2}{m_b^2} \langle \sigma v \rangle_{j,k,l} Y_j Y_k Y_l \, ,
    \label{eq:nuclear network}
\end{equation}
where the sums run over all possible reactions that include $(A_i,Z_i)$ in the initial or in the final state: 
the first sum contains decays, photodisintegrations and semi-leptonic processes, as electron, positron or neutrino captures; the second and the third ones include nuclear fusions with two and three reactants, respectively. In this context, a three-body reaction is a sequence of two-body reactions with an intermediate state with an extremely short lifetime.
The factors $\mathcal{N}^i_{\dots}$ account for multiplicity effects in the case of identical particles: if $N_m$ represents the number of $m$ nuclei involved in a specific reaction with sign (i.e., $N_m>0$ for creation and $N_m<0$ for destruction), then $\mathcal{N}^{i}_{j} = N_i$, $\mathcal{N}^{i}_{j,k} = N_i/(|N_j|!|N_k|!)$ and $\mathcal{N}^{i}_{j,k,l} = N_i/(|N_j|! |N_k|! |N_l|!)$ (for identical reactants, double counting must be avoided so that $N_j +N_k=2 $ and $N_j +N_k + N_l=3 $ for two- and three-body reactions, respectively. For example, for $\alpha+\alpha+\alpha \rightarrow{}^{12}C$, $N_j=3$ and $N_k=N_l=0$).
The calculation of the reaction rates and of the effective decay constants require the knowledge of the evolution of the fluid density and temperature, as well as information about the neutrino irradiation fluxes. 
In actual computations, the evolution of the matter density is usually prescribed or extracted from hydrodynamics simulations. The subsequent evolution of the temperature and $Y_e$ are then self-consistently determined from the detailed abundances, from the EOS of the plasma, and assuming that the expansion proceeds adiabatically, unless for nuclear energy generation and neutrino leakage or irradiation.

We will now describe in more details the most relevant features of the nucleosynthesis happening in a fluid element expanding after a compact binary mergers, according to the outcome of nuclear network calculations. 
For the bulk of the neutron-rich ejecta, characterized by low specific entropy, we can identify four phases:
\begin{itemize}
    \item the initial NSE phase;
    \item the $r$-process nucleosynthesis phase;
    \item the neutron freeze-out phase;
    \item the decay phase. 
\end{itemize}
The different phases and their properties can be observed in Figures \ref{fig: skynet trajectory}, \ref{fig: skynet chart 1} and \ref{fig: skynet chart 2}.

\subsubsection{\textit{The NSE phase}}
The NSE phase (top panel of Figure~\ref{fig: skynet chart 1}) lasts as long as the temperature in the plasma
stays above $T \approx 4$-$5~{\rm GK}$. It is important to notice that a fluid elements can enter and exit the NSE phase several times before eventually decreasing its temperature below the NSE threshold due to hydrodynamics processes and intense nuclear heating. In this case, only the conditions at the last NSE freeze-out influence the subsequent evolution.
During NSE, since $(p,\gamma)$-$(\gamma,p)$ and $(n,\gamma)$-$(\gamma,n)$ reactions are all in equilibrium, the recursive application of the corresponding equilibrium relation among the relativistic chemical potentials, namely $\mu_{(A,Z)} = \mu_p + \mu_{(A-1,Z-1)}$ and $\mu_{(A,Z)} = \mu_n + \mu_{(A-1,Z)}$, yields to $\mu_{(A,Z)} = Z \mu_p  + N \mu_n $.
For particles described by the Maxwell-Boltzmann statistics, once $\rho$, $T$, and the abundances of free $n$'s and $p$'s ($Y_{n,p}$), are provided, the abundance $Y_{(A,Z)}$ of any nucleus $(A,Z)$ in NSE is given by \cite{Clayton:1983}:
\begin{equation}
Y_{(A,Z)} = Y_p^Z Y_n^{(A-Z)} \frac{G_{(A,Z)}(T) A^{3/2}}{2^A} \left( \frac{\rho}{m_b} \right)^{A-1} \left( \frac{2 \pi \hbar^2 }{ m_{\rm b} k_{\rm B}T }\right)^{3(A-1)/2} e^{B(A,Z)/{k_{\rm B}T}} \, ,
\label{eq: NSE abundance}
\end{equation}
where 
$B_{(A,Z)}$ is the nucleus binding energy. $Y_{n,p}$ are ultimately set by requiring baryon conservation and charge neutrality, i.e.
$ 1 = \sum_{(A,Z)} Y_{(A,Z)} A$ 
and $Y_e = \sum_{(A,Z)} Y_{(A,Z)} Z$, respectively.
Very high densities and not too large temperatures favor large nuclei (as it happens in the crust of cold NSs), while photodestruction in hot environments produces light nuclei and ultimately free protons and neutrons (as in the mantle above proto-neutron stars in core-collapse supernovae (CCSNe) or in the cores of merging NSs). For intermediate regimes, the nuclear binding energy favors the most tightly bound nuclei, i.e. iron group nuclei or $\alpha$ particles among the light nuclei, and the temperature regulates the width of the distribution.

For a given plasma configuration (i.e., for a given set of $\rho$, $T$ and $Y_e$) the NSE condition determines the nuclear abundances according to Eq.~(\ref{eq: NSE abundance}) without the need of solving Eqs.~(\ref{eq:nuclear network}). However, the abundances change as a function of time due to the temporal evolution of the expanding and cooling plasma, $\rho = \rho(t)$ and $T = T(t)$. Moreover, weak interactions are out of equilibrium and Eq.~(\ref{eq:nuclear network}) and the charge neutrality conditions reduce to an equation for $Y_e$:
\begin{equation}
\frac{{\rm d}Y_e}{{\rm d}t} = \sum_{i} \left(  \lambda_{e^+,i} - \lambda_{e^-,i}
+ \lambda_{\nu_e,i} - \lambda_{\bar{\nu}_e,i} + \lambda_{\beta^{-},i} - \lambda_{\beta^{+},i} \right) Y_{i} \, ,
\end{equation}
where the sum runs over all nuclei and the rates span all possible (if any) semi-leptonic reactions involving each nucleus $i$.

As visible in the top panel of Figure~\ref{fig: skynet chart 1}, at NSE freeze-out the composition of the expanding plasma is characterized by a distribution of seed nuclei, peaking around the iron group, $\alpha$ particles and free neutrons.
Depending mainly on the initial $Y_e \lesssim 0.4$, the seed nuclei are possibly very or extremely neutron-rich, with $A \sim 60-100$. A high fraction of free neutron is also expected.

If the fluid is radiation dominated, its specific entropy can be approximated by the entropy of the photon gas, see Eq.~(\ref{eq: density photon-dominated gas}).
It is very insightful to substitute this expression inside the expression of the abundances in NSE conditions, Eq. (\ref{eq: NSE abundance}), to express it also in terms of the entropy: 
\begin{equation}
Y_{(A,Z)} \propto Y_p^Z Y_n^{(A-Z)} \frac{A^{3/2}}{2^A~s^{A-1}} \left( \frac{ k_{\rm B}T }{m_{\rm b} c^2}\right)^{3(A-1)/2} e^{B(A,Z)/{k_{\rm B}T}} \, .
\label{eq: approximated NSE abundance}
\end{equation}
Since $Y_{(A,Z)} \propto s^{-(A-1)}$, we recover the result that in the case of matter with high specific entropy an $\alpha$-rich freeze-out is obtained.

\subsubsection{\textit{The $r$-process nucleosynthesis phase}}
Once $T \lesssim 4~{\rm GK}$, the first reactions that run out of equilibrium are charged nuclear reactions involving the less abundant nuclei. 
Under the assumption that neutron captures, photodestructions and $\beta^-$ decays (possibly emitting $j$ delayed neutrons, with $j=0$ being the classical $\beta^{-}$ decay) are the most relevant reactions, Eq.(\ref{eq:nuclear network}) becomes:
\begin{eqnarray}
    \frac{{\rm d} Y_{(A,Z)}}{{\rm d} t} & \approx & 
    n_n \langle \sigma v \rangle_{(A-1,Z)(n,\gamma)(A,Z)} Y_{(A-1,Z)} 
    + \lambda_{(A+1,Z)(\gamma,n)(A,Z)} Y_{(A+1,Z)} + \nonumber \\
    & & - \left[ n_n \langle \sigma v \rangle_{(A,Z)(n,\gamma)(A+1,Z)} 
    + \lambda_{(A,Z)(\gamma,n)(A-1,Z)} \right] Y_{(A,Z)} +  \nonumber \\
    & & + \sum_{j=0}^{J} \lambda_{(A+j,Z-1)\rightarrow (A,Z)+e^- + \bar{\nu}_e + j~n}~Y_{(A+j,Z-1)} + \nonumber \\
        & & - \sum_{j=0}^{J} \lambda_{(A,Z)\rightarrow (A-j,Z+1)+e^- + \bar{\nu}_e + j~n}~Y_{(A,Z)} \, .
    \label{eq:simplified network}
\end{eqnarray}
In the first line, we have considered the creation of $(A,Z)$ nuclei through $(n,\gamma)$ reactions on $(A-1,Z)$ nuclei and $(\gamma,n)$ reactions on $(A+1,Z)$ nuclei.
In the second line, we have considered the destruction of $(A,Z)$ nuclei through $(n,\gamma)$ and $(\gamma,n)$ reactions.
In the third and forth lines, we have taken into account $\beta^-$ decays that can include a $(A,Z)$ nucleus in the final or initial state, respectively, possibly through the additional delayed emission of $0 < j \leq J$ neutrons.

As we have seen in the previous section, if the temperature and density are large enough, the timescale of $\beta^-$ decays are much longer than the timescales of $(n,\gamma)$ and $(\gamma,n)$ reactions. The latter absorption processes connect nuclei among the same isotopic chain ($Z$ is fixed) and their equilibrium ensures an almost steady free neutron fraction, while the former decays connect nuclei of contiguous isotopic chains ($Z-1$ and $Z$, and $Z$ to $Z+1$).
Then, the evolution predicted by Eq.~(\ref{eq:simplified network}) can be split into two separate problems: the much faster $(n,\gamma)$-$(n,\gamma)$ equilibrium inside each isotopic chain; the slower flow through different isotopic chains driven by $\beta^-$ decays.
In the following we will closely analyse each of the two problems separately.

We start by consider an isotopic chain characterized by a specific $Z$. The equilibrium condition inside the chain (i.e. $\mu_{(A+1,Z)} = \mu_{(A,Z)} + \mu_{n}$) allows to write an equation for the abundances of two adjacent nuclei in the chain, starting from Eq.(\ref{eq: chain abundaces}) and simply noticing that $Y_{(A+1,Z)}/Y_{(A,Z)} = n_{(A+1,Z)}/n_{(A,Z)}$. Evaluated the resulting expression for typical magnitudes, we obtain:
\begin{equation}
    \frac{Y_{(A+1,Z)}}{Y_{(A,Z)}} \approx 5.71 \times 10^3
    \left( \frac{n_n}{8 \times 10^{24} {\rm cm^{-3}}} \right)
    \left( \frac{k_B T}{0.1 {\rm MeV}} \right)^{-3/2}
    \left( \frac{\exp{\left( S_n/k_{\rm B} T \right) }}{\exp{(30)}} \right) \, .
    \label{eq: abundance in ng-gn equilibrium}
\end{equation}
We notice that the abundance ratio depends only on $n_n$, $T$ and on $S_n$. The latter introduces a dependence on the nuclear masses.
Also in this case, the numerical value of this ratio is dominated by the exponential factor whose argument compares the neutron separation energy with the plasma temperature. Close to the valley of stability $S_n \approx 15{\rm MeV} \gg k_{\rm B} T$ and $Y_{(A+1,Z)} \gg Y_{(A,Z)}$, i.e. abundances increase steeply moving toward neutron richer nuclei. Assuming $k_{\rm B}T = 0.1 ~{\rm MeV}$ and $S_n = 3 {\rm MeV}$, the above estimate still implies $Y_{(A+1,Z)} \gg Y_{(A,Z)}$. For $S_n = 2 {\rm MeV}$, the numerical prefactor decreases to 0.26, meaning that $Y_{(A+1,Z)} < Y_{(A,Z)}$. Since $S_n \rightarrow 0$ at the neutron drip line, there exists always a turning point in the isotopic chain, where $Y_{(A+1,Z)}$ stops to increase with respect to $Y_{(A,Z)}$ before decreasing when approaching the neutron drip line. 
We estimate this point by requiring $Y_{(A+1,Z)}/Y_{(A,Z)}=1$. Assuming a certain $n_n$ and temperature $T$, this condition translates in a reference neutron separation energy $S_{n}^0$:
\begin{eqnarray}
    S_{n}^0 & = & k_{\rm B} T \ln{\left( \frac{2}{n_n} \left( \frac{m_{\rm B} k_{\rm B}T}{2 \pi \hbar^2} \right)^{3/2} \right) } \approx 2.14~{\rm MeV} \left( \frac{k_{\rm B}T}{0.1~{\rm MeV}}\right) \times \nonumber \\ 
    &  & 
    \left[ 1 - 0.047~\ln \left( \frac{n_n}{8 \times 10^{24} {\rm cm^{-3}}}\right) + 0.070~\ln \left( \frac{k_{\rm B}T}{0.1{\rm MeV}} \right) \right] \, .
    \label{eq: S_n^0}
\end{eqnarray}
The value of $S_{n}^0$ depends on $n_n$ and $T$, but not on $Z$. So, at any given time, all populated isotopic chains have their abundance peaks at nuclei characterized by the same $S_{n}^0$ and the conditions $S_{n}(A,Z) \gtrsim S_{n}^0$ defines the $r$-process path.
A better approximation can be done by considering that, due to nuclear paring effect, even neutron numbers are favored. Indeed, while $S_{n}$ decreases for increasing $D$, but with a even-odd modulation, $S_{2n}$ defined as $S_{2n}(A+2,Z) = (m_{A,Z} + 2 m_n - m_{(A+2,Z)})c^2 $ and called the \textit{two neutron separation energy}, decreases in a smoother way, with a sudden decrease at the magic neutron numbers.
By considering the effective equilibrium $(A+2,Z) + \gamma \leftrightarrow (A,Z)+2n$ and the corresponding relation on the chemical potentials, $\mu_{(A+2,Z)}= \mu_{(A,Z)} + 2 \mu_n$, the $r$-process path can be defined as the set of nuclei belonging to different isotopic chains for which $S_{2n}(A,Z)$ is closer to $2 S^0_{n}(n_n,T)$ along each specific chain. 

Since the relative distribution of abundances inside an isotopic chain is set by the fast $(n,\gamma)$-$(\gamma,n)$ equilibrium and determined by Eq.~(\ref{eq: abundance in ng-gn equilibrium}), it is useful to consider the total abundance along the chain, $Y_{Z} \equiv \sum_{A} Y_{(A,Z)}$. 
Starting from Eq.~(\ref{eq:simplified network}) and considering only $\beta^-$ decays without delayed neutron emission ($J=0$), we obtain an evolution equation for $Y_{Z}$:
\begin{equation}
    \frac{{\rm d} Y_{Z}}{{\rm d} t} \approx \tilde{\lambda}_{\beta,(Z-1)} Y_{Z-1} - \tilde{\lambda}_{\beta,Z} Y_{Z}
\end{equation}
where $\tilde{\lambda}_{\beta,Z}$'s are the effective, abundance-weighted $\beta$ decay rates of the whole chains: 
\begin{equation}
\tilde{\lambda}_{\beta,Z} \equiv \left( \sum_A \lambda_{\beta,(Z,A)} Y_{(A,Z)} \right) / \left( \sum_A Y_{(A,Z)} \right) \, .     
\end{equation}
Under the assumption that the amount of free neutrons and the temperature are high enough, the duration of the $r$-process becomes comparable or even larger than the $\beta$ decay lifetime (Eq.~\ref{eq: tau_beta estimate}) over the most relevant part of the nuclear chart, and the chain abundances $Y_Z$ tend toward an equilibrium configuration, i.e. ${\rm d} Y_{Z}/{\rm d} t \approx 0 $, which translates into:
\begin{equation}
    \tilde{\lambda}_Z Y_{Z} = \tilde{\lambda}_{Z-1} Y_{Z-1} \approx {\rm const} \, .
\end{equation}
This condition is knows as \textit{steady $\beta$-flow approximation} \cite{Kratz.etal:1993,Freiburghaus.etal:1999a} and it implies that the abundance of a chain is proportional to its effective $\beta$ decay lifetime. 
Since the nuclei with magic neutron numbers $N=50,82,126$ (or just above them) and closer to the valley of stability have the longest $\beta$ decay lifetime, it is expected that maxima in the abundances will occur at the top end of the kinks in the $r$-process path corresponding to the neutron shell closures. This is what quantitatively defines the location and the relative importance of the waiting points inside the $r$-process path.

The duration of the $r$-process nucleosynthesis phase crucially depends on the availability of free neutrons. Under the assumption that neutron captures and photodestructions are the most relevant reactions, the timescale over which the neutron abundance changes, $\tau_n$, can be evaluated from Eq.\ref{eq:simplified network} as:
\begin{equation}
    \frac{1}{\tau_n} \equiv \frac{1}{Y_n} \left| \frac{{\rm d}Y_n}{{\rm d}t} \right| \approx \frac{1}{Y_n/Y_{\rm seed}} \left( \frac{1}{\tau_{(n,\gamma)}} - \frac{1}{\tau_{(\gamma,n)}} \right) \, .
    \label{eq: neutron timescale}
\end{equation}
The relevant timescales $\tau_{(n,\gamma)}$ and $\tau_{(\gamma,n)}$ can be estimated through Eqs~(\ref{eq:tau_n_gamma estimate}) and (\ref{eq: tau_g_n estimate}) as we did in the previous section, but they can also be computed more rigorously from the reaction rates that enter Eq.(\ref{eq:nuclear network}):
\begin{eqnarray}
    \frac{1}{\tau_{(n,\gamma)}} \equiv \left( \sum\limits_{(A,Z)} Y_{A,Z}~n_n~\langle \sigma v \rangle_{(A,Z)(n,\gamma)(A+1,Z)} \right)/ \sum\limits_{(A,Z)} Y_{(A,Z)} \, , \\
    \frac{1}{\tau_{(\gamma,n)}} \equiv \left( \sum\limits_{(A,Z)} Y_{A,Z}~\lambda_{(A,Z)(\gamma,n)(A-1,Z)} \right) / \sum\limits_{(A,Z)} Y_{(A,Z)} \, .
    \label{eq: tau n g network}
\end{eqnarray}
As long as the neutron-to-seed ratio is large, $\tau_n$ is relatively long and there is enough time for many isotopic chain (with increasing $Z$) to be reached and for the $\beta$-flow equilibrium to establish.
Detailed calculation show that, for typical conditions in compact binary merger ejecta, this $r$-process phase can last up to $\sim$1 second after the merger (see for example the bottom panel of Figure~\ref{fig: skynet chart 1}).
Eventually, the $r$-process path can reach the neutron magic number $N=184$. Above that, fission becomes the dominant nuclear process. 
Typical super-heavy, neutron-rich nuclei have $(A,Z) \sim (250,100) $, while daughter nuclei can be approximated by distribution around the double-magic number nucleus $(A_{1,2},Z_{1,2}) \sim (132,50)$.
Thanks to fission and depending on the lighter fission fragment distribution, the abundances of heavy elements double and specific region of the nuclear chart start to populate. If the initial neutron richness is high enough (in particular, when the initial $Y_e \lesssim 0.1$) fission becomes relevant while $(n,\gamma)$ and $(\gamma,n)$ reactions are still in equilibrium. The lighter nuclei resulting from the fission still capture many neutrons, reaching again $N=184$ and many fission cycles occur \cite{Korobkin.etal:2012,Mendoza.etal:2015,Lippuner.Roberts:2015}.

The possibility of reaching $(n,\gamma)$-$(\gamma,n)$ equilibrium and its duration are critically related with the temperature evolution. For a radiation dominated plasma, if the expansion proceeded adiabatically the temperature would simply evolve as $T(t) \approx T_{\rm hom} (t_{\rm hom}/t)$, with $T_{\rm hom}$ of the order of a few GK. Then, within $\sim$ 0.1 s, the temperature would drop below 1GK. 
While this relation describes in good approximation the evolution of matter temperature before and after the $r$-process, it is rather inaccurate \textit{during} it. Indeed, during the $r$-process the energy released by nuclear reactions and the matter density are often large enough such that matter is significantly re-heated and temperature becomes again larger than 1 GK for a much longer time (up to $\sim$ 1 s), before dropping again as predicted by the adiabatic expansion law, $T \propto t^{-1}$. To estimate if the specific nuclear heating $\dot{e}_{\rm nucl}$ (nuclear energy per unit mass, per unit time) can affect significantly the matter temperature, we compare the density of energy of the radiation field with the energy released by the nuclear decays during a time comparable to the expansion timescale $\tau$, i.e. $\pi^2 k_{\rm B}^4 T_{\rm max}^4/ 15 (\hbar c)^3 \sim \dot{e}_{\rm nucl} \rho \tau $,
and we estimate $\dot{e}_{\rm nucl} \sim S_n Y_n / (m_b \Delta t_{r{\rm -proc}}) $ \cite{Mendoza.etal:2015}.
By solving for $T_{\rm max}$, we can estimate the maximum temperature reached by matter due to the nuclear heating. Assuming that around $t \approx 0.1 {\rm s}$ the density has decreased down to $10^5{\rm g~cm^{-3}}$ and the $(n,\gamma)$-$(\gamma,n)$ equilibrium lasts for $\Delta t_{r{\rm -proc}} \sim 1~{\rm s}$:
\begin{equation}
    T_{\rm max} \approx 0.75~{\rm GK} 
    \left( \frac{Y_n~\rho}{0.5 \times 10^5~{\rm g~cm^{-3}}} \right)^{1/4}
    \left( \frac{S_n}{5~{\rm MeV}} \right)^{1/4}
    \left( \frac{\tau}{10~{\rm ms}} \right)^{1/4}
    \left( \frac{\Delta t_{r{\rm -proc}}}{1~{\rm s}} \right)^{-1/4}   \, . 
    \label{eq: max temp neutron capture phase}
\end{equation}
When $T_{\rm max} \gtrsim 0.7~{\rm GK}$ the temperature is large enough to guarantee $(n,\gamma)$-$(\gamma,n)$ equilibrium during most of the neutron captures. This is defined as the \textit{hot r-process}. If, on the other hand, $T_{\rm max} \lesssim 0.3~{\rm GK}$, the density is large enough to sustain the neutron captures, but not $(n,\gamma)$-$(\gamma,n)$ equilibrium. In this latter case, the neutron density decreases much faster due to neutron consumption and this is referred as \textit{cold r-process}. 
It is important to stress again that the relevant $r$-process regime (hot or cold) does not depend on the absolute peak temperature, but on the maximum temperature during the re-heating phase occurring when seed nuclei capture neutrons.
Other factors, as for example the expansion timescale, determined if the $r$-process happens in hot or cold conditions.

\subsubsection{\textit{The neutron freeze-out and the decay phases}} 
As the $r$-process proceeds, and matter expands and cools below $\sim$1GK, $\tau_{(\gamma,n)} \gg \tau_{(n,\gamma)}$ and photodestruction becomes inefficient in keeping a high neutron density outside the heavy nuclei, while the still effective neutron captures and $\beta^-$ decays produce more and more heavy nuclei. 
Once $Y_n/Y_{\rm seed} \sim 1$, according to Eqs.~(\ref{eq:tau_n_gamma estimate 2}) and (\ref{eq: neutron timescale}), it happens that $\tau_n \sim Y_n/Y_{\rm seed} \tau_{(n,\gamma)} \ll 1 {\rm s}$: the free neutron lifetime suddenly decreases because the seed nuclei start to compete for the few available neutrons. This phase is called \textit{neutron freeze-out} and it is characterized by a sudden drop of the neutron density and of the neutron-to-seed ratio, as visible both and in Figure~\ref{fig: skynet trajectory} and in the the top panel of Figure~\ref{fig: skynet chart 2}. 
After this drop, the $\beta^-$ decays (often followed by the emission of free neutrons) start to compete with the neutron capture, since $\tau_{(n,\gamma)}$ becomes comparable to $\tau_{\beta}$. 
The latter can be estimated as Eq.(\ref{eq: tau_beta estimate}), but also as:
\begin{equation}
    \frac{1}{\tau_{\beta}} \equiv \left( \sum\limits_{(A,Z),~A > 4} \sum\limits_{j=1,J} Y_{(A,Z)}
    \lambda_{(A,Z) \rightarrow (A-j,Z+1)+e^{-} + \bar{\nu}_e + jn} \right) /
    \left( \sum\limits_{(A,Z),~A > 4} Y_{(A,Z)} \right) \, .
    \label{eq: tau beta network}
\end{equation}
The presence of freshly emitted free neutrons provides a new source of neutrons available to be captured also at later time. The result is that, while the nuclei that were located along the $r$-process path during the $(n,\gamma)$-$(\gamma,n)$ equilibrium decay collectively toward the valley of stability, the competition between the neutron captures and the $\beta$ decays smooths the $r$-process abundances. In particular, it removes the strong oscillations in the mass number that characterize the abundances just before neutron freeze-out, due pairing and collective effects in the nuclear properties. 

After a timescale ranging from a few up to a few tens of seconds (depending on the environment and increasing for decreasing initial $Y_e$)
the neutron density has decreased such that also the neutron captures become negligible. Most of the matter is still in the form of unstable nuclei, with neutron excess $D \sim 5$-$10$. 
At this stage $\beta^{-}$ decays and, depending on how extended the $r$-process path is, $\alpha$ decays are the most relevant nuclear processes that bring the abundances toward the valley of stability. 
While most of the nuclei have reached stable configurations within a few tens of days, a few heavy isotopes have longer half-lives, extending above $10^7$ yrs for 14 isotopes and above $10^9$ yrs for 7 isotopes. 
Since all these reactions liberate nuclear energy, they can heat up matter or become a distinct source of  radiation. The detection of the various, distinct signals is thus a clear indication of the $r$-process nucleosynthesis occurring inside the ejecta.

\subsubsection{\textit{The $r$-process peaks and the s-process nucleosynthesis}}

During the $(n,\gamma)$-$(\gamma,n)$ equilibrium the $r$-process path is characterized by the presence of waiting points, where matter accumulates due to the longer lifetimes occurring in correspondence of the neutron magic numbers $N=50,82,126$.
Waiting points are identified by neutron close shells, but they can span a large range of $Z$'s and they can also be influenced by the presence of the proton magic numbers $Z=28,50$ (see the bottom panel of Figure~\ref{fig: skynet chart 1}).
Since at neutron freeze out nuclei are still very neutron-rich (typically $Z/A \sim 0.5$), these waiting point nuclei are characterized by  $A \sim 80,130,195$. When decaying back to stability, nuclear abundances proceed along $A \approx {\rm const}$ paths, i.e. they keep their mass number approximately constant while increasing their $Z$. However, the residual delayed neutron emissions and neutron captures can further change $A$ by a few units during the process. Due to the decreasing neutron density, the effect due to $\beta$ decays eventually takes over.
The result is that the final abundance pattern is characterized by three peaks (known as the \textit{r-process peaks}): the first peak is located around $A \approx 80$ and covers the range $ A \approx 72-94$, the second around $A \approx 130$ with range $A \approx 111-145$, and the third around $A \approx 195 $ with range $A \approx 188-206$. Due to the $\beta$ decays, the nuclei neutron content has globally decreased, the $Y_e$ has increased toward $\lesssim 0.4$. As a function of $Z$, the peaks are located around $Z \approx 35$ (e.g. Selenium, Bromine, Krypton) for the first peak, $Z \approx 53$ (e.g. Tellurium, Iodine, Xenon) for the second peak, and $Z \approx 77 $ (e.g. Osmium, Iridium, Platinum) for the third peak. Thus, due to the combination of multiple neutron captures followed by many $\beta$ decays, the final peaks are shifted toward the left by several units with respect to the nuclei characterized by magic neutron numbers along the valley of stability.

However, when considering the element above Zn, the abundances that we observe inside the solar system reveal, close to the $r$-process peaks, other peaks shifted toward the right by a few units in the mass number and happening precisely in correspondence of the magic neutron numbers. Then, differently from what happens in the $r$-process, the production of these peaks must make place close to the valley of stability, in such a way that a neutron capture on a stable or long-lived nucleus is followed by a $\beta$ decay before another neutron can be captured, meaning that $t_{(n,\gamma)} \gg t_{\beta}$. 
This is possible if the neutron densities are in the range $10^{6-11}{\rm cm^{-3}}$, so several orders of magnitudes lower than the ones required by the $r$-process. This kind of nucleosynthesis is called slow neutron capture process ($s$-process) and it is thought to happen inside low and intermediate mass stars (starting their life with a mass between 0.6 and 10 \msun) during their asymptotic giant branch (AGB) phase.
Starting from iron group nuclei produced in a previous SN explosion and already present inside the star, the $s$-process proceeds through the $s$-process path up to lead on a timescale of several thousands of years.
The abundances of the $s$-process nucleosynthesis are well understood since they rely on the well-known properties of nuclei at or very close to the valley of stability. Thus, the solar $r$-process abundances are defined as the residual of the solar abundances, once the $s$-process contribution has been removed, see e.g. \cite{Prantzos.etal:2020}.
Most of the heavy elements receive a contribution from both the $s$- and the $r$-process. It is important here to stress that information about the isotopic composition is in general harder to obtain, while astrophysical observations often provide information on the elemental composition. So, an element usually receives contributions from different isotopes and these isotopes can be synthesised in different environments. Nevertheless the structure of the valley of stability is such that some stable isotopes can only be produced in the $s$-process or in the $r$-process. In the first case, this happens when a stable isotopes is shielded by another stable isotopes along a $\beta$ decay line. In the second case, this happens when a stable nucleus can be reached by a neutron capture and a subsequent $\beta$ decay sequence only from a nucleus that has a very short lifetime or that is outside the $s$-process path. 
When this happens for all or for the most relevant isotopes of an element, this element is called $s$-only or $r$-only element, respectively.
A $r$-only isotope has the advantage of allowing the study the contribution of the $r$-process independently from possible contamination from the $s$-process.
Europium and elements above lead (including Uranium, Thorium and Plutonium) are among the $r$-only elements. 

\subsubsection{\textit{Nucleosynthesis in high entropy and fast expanding ejecta}}

Not all the ejecta are characterized by low or moderate entropy, or by expansion timescales ranging from a few ms up to several tens of ms. 
In particular, detailed merger models show the presence of high entropy and/or extremely fast expanding tails in the ejecta distribution.
Moreover, it is not always guaranteed that the peak temperature will be above 4GK or that the temperature will be high enough during the $r$-process phase to ensure $(n,\gamma)$-$(\gamma,n)$ equilibrium.
The nucleosynthesis in these conditions can be significantly different than the one described above.

For example, if the expansion timescale $\tau$ becomes of the order of (or even smaller than) 1 ms, the re-heating during the $r$-process phase could become inefficient and $(n,\gamma)$-$(\gamma,n)$ equilibrium cannot establish, see Eq.(\ref{eq: max temp neutron capture phase}). In this case, a cold $r$-process happens. Moreover, despite the possible large abundance of free neutrons and the availability of seed nuclei, the decrease in the neutron density is so fast that neutron captures become inefficient too early and the $r$-process does not proceed much. A large fraction of the free neutrons are not captured by nuclei and decay into protons. The resulting abundances are very different from the one observed in a full $r$-process.
According to Eq.(\ref{eq: expansion timescale VS velocity}), for the dynamical ejecta emerging from BNS mergers, this is the case for matter expanding at $v\sim 0.6-0.8 c$. Such ejecta are observed in some models in the high speed tail, especially if the EOS of nuclear matter is rather soft and the shocks produced by the NS collision are very violent.

Finally, if the specific entropy is large enough ($s \gtrsim 80~k_{\rm B}{\rm baryon^{-1}}$), the NSE ends with a $\alpha$-rich freeze-out (see e.g. \cite{Just.etal:2015} and reference therein).
If, additionally, $Y_e \lesssim 0.4$, there is a significant abundance of free neutrons that are not bound inside the very abundant $\alpha$ particles and inside the fewer seed nuclei. Thus, in these conditions the neutron-to-seed ratio is still very high and the $r$-process nucleosynthesis can still occur to produce heavy $r$-process nuclei. The fundamental different is that the final abundances are overall dominated by the $\alpha$ particles emerging from the NSE freeze-out, see e.g. \cite{Qian.Woosley:1996,Hoffman.etal:1997}.

\subsection{\textit{Nuclear physics input and detailed network calculations}}

Equilibrium argument are very useful to get a qualitative understanding of the main features of the $r$-process nucleosynthesis.
However, only the numerical solution of Eqs.(\ref{eq:nuclear network}) provides detailed nucleosynthesis yields. 
In nuclear network calculations abundances are typically initialized in NSE conditions at high temperatures ($T \sim 6-8{\rm GK}$) and then evolved consistently following the chemical evolution of the ejecta during their expansion.
For $r$-process nucleosynthesis calculations more than 7000 nuclei are necessary, ranging from free neutrons and protons up to very neutron-rich transuranic elements (e.g. Curium), and covering the whole neutron-rich side of the valley of stability.
The reaction rates needed by the network require a vast amount of nuclear physics knowledge. In the following, we briefly discuss the most relevant inputs, namely the nuclear masses, the $\beta$-decay half lives, the neutron captures rates and the fission physics. For more detailed information we address the reader to Ref~\cite{Cowan.etal:2019}. 

\textbf{Nuclear masses.} The most basic nuclear property are the masses of all nuclei, since they determines the threshold energies for all relevant reactions and, in particular, the neutron separation energies. While the masses of stable nuclei and of nuclei close to the valley of stability are experimentally well known, only theoretical values predicted by nuclear mass models are available for exotic neutron-rich nuclei. Nuclear mass models are tuned on experimentally known masses and then extrapolated, so that the uncertainties grow moving toward the neutron drip line and just above the shell closure points, where correlations and deformation may be very relevant, but difficult to be taken into account. 

\textbf{Beta decay rates.} Also the values of $\beta$ decay half-lives are experimentally unknown far from the valley of stability and theoretical calculations are necessary. Their values are crucial, since they determine the matter flow between different isotopic chains. In particular, the most relevant decay rates are those of nuclei at the neutron magic numbers, since they are the waiting points of the $r$-process path. The calculation of these decay rates dependes on the reaction $Q$-value (and thus on the mass model), on the transition strength and, in particular, on its energy distribution. If the transition leads to a final state whose energy is above the neutron separation energy, the emission of one or more neutrons in the final state ($\beta$-delayed neutron emission) is possible. This emission is relevant after the neutron freeze-out phase. Also in this case, shell effects and nucleon correlations are possibly very relevant but difficult to model.

\textbf{Neutron capture rates.}
While neutron captures are essential to establish $(n,\gamma)$-$(\gamma,n)$ equilibrium, the corresponding QSE condition makes the detailed knowledge of their rates not very relevant in modeling the $r$-process. It is after neutron freeze-out, when the temperature drops below 1 GK and the photodisintegration becomes less and less relevant, that a detailed knowledge of the relevant neutron capture rates becomes more important. At this stage, nuclei are still very exotic and theoretical calculations are once again necessary. They are usually computed within the statistical model of nuclei, which is done using the Hauser-Feshbach approach and modeling the nuclear energy density, the $\gamma$-strength function for the decay of the compound nucleus, and the potentials of several light particles.  

\textbf{Nuclear fission.}
The fission properties of super-heavy nuclei ($A \lesssim 280$) are very relevant in calculations involving very neutron-rich matter (at least for initial $Y_e \lesssim 0.10$), but they are very uncertain. The theoretical description of fission is highly non-trivial, and it mainly depends on the fission barriers, defined as the energy required for a nucleus to undergo fission.
This energy is necessary to deform the nucleus such that the transition to the fission fragments states becomes energetically favored. If the transition to the final state does not require any additional energy, the fission is spontaneous. If additional energy is provided by the interaction or the absorption of another particle, then the fission is induced. All possible fission channels (e.g., spontaneous, neutron-induced, $\gamma$-induced) and all the other competing reactions (e.g. $\alpha$ and $\beta$ decays) need to be included in the model to provide reliable predictions.
In the case of a very neutron-rich environment, as the one typically expected in the ejecta of compact binary mergers, neutron-induced fission is usually the most relevant channel.
The very uncertain fission fragment distributions are also very important as they influence the abundances around $A \lesssim 140$.
{~}\\

While the basic features of the $r$-process nucleosynthesis in the ejecta of compact binary mergers are robust and, especially for low entropy ejecta, mainly influenced by the initial $Y_e$, the nuclear input physics can significantly change the fine structure of the abundance pattern. 
For example, the $\beta$ decay half-lives of nuclei with $Z \gtrsim 80$ regulate the mass flow toward the magic neutron number $N=126$ (possibly enhancing the amount of nuclei that go though fission cycles) and possibly affect the position of the third $r$-process peak \cite{Eichler.etal:2015,Mendoza.etal:2015}. Moreover, since fission fragments are located around $A \lesssim140$, fission physics can influence the width of the second $r$-process peak, as well as the abundances of the rear-earth elements, between Lanthanium and Lutetium \cite{Goriely.etal:2013}.

\subsection{\textit{Detailed network calculations
and nucleosynthesis yields from compact binary mergers}}

\begin{figure}[t!]
    \centering
    \includegraphics[width=0.49\linewidth]{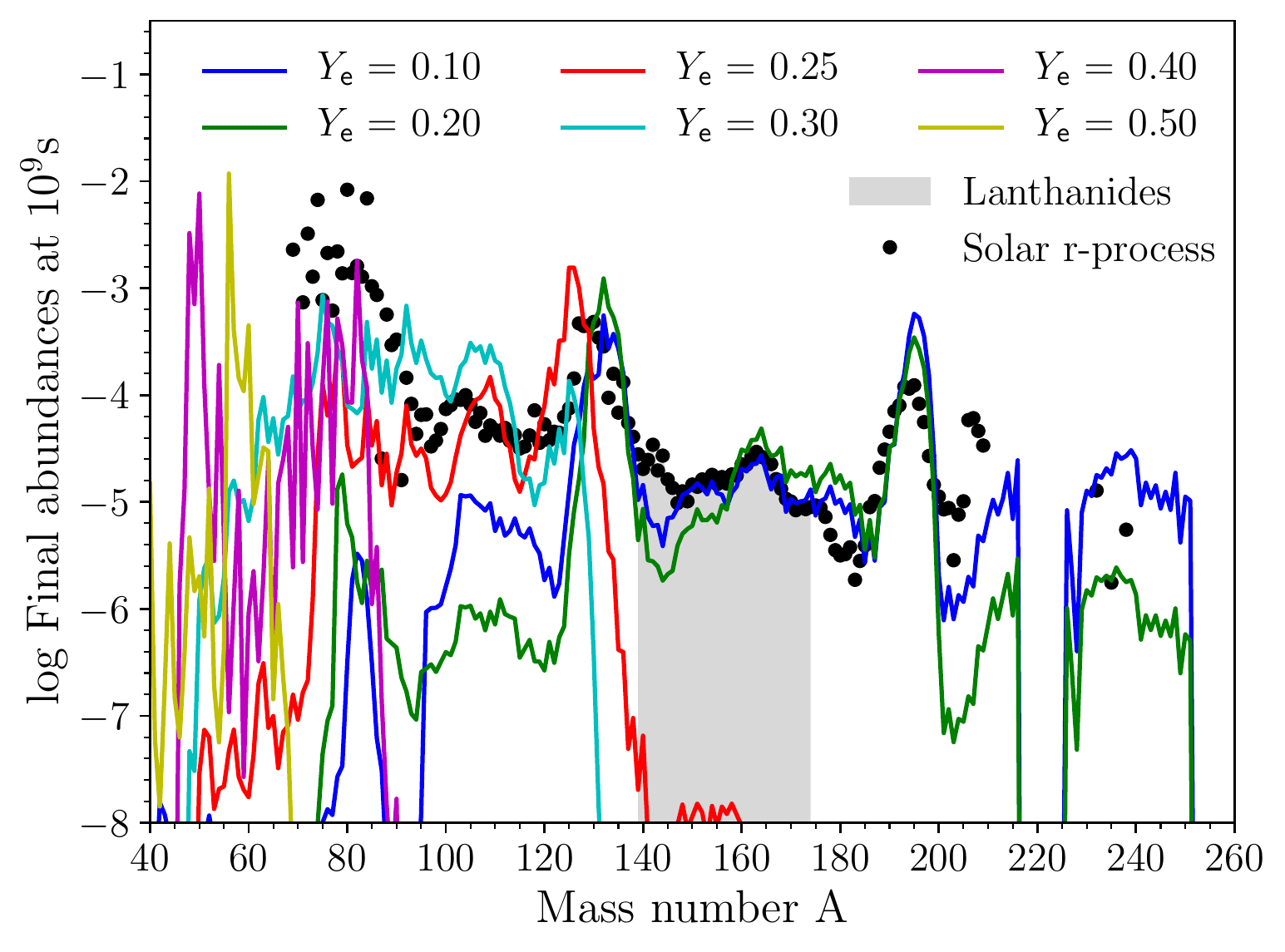}
    \includegraphics[width=0.49\linewidth]{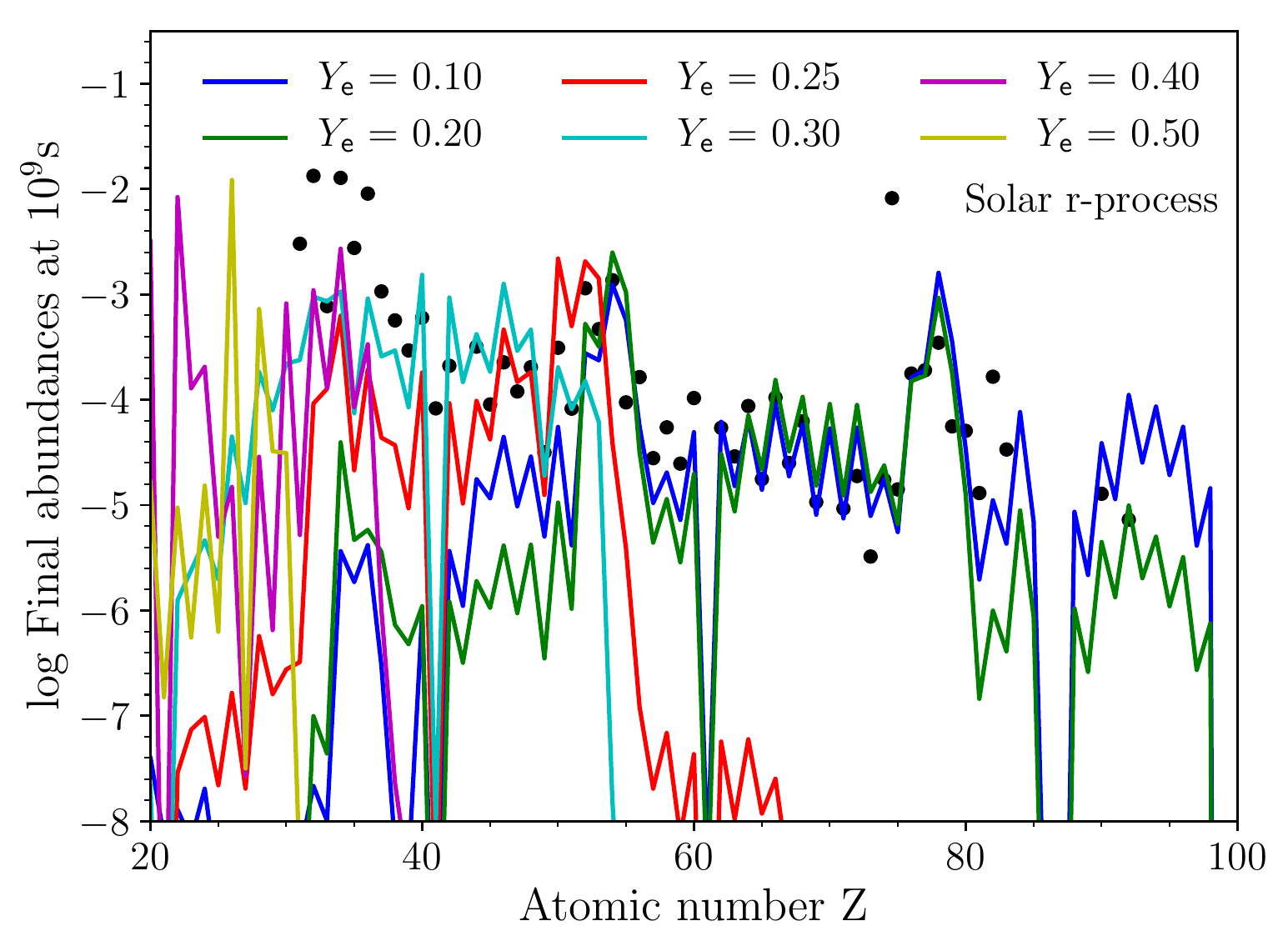}    
    \caption{Abundances as a function of the mass number ($A$, left panel) and of the atomic number ($Z$, right panel) at 10$^9$ seconds after merger for trajectories characterized by $s \approx 11k_{\rm B}{\rm baryon^{-1}}$ and $\tau \approx 11~{\rm ms}$, but for different initial $Y_e$'s, computed using the SkyNet nuclear network \protect\cite{Lippuner.Roberts:2017}. Black dots represent the Solar $r$-process residual, as reported by \protect\cite{Prantzos.etal:2020} (Courtesy of D. Vescovi).}
    \label{fig: skynet ye abundances A}
\end{figure}

Once all the relevant nuclear input physics has been considered, nuclear network calculations can predict with accuracy the distribution of the yields synthesised by a specific fluid elements. 
Nuclear network calculations for fluid elements expanding adiabatically show that the nucleosynthesis outcome depends mainly on the three parameters that we have used to described the properties of the ejecta at the onset of the homologous expansion phase, namely the electron fraction, the specific entropy and the expansion timescale (see, e.g., \cite{Lippuner.Roberts:2015} for a systematic study).
As representative cases for the abundances obtained at $\sim 30{\rm y}$ after merger, in Figures~(\ref{fig: skynet ye abundances A}) 
we present the abundance patterns obtained by detailed nuclear network calculations. 
All trajectories have the same specific entropy and expansion timescales ($s\approx 11k_{\rm B}{\rm baryon^{-1}}$ and $\tau \approx 11~{\rm ms}$), but differ because of their initial electron fraction. 
In all cases, we compare the calculation outcome with the solar $r$-process residual. For $Y_e \lesssim 0.2$ all elements between the second and the third $r$-process peaks are synthesized with a pattern that well reproduces the solar one. Here the observed differences in the shape of the rare-earth peak could be mostly ascribed to the nuclear physics input. This nucleosynthesis outcome is often referred as the "strong $r$-process". For very low electron fractions ($Y_e \lesssim 10$) the nucleosynthesis proceeds up to Ur and Th, and the abundance pattern is very robust, due to the extremely long $r$-process path and to the effect of several fission cycles, while for $0.1 \lesssim Y_e \lesssim 0.2$ the production of actinides decreases. Around $Y_{e,{\rm crit}} \approx 0.23-0.24$, the neutron-to-seed ratio at NSE freeze out is only of a few, the production of nuclei above the second $r$-process peak is inhibited, and the abundances move toward the first $r$-process peak for increasing $Y_e$. This nucleosynthesis outcome is sometimes called "weak $r$-process".


\subsubsection{\textit{Nucleosynthesis yields from compact binary mergers}}

Detailed abundances in the ejecta from compact binary mergers reflect the ranges and the relative relevance of the distributions of entropy, expansion timescale and electron fraction that emerge from the merger dynamics. Different abundance patterns characterize the different ejection channels and keep an imprint of the merger nature, dynamics and aftermath.

\begin{figure}[t!]
    \centering
    \includegraphics[width=0.49 \linewidth]{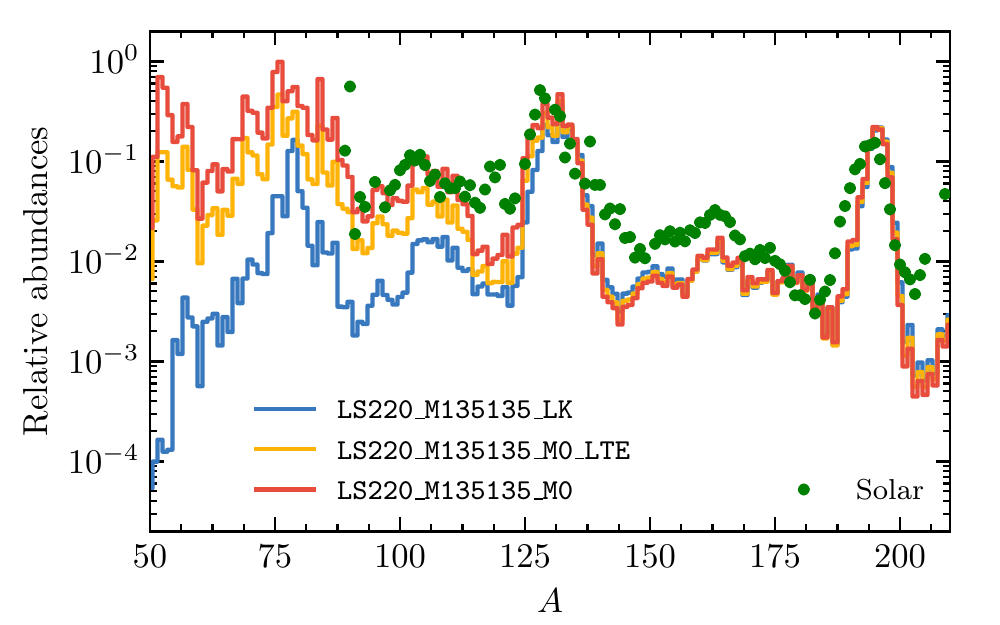}
    \includegraphics[width=0.49 \linewidth]{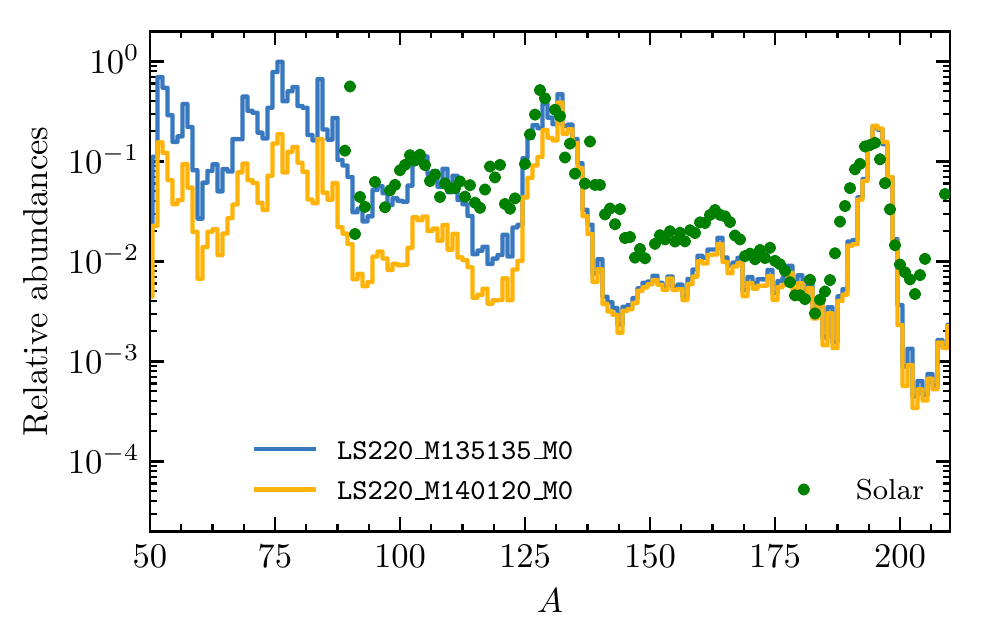}
    \caption{Mass integrated abundances in the dynamical ejecta obtained by simulations of BNS merger models. In the left panel, the different curves represent different neutrino treatment,with the blue curve not including neutrino irradiation, while the other two including it.
    In the right panel, neutrino irradiation is always accounted for, but the different curves show the difference between an equal (blue) and an unequal (yellow) mass mergers.
    Figures taken from \protect\cite{Radice.etal:2018}.}
    \label{fig: dynamical ejecta abundances}
\end{figure}

\begin{figure}[t!]
    \centering
    \includegraphics[width=0.49 \linewidth]{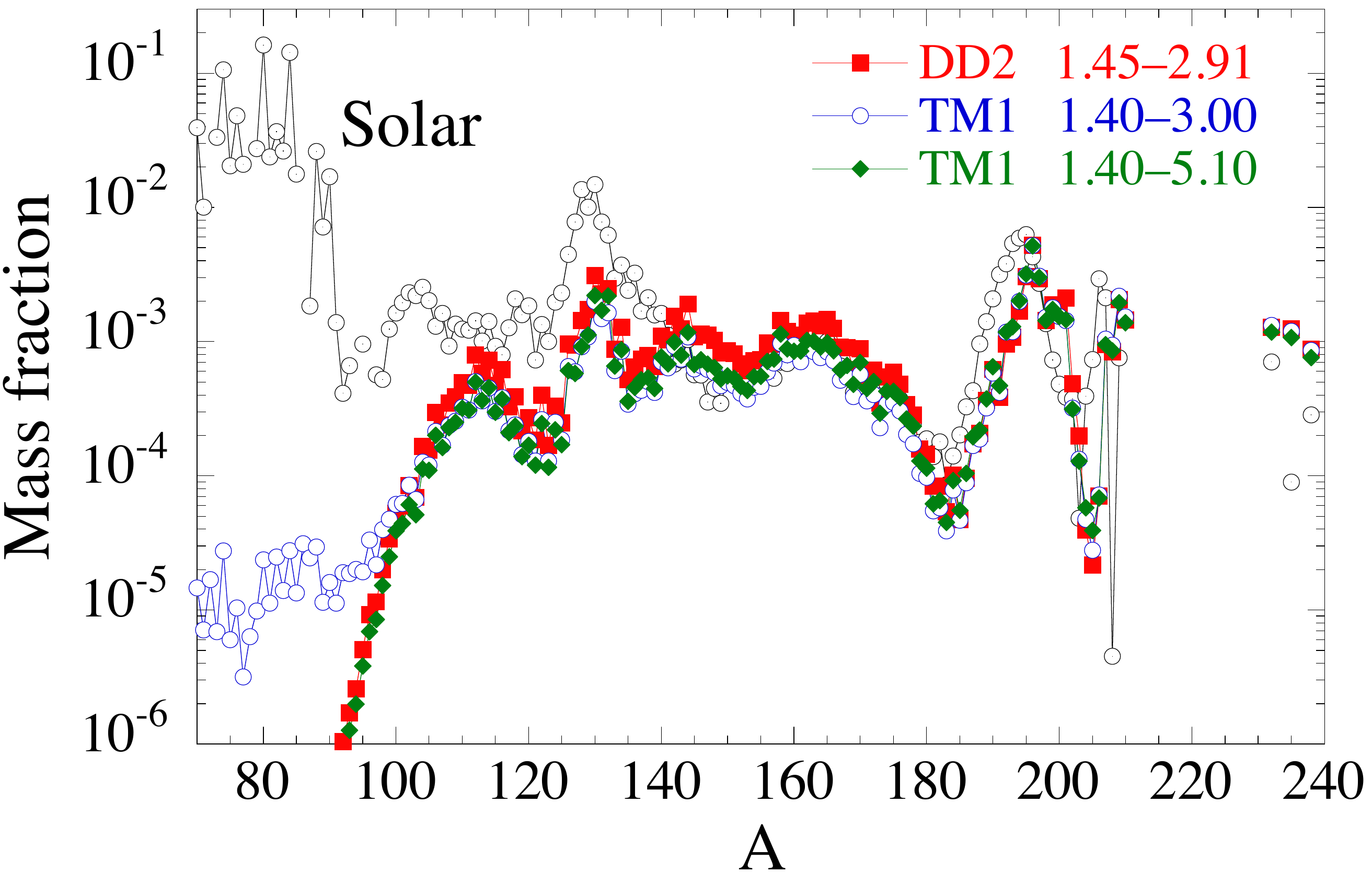}
    \includegraphics[width=0.49 \linewidth]{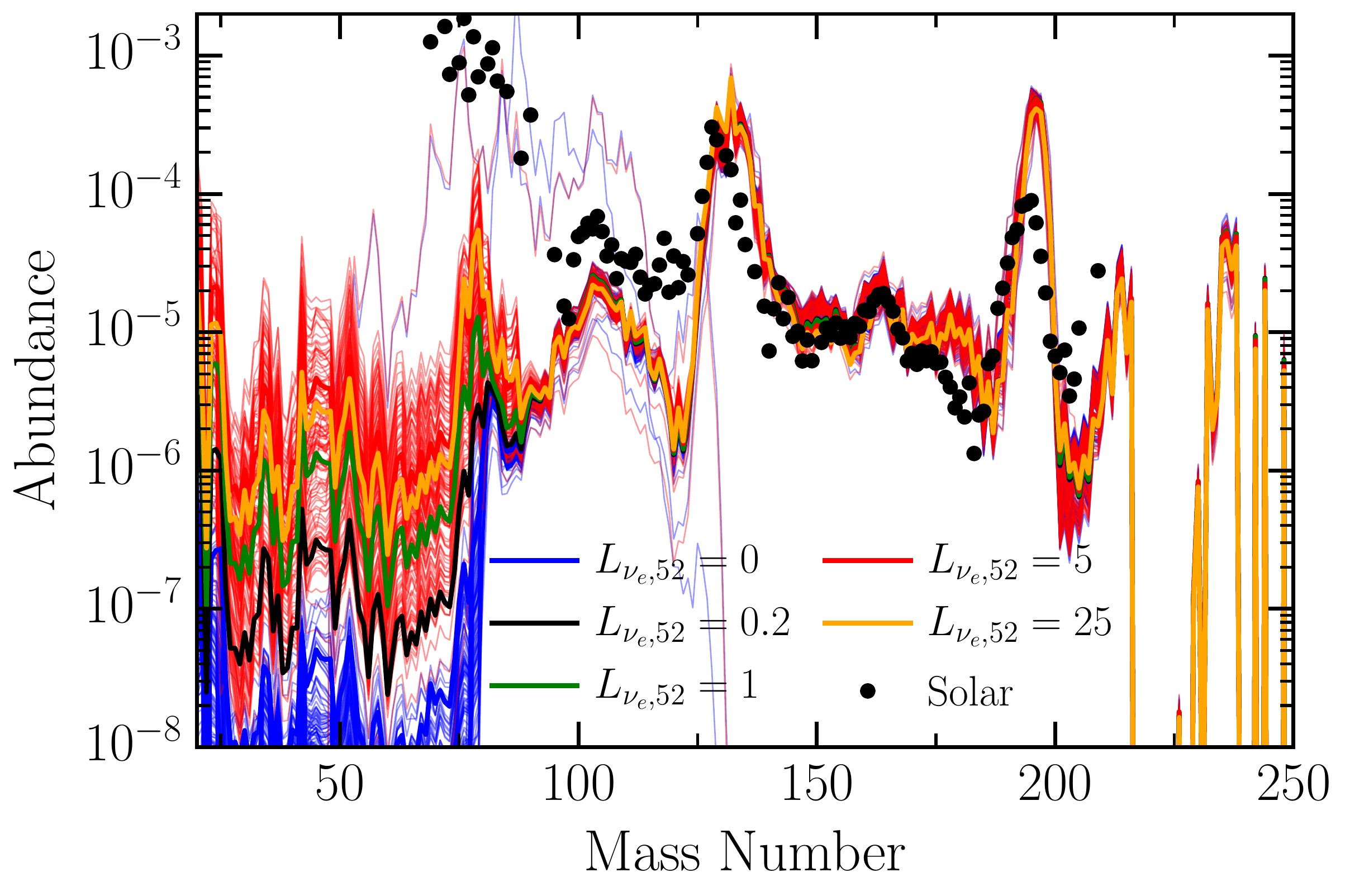}
    \caption{Mass integrated yields in the dynamical ejecta obtained by simulations of BHNS merger. In the left panel, mass fractions are presented for different combinations of NS and BH masses. In the right panel, the different curves represent abundances obtained for different intensities of the neutrino luminosities. Figures taken from \protect\cite{Just.etal:2015} (left panel) and \protect\cite{Roberts.etal:2017} (right panel).}
    \label{fig: dynamical BHNS ejecta abundances}
\end{figure}

\textbf{Dynamical ejecta}. The low-entropy, equatorial dynamical ejecta have $Y_e \lesssim 0.2$ and produce robustly all elements between the second and the third $r$-process peak. In addition, if $Y_e \lesssim 0.1$, significant abundances of translead nuclei and actanides are produced in relative abundances that can be comparable or even larger than the solar ones. This is the case especially for BHNS mergers and BNS mergers characterized by very different NS masses.
In the case of BNS mergers of comparable NS masses, at polar latitudes the entropy increases but stays on average below 20 $k_{\rm B} {\rm baryon^{-1}}$, while $Y_e \gtrsim 0.25$ and the production of heavy $r$-process elements is suppressed. In this case, the $r$-process nucleosynthesis does not proceed up to very high mass numbers, but produces elements from the first up to the beginning of the second peak.
The relative importance of the two contributions (polar and equatorial) to the dynamical ejecta depends on the binary mass ratio and on the properties of the nuclear EOS. However, the equatorial component is expected to be always present and overall dominant. Representative results for mass integrated abundances in the dynamical ejecta of BNS and BHNS simulations are presented in Figures~\ref{fig: dynamical ejecta abundances} and \ref{fig: dynamical BHNS ejecta abundances} and they show that the dynamical ejecta can produce a significant fraction of all $r$-process nuclei from the first to the third peaks, see e.g.~\cite{Wetal:2014,Radice.etal:2018}.

\begin{figure}
    \centering
    \includegraphics[width=0.54\linewidth]{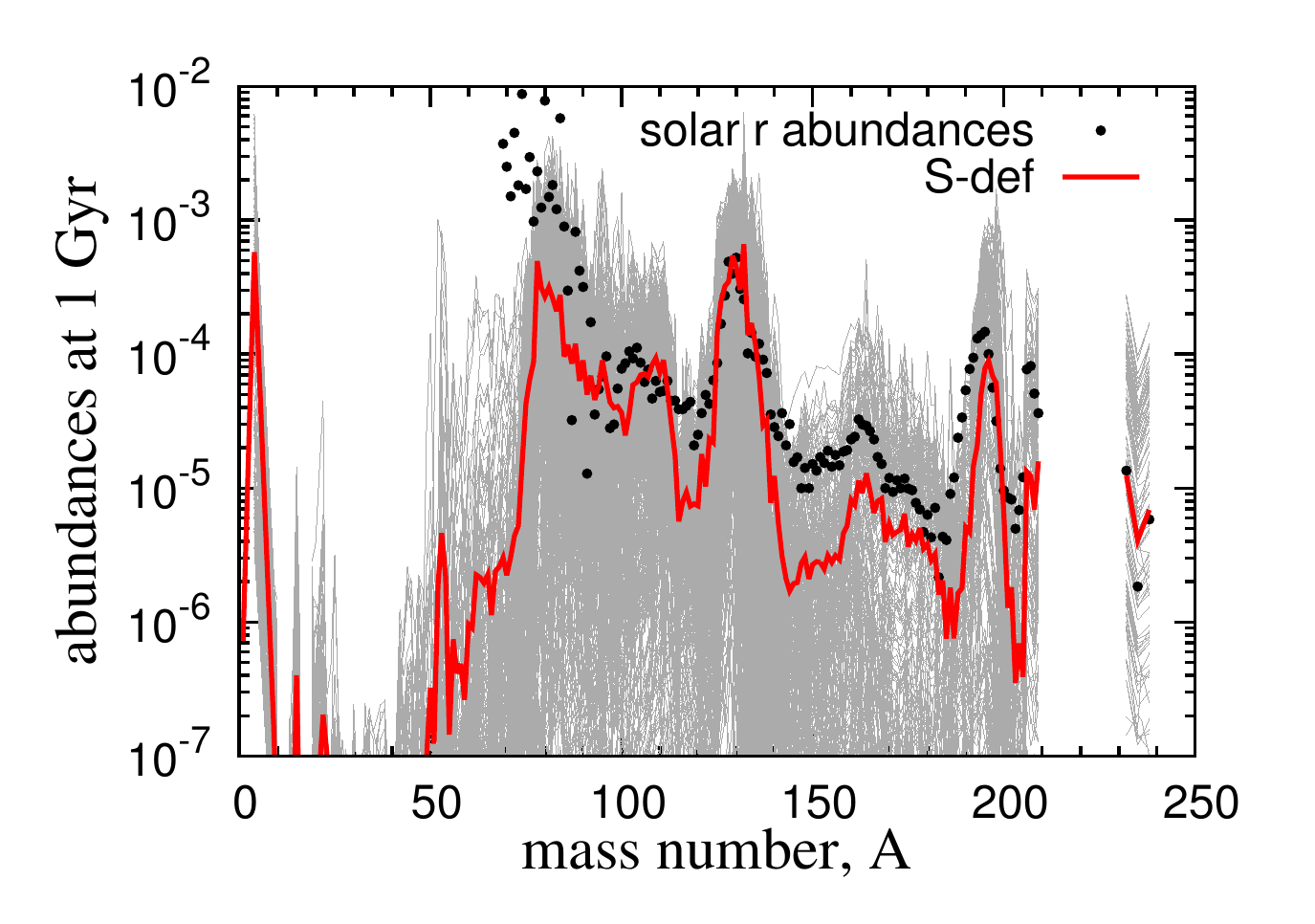}        
    \includegraphics[width=0.45\linewidth]{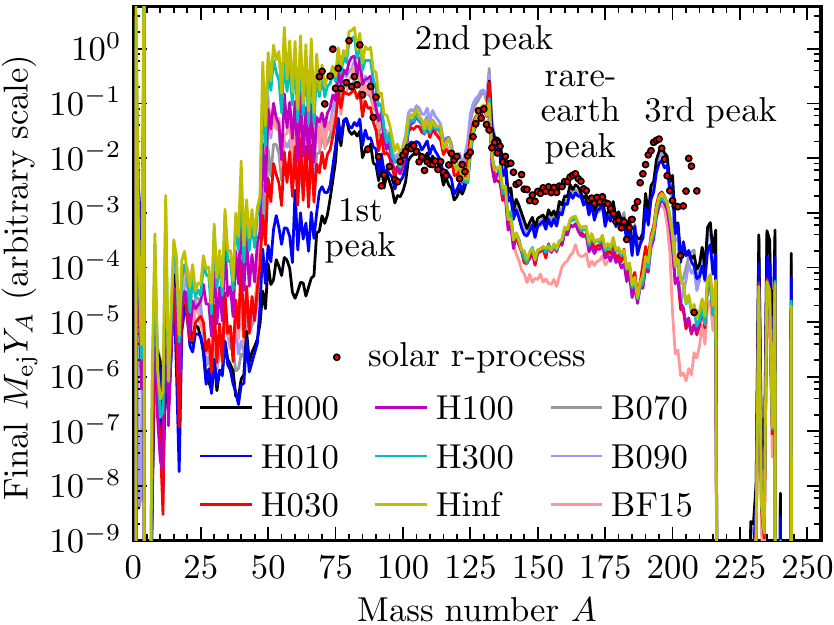}
    \caption{Mass integrated nucleosynthesis results (as a function of the mass number $A$) for the disk-wind ejecta obtained in the case of a BH-disk systems (left panel) and of a massive NS remnant (right panel, with different colors corresponding to different NS lifetimes). 
    Figures taken from \protect\cite{Wu.etal:2016} (left panel) and \protect\cite{Lippuner.etal:2017} (right panel).}
    \label{fig: disk-wind ejecta abundances}
\end{figure}

\textbf{Disk wind ejecta from BH-torus systems}. For a BHNS merger or for a BNS merger whose the central remnant collapses to a BH within a few dynamical timescales, the ejecta from disk winds are dominated by the viscous component, have low entropy and cover a broad range of electron fractions across the critical value $Y_{e,{\rm crit}} \approx 0.23-0.25 $.
In these conditions, the production of all heavy elements between the first and the third $r$-process peak is expected
(see e.g. \cite{Just.etal:2015,Wu.etal:2016}). Due to the smaller effect of neutrino irradiation, the synthesis of elements between the second and the third $r$-process peaks is significant while the production of the first peak elements is below the solar ratio.
The angular distribution of the abundances is in this case rather insensitive to the latitude: high-$Y_e$ matter at polar latitude possibly synthesize only elements below the second $r$-process peak, but this $\nu$-driven component is very subdominant for BH-torus systems, due to the lower neutrino luminosities.
Nucleosynthesis results for the viscous ejecta from BH-torii systems are presented in the left panel of Figure~\ref{fig: disk-wind ejecta abundances}.

\textbf{Disk wind ejecta from systems hosting a massive NS}. If the massive NS resulting from the merger of two NSs is not a short-lived one, a larger variability is expected, reflecting the presence of several components in the disk winds and the impact of neutrino irradiation \cite{Lippuner.etal:2017}.
If the remnant survives for a timescale comparable to the disk lifetime ($t_{\rm acc}$, see Eq.(\ref{eq: disk timescale})), neutrino irradiation progressively shift the $Y_e$ distribution of the viscous ejecta toward larger values $\gtrsim Y_{e,{\rm crit}}$, compared to the BH-torus system.
While the production of both light and heavy $r$-process elements is forseen, the relative importance of the former with respect to the latter increases as a function of the remnant lifetime. 
The presence of a very long-lived massive NS (collapsing on a timescale $\gg t_{\rm acc}$) could possibly prevent the formation of the second peak and of all the elements above it in the bulk of the ejecta, producing a significant amount of light $r$-process elements.
The presence of a massive NS causes also disk wind ejecta emerging from distinct portions of the solid angle to be characterized by different nucleosynthesis patterns.
As in the case of the dynamical ejecta, matter expelled at high (polar) latitudes is more efficiently irradiated by neutrinos and for it the production of heavy $r$-process elements at and beyond the second $r$-process peak becomes soon inefficient, while elements characterized by $75 \lesssim A \lesssim 120$ and $33 \lesssim Z \lesssim 55$ are eventually produced in solar proportions even for ejecta expelled within the first few tens of milliseconds \cite{Perego.etal:2014,Martin.etal:2015}.
Nucleosynthesis yields obtained for the neutrino- and the viscosity-driven ejecta from merger remnant hosting a massive NS in the center are presented in the right panel of Figure~\ref{fig: disk-wind ejecta abundances}.\\

\textbf{High entropy, high velocity tails.} In the case of shock-heated ejecta and ejecta expelled from strong magnetic fields in torii around BHs, a high entropy tail is observed in the ejecta distribution. In the first case, the ejecta also expand with a high velocity, such that $\tau \lesssim 2{\rm ms}$. In this case, assuming $Y_e \lesssim 0.4$ (but often $Y_e < 0.3$) the $r$-process nucleosynthesis proceeds in $\alpha$-rich freeze-out conditions and it synthesises a significant fraction of nuclei up to lantanides ($Y_e \lesssim 0.4 $) or even actinides ($Y_e \lesssim 0.2 $). Differently from the low entropy, neutron rich ejecta, these matter present also a significant amount of H and He in the final composition.


\section{\textit{Observables of compact binary merger nucleosynthesis}}

Binary compact mergers have long been thought as promising astrophysical sites for $r$-process nucleosynthesis. While the expected conditions of the ejecta and the outcome of detailed numerical models indicate that the production of heavy $r$-process elements above the iron group is robust, observational evidences are crucial to validate our models and to discriminate between the many theoretical uncertainties that still affect our theoretical understanding. 

In the following we will discuss two major observables and their relation with the outcome of $r$-process nucleosynthesis in compact binary mergers: the electromagnetic transient called kilonova and the evolution of the chemical abundances of $r$-process elements in the stars of our Galaxy and in its satellites.

\subsection{\textit{Electromagnetic signatures of $r$-process nucleosynthesis in compact binary mergers}}

\subsubsection{\textit{What is a kilonova?}}

Starting from a few seconds and continuing for several hundreds of days after merger, a large amount of nuclear energy is released by the combination of $\beta$ decays, $\alpha$ decays and fission processes that follow the $r$-process nucleosynthesis in the ejecta of a compact binary merger. This energy heats up the expanding matter and produces a nuclear powered transient called kilonova \cite{Li.Paczynski:1998}.
Depending on the ejecta properties, a kilonova is expected to have its peak luminosity between a few hours and several days after merger in the UV/optical/near-IR frequences, with a fast declining luminosity. For a detailed treatment of this transient we refer to the dedicated Chapter and to recent reviews, e.g. \cite{Fernandez.Metzger:2016,Metzger:2019,Cowan.etal:2019}.

The nuclear energy associated with each decay is distributed among the daughter nuclei and the other particles in the final state (electrons, photons and neutrinos), in a way that primarily depends on the nature of the decay. Typical $Q$-values are of the order of 1-100 MeV, much larger than the decreasing matter temperature in the expanding ejecta. 
Indeed, if the temperature at the end of the $r$-process nucleosynthesis ($t_{r-{\rm proc}} \sim 1~{\rm s}$) is 
$k_{\rm B} T_{r-{\rm proc}} \sim 0.1 {\rm MeV}$, then:
\begin{equation}
k_{\rm B}T(t) \sim k_{\rm B}T_{r-{\rm proc}} 
\left( \frac{t}{t_{r-{\rm proc}}} \right)^{-1} 
\approx 1.16~{\rm eV} 
\left( \frac{k_{\rm B}T_{r-{\rm proc}}}{0.1~{\rm MeV}} \right)
\left( \frac{t_{r-{\rm proc}}}{1~{\rm s}} \right)
\left( \frac{t}{1~{\rm d}} \right)^{-1} \, .
\label{eq:kilonova temperature}
\end{equation}
While the energy emitted in neutrinos is always lost, several processes can thermalize at least a fraction of the released nuclear energy, making it available for the kilonova.
The intensity of the nuclear heat and the efficiency of the thermalization processes are maximal immediately after the $r$-process nucleosynthesis. However matter is initially very opaque to photons and the timescale for photon diffusion is much larger than the dynamical timescale over which the ejecta expand. The photon optical depth depends on the matter density profile and on the photon opacity. It is thus necessary to wait for the density to drop such that thermal photons can efficiently diffuse and be emitted at the photosphere.

\subsubsection{\textit{$r$-process nucleosynthesis and kilonovae}}

If this is the mechanism behind kilonovae, $r$-process nucleosynthesis can influence these transients mainly through three aspects: the nuclear heating, the thermalization efficiency and the photon opacity. In the following we will analyze the main features of each of them.

\textbf{Nuclear heating}. On the kilonova timescale nuclear abundances in the ejecta change in time according to a set of decays and fission reactions. 
Each reaction is characterized by an exponential behavior, $N_i(t) = N_{i,0}\exp(-\lambda_i t)$ where $N_{i}(t)$ is the number of parent nuclei at time $t$ and $\lambda_i$ the reaction rate.
The specific nuclear heating rate as a function of time can be computed as: 
\begin{equation}
\dot{e}_{\rm nucl}(t) = 
\frac{\sum_{i} Q_i \lambda_i N_i(t)}{m_{\rm ej}} \approx
\dot{e}_{e^{-}}(t) +
\dot{e}_{\nu}(t) +
\dot{e}_{\alpha}(t) + 
\dot{e}_{\gamma}(t) + 
\dot{e}_{\rm fission}(t) \, ,
\label{eq:general heating rate}
\end{equation}
where $Q_i$ is the $Q$-value of each reaction $i$ and $m_{\rm ej}$ is the ejecta mass. In the last step we have explicitly indicated the decay particles whose kinetic energy provide the available nuclear energy, neglecting the kinetic energy of the daughter nuclei in the case of $\alpha$ and $\beta$ decays.
Detailed network calculations (e.g., \cite{Metzger.etal:2010,Korobkin.etal:2012,Lippuner.Roberts:2015}) show that the dominant contribution to $\dot{e}_{\rm nucl}$ can be approximately described by a power-law dependence in time, with possible corrections in the form of exponential terms:
\begin{equation}
\dot{e}_{\rm nucl}(t) \approx \dot{e}_{\rm nucl,0}~(t/t_0)^{-\alpha} + \sum_{i=1}^{N} \beta_i \exp(-\gamma_i~t) \, .
\label{eq:heating parametrization}
\end{equation}
with $t_0$ being a reference timescale.
The precise values of $\dot{e}_{\rm nucl,0}$, $\alpha$, $\beta_i$ and $\gamma_i$ depend on the initial ejecta properties, i.e. on $(Y_e,s,\tau)$ of the specific trajectory, and on the nuclear input physics. For low entropy ejecta, $Y_e$ is the dominant parameter. 
For $Y_e \lesssim 0.25$,
$\alpha \approx 1.3$ with a possible variation interval $1.1-1.4$, while different mass models give $\dot{e}_{\rm nucl,0} \sim 10^{16}-10^{17}{\rm erg~s^{-1}~g^{-1}}$ (assuming $t_0=1~{\rm s}$) \cite{Rosswog.etal:2017}. This is mainly due to the fact that, especially for very low $Y_e$, translead nuclei tend to decay trough $\alpha$ decays and fission, which are very sensitive to the specific mass model \cite{Barnes.etal:2016,Rosswog.etal:2017,Wu.etal:2019}.  
The dominant power-law term in Eq.~(\ref{eq:heating parametrization}) can be understood by considering that in most of the cases (and especially for $Y_e \lesssim 0.2$, when strong $r$-process nucleosynthesis occurs) a large statistical ensemble of nuclei is produce \cite{Metzger.etal:2010}. As explained in \cite{Metzger.etal:2010}, assuming $\beta^-$ decay to be the dominant decay channel, along an isotopic chain (i.e. for fixed $Z$), the $Q$-value is roughly proportional to the neutron excess $D$.  Since $\lambda_\beta \propto Q^{5}$, then $\lambda_\beta \propto D^{5}$.
The nuclei at the end of the $r$-process distribute along the $r$-process path, characterized by $S_n \approx S_n^0$, see Eq.(\ref{eq: S_n^0}). 
According to the $S_n$ distribution on the nuclear chart, at high $N$ a larger $D$ is required to fulfill the $r$-process path condition and, for a given time $t$, the nuclear distribution is such that the number of nuclei per interval of $D$ is approximately constant within the relevant interval of neutron excess:
\begin{equation}
\frac{{\rm d} N}{{\rm d} D} \approx {\rm const} \rightarrow
\frac{{\rm d} N}{{\rm d} \lambda} = \frac{{\rm d} N}{{\rm d} D} \frac{{\rm d} D}{{\rm d} \lambda} = C \lambda^{-4/5} \, ,
\end{equation}
with $C$ being a constant.
Due to the large number of reactions and nuclei involved, Eq.(\ref{eq:general heating rate}) can be converted in an integral over $\lambda$:
\begin{equation}
    \dot{e}_{\rm nucl} \propto \int_0^{+\infty} 
    \lambda^{1/5} \lambda \exp{(- \lambda t)} \lambda^{-4/5} {\rm d}\lambda \propto t^{-7/5} \, .
\end{equation}
While this argument explains well the presence of the power-law and provides a good estimate of its slope, many details can affect the precise calculation, accounting for the lower exponent obtained in detailed calculations.

The power law term in Eq.(\ref{eq:heating parametrization}) is very robust for ejecta characterized by $Y_e \lesssim 0.25$ and undergoing strong $r$-process nucleosynthesis. For $Y_e \gtrsim 0.25$, the decay of a few nuclei dominate the heating rate and deviations from the power-low behavior are accounted by the exponential terms in Eq.(\ref{eq:heating parametrization}). These nuclei (including for example neutron-rich isotopes of Kr, Rb, Br, Sr, Ce, As, and Ge, see e.g. \cite{Martin.etal:2015}) are usually below the second $r$-process peak and are characterized by large $Q$-values and half-lives of a few hours. Thus the heating rate is enhanced by a factor of a few within the first day and decreased by a factor $\sim 2$ at later time with respect to the $Y_e \lesssim 0.25$ case.
However, since the exponential corrections depend significantly on the ejecta distribution in the $(Y_e,s,\tau)$ space and the latter is usually rather broad in $Y_e$, a power law behavior is partially recovered and $\dot{e}_{\rm nucl}$ varies only by a factor of a few for relevant ejecta conditions during the first week after merger (e.g. \cite{Lippuner.Roberts:2015}).
At late time (between 10 and several hundreds days), independently from the initial $Y_e$, the nuclear distribution is very close to the valley of stability and only very few nuclei have the right lifetime to decay within this time window. Thus, the heating rate becomes possibly very sensitive to the details of the abundance distribution through emerging bumps in its time evolution. 
Due to the temporal evolution of the thermalization efficiency (see below) and depending on the detailed yields, heavy nuclei with $\beta$ decay half-lives around 14 days can be relevant for the light curve behavior at a few weeks, while a few actinides and transuranium elements with $\alpha$-decay half-lives of several tens of days can affect the kilonova emission up to hundreds of day after the merger \cite{Wu.etal:2019,Zhu.etal:2018}.  

\textbf{Thermalization efficiency}. 
Thermalization efficiency is defined as the ratio of the total energy released by all radioactive processes, $\dot{e}_{\rm nucl}$, to the energy effectively transferred to the ejecta, $\dot{e}_{\rm th}$, i.e. $ f_{\rm th} \equiv \dot{e}_{\rm th}/\dot{e}_{\rm nucl} $.
Since both these quantities changes with time, also $f_{\rm th}$ is a function of time. Detailed studies on $f_{\rm th}$ can be found, e.g., in Refs.~\cite{Barnes.etal:2016,Hotokezaka.etal:2016}.

First of all, the thermalization efficiency depends on how relevant each decay channel is and on how energetic the particles in the final state are.
These two points ultimately relate with the physics of the decays and with the actual abundances in the ejecta.
Beta decay is the dominant decay channel for nuclei $A \lesssim 200$ and it is relevant for all $Y_e$ conditions, especially for $Y_e \gtrsim 0.25$.
The $Q$-values of the $\beta$ decays relevant for the kilonova emission are of the order a 1-2 MeV and most of this energy is emitted in form of $\gamma$ rays ($\sim45 \%$, emitted from the excited daughter nucleus) while $e^{-}$'s takes usually 20\%, the rest being lost in neutrinos. The $Q$-values and the $e^{-}$ energies are larger for larger $Y_e$.
Alpha decay is the dominant decay channel for $A \gtrsim 200$. Thus, it affects significantly the heating rate if actinides are produced, i.e. if $Y_e \lesssim 0.20$. 
For these decays, typical $Q$-values are in the range 5-9 MeV. Most of this energy is taken by the $\alpha$ particle while the daughter nucleus de-excitation is negligible.
Fission is very effective for super-heavy nuclei ($A \gtrsim 250$) produced for $Y_e \lesssim 0.10$.
The $Q$-value of these reactions can be approximated by the kinetic energy of the fission fragments and the latter can be estimated as the Coulomb repulsion energy of the two fragments at their formation,
$Q_{\rm fission} \sim E_{\rm Coul} = Z_1 Z_2 e^2 / ( r_0 (A_1^{1/3} + A_2^{1/3}))$,
where we have approximated the radius of a nucleus of mass number $A$ as $r_0 A^{1/3}$ with $r_0 \approx 1.8~{\rm fm}$. For typical super-heavy nuclei with $(A,Z) \sim (250,100) $, and daughter nuclei with $(A_{1,2},Z_{1,2}) \sim (132,50)$, $Q$ is $\sim 100~{\rm MeV}$.

Additionally, $f_{\rm th}$ depends on how efficiently the different final state particles termalize in the plasma.
This in turn depends on the physical processes providing thermalization and on the density of the medium, since $f_{\rm th}$ is larger in high density ejecta (e.g. in more massive and slower ejecta).
High energy charged particles ($e^{-}$'s, $\alpha$'s and fission fragments) lose their energy in the ambient plasma through Coulomb interactions with the free and the atomic $e^-$'s. In the latter case, they can excite or even ionize atoms and these are the most efficient thermalization mechanisms. Each of these distant interactions transfers a relatively small amount of energy to other electrons. Then many interactions are required to thermalize a single projectile particle, but due to the low transferred energy the target electrons in the final state thermalize very rapidly. Due to its $Z_1 Z_2$ dependence, Coulomb interactions with ions are relevant only for fission fragments, while strong nuclear interactions with other nuclei are negligible.
The cumulative nature of the Coulomb processes that affect charged particles allows to thermalize efficiently at least a fraction of the available energy.
Also $\gamma$-rays lose their energy in the plasma by interacting with electrons through photoionization and Compton scattering. Due to the relatively high ionization thresholds of heavy elements ($\sim$100~keV), photoionization is the most relevant process up to $\sim 1~{\rm MeV}$, while for higher photon energies the Compton scattering becomes dominant. 

Any thermalization process is efficient as long the timescale over which it acts is smaller than the ejecta expansion timescale.
Due to the relatively low opacity provided by Compton scattering and photoionization, high energy photons stop to thermalize when matter become transparent to them, and for typical kilonova conditions this happens within the first day after merger ($t_{{\rm ineff},\gamma} \sim 1~{\rm day}$, where $t_{\rm ineff}$ is the timescale when thermalization becomes inefficient). For supra-thermal electrons and $\alpha$ particles, an efficient thermalization can occur up to several days after the merger ($t_{\rm ineff} \sim 8~{\rm day}$), while fission fragments (again due to the $ \propto Z_1Z_2$ dependence of the Coulomb interaction) thermalize efficiently up to a few weeks ($t_{\rm ineff} \sim 16~{\rm day}$).
Detailed calculations (e.g. \cite{Barnes.etal:2016}) show that $f_{\rm th}(t)$ decreases from $0.5-0.6$ during the first day down to 0.1 around 10 days after merger. 

\textbf{Atomic opacity}.
The photon opacity, $\kappa_{\gamma}$, quantifies the degree of transparency of matter to electromagnetic radiation. In particular, it can be understood as the cross section ($\sigma$) per unit mass of a fluid element to radiation: $ \kappa_{\gamma} = n \sigma / \rho$, where $\rho$ is the matter density and $n$ the target particle density.
In general, $\kappa_{\gamma}$ depends on the energy of the incident photon and on its complex interactions with the electron structure of the atom, in all its possible ionization states. Thus information about the composition and the ionization degree of each species is crucial to take properly into account the most relevant atomic opacities.
While the ejecta expand and cool, electrons recombine with atoms to form ions and neutral atoms. First ionization energies vary between 3 and 25 eV, while the innermost electrons of heavy elements have ionization energy of $\sim 100 {\rm keV}$. Given the expected temperature, Eq.(\ref{eq:kilonova temperature}), most of the electrons have recombined at the time of the kilonova emission.
and the abundance of free electron is $Y_{e,{\rm free}} \sim$ a few $0.01$, decreasing as a function of time due to electron recombination.
Despite the fact that plasma collisions are not effective enough to maintain thermodynamics equilibrium during the ejecta expansion, due to the high opacity expected in the ejecta of compact binary mergers, radiation can effectively drive the ion abundances towards local thermodynamical equilibrium (LTE), at least during the first days after merger (e.g. \cite{Kasen.etal:2017} and references therein).

For the photon energy interval relevant for kilonovae, bound-bound transitions are the most important atomic processes, followed by bound-free and free-free opacities, usually smaller by several orders of magnitudes. 
Detailed and exhaustive experimental values of $\kappa_{\gamma}$ for heavy elements, for a broad range of photon energies and in relevant thermodynamics conditions, are mostly missing, and the recourse to sophisticated, but still uncertain, atomic physics calculations is necessary
(e.g. \cite{Kasen.etal:2013,Tanaka.etal:2013}).
The energy-independent electron scattering opacity (Thomson's opacity), $\kappa_{\gamma,{\rm Th}} = 
    8 \pi/3
    \left( \alpha \hbar c/m_e c^2 \right)^2 
    n_{e,{\rm free}}/\rho \approx 3.97 \times 10^{-3}~{\rm cm^2~g^{-1}} 
    \left( Y_{e,{\rm free}}/0.01 \right)$,
where $\alpha=1/137$ is the electromagnetic coupling constant, sets a lower limit that becomes relevant only for photon wavelengths $\lambda_\gamma \gtrsim 10^4 \angs$. 
Energy- and composition-averaged values of the opacity (sometimes called gray opacity) can be use to roughly characterize the global plasma behavior. For example, for the Planck mean opacity $\kappa_{\gamma}(\lambda_\gamma)$ is averaged over the Planck distribution function, while for the Rossland mean opacity $1 / \kappa_{\gamma}(\lambda_\gamma)$ is averaged over the temperature derivative of the Planck distribution function.

When the ejecta contain light $r$-process elements ($A \lesssim 140$), bound-bound transition involve $d$-shell valence electrons. For $\lambda_{\gamma} \gtrsim 10^3 \angs$ the opacity strongly decreases with the photon wavelength ($\kappa_\gamma(\lambda_{\gamma}=10^3 \angs) \sim 10^{3} {\rm cm^2~g^{-1}}$ while $\kappa_\gamma(_{\gamma}=10^{4} \angs) \sim 10^{-3}{\rm cm^2~g^{-1}}$) and typical gray opacities are $\lesssim 1 {\rm cm^2~g^{-1}}$. 
If lanthanides and/or actinides are also present, the opening of the electron $f$-shell increases enormously the number of possible transitions and the bound-bound opacity is characterized by a forest of lines, Doppler-broadened by the large expansion velocity. The spectral opacity still decreases with the photon wavelength, but much mildly ($\kappa_\gamma(10^3 \angs) \sim 10^{3} {\rm cm^{2}~g^{-1}}$, but $\kappa_\gamma(2.5 \times 10^{4} \angs) \sim 10^{-3} {\rm cm^2~g^{-1}}$), and typical gray opacities are in this case $\gtrsim 10~{\rm cm^2~g^{-1}}$. We stress that even a small amount of lanthanides or actinides ($ X_{\rm La's} + X_{\rm Ac's} \lesssim 10^{-4}$) can already change the global matter opacity.

\subsubsection{\textit{Modeling kilonovae}}

Kilonova modeling is an extremely challenge task. State-of-the-art models require the solution of the photon radiative transfer equation in the expanding ejecta, possibly in multiple dimensions, e.g. \cite{Kasen.etal:2017,Tanaka.etal:2017,Wollaeger.etal:2018}. In addition to physically motivated density and temperature profiles, theyse models also require detailed information on the nuclear composition, on the ion abundances and on the spectral opacities at every time and everywhere in the ejecta. 
These models are able to provide light curves and spectra from a given configuration, at different epochs after merger. 
If the measurement or computation of all relevant transition lines is one of the major theoretical uncertainties, also their treatment in radiative transfer codes and in particular their translation in an effective opacity (``effective" because it refers to a discretization procedure involving finite wavelength interval, where many single lines are present) is not obvious. 

In the following, we will derive fundamental scaling relations, based on a simplified analytical model, which are in qualitative agreement with more detailed models. 
In addition to provide the fundamental scales of the problem, they also highlight the impact of nuclear physics input and the variety implied by the different ejecta conditions expected in compact binary ejecta.
To model the ejecta we consider a spherically symmetric distribution of total mass $m_{\rm ej}$ and average speed $\langle v_{\rm ej} \rangle $, characterized by a grey opacity $\kappa_{\gamma}$ and expanding homologously. 
Matter at the outer edge is moving at velocity $v_{\rm ej,max}$, and
at each time $t$ its radial position is $R_{\rm max} = v_{\rm ej,max} t$. 
Any internal shell of mass $\delta m$ is expanding at a speed $v_{\rm ej} < v_{\rm ej,max}$, constant in time and proportional to the radius, such that its radial position evolves according to $R=v_{\rm ej}t$. We further define $m_{\rm env}$ the mass of the envelope above $\delta m$ and $\Delta R = R_{\rm max} - R$ its radial thickness.
The expansion timescale of this envelope can be computed as $t_{\rm exp} \sim \Delta R/v_{\rm ej}$.
Thermal photons produced inside the envelope will contribute to the kilonova if $t_{\rm exp} \sim t_{\rm diff}$, where $t_{\rm diff}$ is the photon diffusion timescale. 
The latter can be determined from random-walk arguments starting from the photon optical depth ($\tau_\gamma$) and mean free path ($\ell_\gamma = 1/ \left( \kappa_{\gamma} \rho \right)$) as $t_{\rm diff} \sim \tau_{\gamma}^2 \ell_{\gamma}/c $.
The optical depth is defined as the integral of the photon inverse mean path, $\ell_{\gamma}^{-1}$, along an outgoing radial path. In words, 
$\tau_{\gamma}$ counts the average number of interactions that a photon experiences before being emitted at the photosphere and it can be approximated by $\tau_{\gamma} \sim \langle \rho \rangle \kappa \Delta r $, where $\langle \rho \rangle$ is the average density experienced by the outgoing photon, $\langle \rho \rangle \sim m_{\rm env}/(4 \pi R^2~\Delta R)$.
By equating the expansion and the diffusion timescale we can determine the time $\tilde{t}(m_{\rm env})$ at which the photons emitted by $m_{\rm env}$ will contribute to the kilonova, and we can estimate the peak time of the kilonova emission $t_{\rm peak}$ by taking $m_{\rm env} \sim m_{\rm ej}$ and $v_{\rm ej} \sim \langle v_{\rm ej} \rangle $:
\begin{equation}
    t_{\rm peak} \sim \sqrt{\frac{m_{\rm ej}\kappa_{\gamma}}{4 \pi \langle v_{\rm ej} \rangle c }} \approx 4.6~{\rm days}
    \left( \frac{\kappa_{\gamma}}{10~{\rm cm^2~g^{-1}}} \right)^{1/2}
    \left( \frac{m_{\rm ej}}{0.01~M_{\odot}} \right)^{1/2}
    \left( \frac{\langle v_{\rm ej} \rangle }{0.1~c} \right)^{-1/2}    
    \, .
    \label{eq: peak time estimate}
\end{equation}
The energy available to power the kilonova at $\tilde{t}$ is the nulcear energy released by $m_{\rm ej}$ and thermalized by the plasma: $L_{\gamma}(\tilde{t}) \approx \dot{e}_0~( \tilde{t}/1~{\rm sec})^{-\alpha}~f_{\rm th}(\tilde{t})~m_{\rm rad}(\tilde{t})$,
with $\alpha = 1.3$ as a typical value. Once again, we can estimate the peak luminosity by taking $\tilde{t} = t_{\rm peak}$ and $m_{\rm env} \sim m_{\rm ej}$ to obtain:
\begin{eqnarray}
    L_{\rm peak} & \sim &  
    2.4 \times 10^{40}~{\rm erg~s^{-1}} 
    \left( \frac{\kappa_{\gamma}}{10~{\rm cm^2~g^{-1}}} \right)^{-\alpha/2}
    \left( \frac{m_{\rm ej}}{0.01~M_{\odot}} \right)^{1-\alpha/2}
 \nonumber \\
    & &     \left( \frac{\langle v_{\rm ej} \rangle}{0.1~c} \right)^{\alpha/2} \left( \frac{\dot{e}_0}{5 \times 10^{16}{\rm erg~s^{-1}~g^{-1}}} \right)
    \left( \frac{f_{\rm th}}{0.5} \right) \, .
    \label{eq: luminosity estimate}
\end{eqnarray}
The radius of the photosphere at $\tilde{t}$ is $R_{\rm ph}(\tilde{t}) \approx v_{\rm ej} \tilde{t}$ and then at the luminosity peak:
\begin{equation}
    R_{\rm ph,peak} \sim 1.26 \times 10^{15}~{\rm cm} 
    \left( \frac{\kappa_{\gamma}}{10~{\rm cm^2~g^{-1}}} \right)^{1/2}
    \left( \frac{m_{\rm ej}}{0.01~M_{\odot}} \right)^{1/2}
    \left( \frac{\langle v_{\rm ej} \rangle }{0.1~c} \right)^{1/2} \, .
    \label{eq: photospheric radius estimate}
\end{equation}
Finally, assuming black body emission and using the Stefan-Boltzmann law, the effective photospheric temperature, $T_{\gamma,{\rm eff}}$, can be determined as $T_{\gamma}^4 = \left( L_{\gamma}/(4 \pi R_{\rm ph}^2 \sigma_{\rm SB}) \right)$.
This expression can be evaluated at the peak:
\begin{eqnarray}
T_{\gamma,{\rm peak}} & \sim & 2.15 \times 10^{3}{\rm K} 
    \left( \frac{\kappa_{\gamma}}{10~{\rm cm^2~g^{-1}}} \right)^{-(\alpha+2)/8}
    \left( \frac{m_{\rm ej}}{0.01~M_{\odot}} \right)^{-\alpha/8}
 \nonumber \\
    & &     \left( \frac{\langle v_{\rm ej} \rangle}{0.1~c} \right)^{(\alpha-2)/8} \left( \frac{\dot{e}_0}{5 \times 10^{16}{\rm erg~s^{-1}~g^{-1}}} \right)^{1/4}
    \left( \frac{f_{\rm th}}{0.5} \right)^{1/4} \, ,
    \label{eq: BB temperature estimate}
\end{eqnarray}
and translated in a peak wavelength
$\lambda_{\gamma,{\rm peak}} = 1.35 \times 10^3 {\rm nm}~(T_{\gamma,{\rm peak}}/2.15 \times 10^3 {\rm K})^{-1}$.
The above estimates have been done assuming $\kappa_{\gamma} \sim 10~{\rm cm^2~g^{-1}}$, i.e. considering ejecta that contain a significant fraction of lanthanides and actinides (meaning that the ejecta contains some matter with initial $Y_e \lesssim 0.25$).
In this case, the luminosity peak is expected to occur around one week and at near-IR wavelengths. The luminosity is more than 100 times larger than the one of a typical nova, but several orders of magnitudes lower than a supernova (SN). Since SNe are also powered by radioactive decays in expelled matter, the mean reason for such a large difference is in the amount of ejecta ($L_{\gamma} \propto m_{\rm ej}^{1-\alpha/2} $), much lower in the case of compact binary mergers.
If the wind ejecta or the dynamical ejecta have been significantly irradiated by neutrinos, the initial $Y_e$ could have increased such that the production of lanthanides is prevented. In that case, $\kappa \lesssim 1 {\rm cm^2~g^{-1}}$ and the lumonisity peak is expected to happen earlier (around 1 day), with a higher luminosity (more than $10^3$ times the one of a nova) and at bluer peak frequency ($\lambda_{\gamma} \sim 500~{\rm nm}$). 

\subsubsection{\textit{GW170817 and its kilonova}}

On August the 17th 2017 and in the subsequent weeks, the first unambiguous kilonova resulting from a compact binary merger was detected, see e.g. \cite{Abbott:2017b}. This kilonova (called AT2017gfo) followed GW170817, the first GW signal compatible with the late inspiral of two NSs \cite{Abbott:2017a}. 
The total mass of the system was 2.74 \msun while the mass ratio, assuming slowly spinning NSs, was measured to be between 0.7 and 1.0. 
Light curves in different photometric bands of this unprecedented UV/visible/IR emission showed an early peak (around 1 day after merger) in the visible frequencies, followed by a later peak (around 5-7 days after merger) in the near-IR, see \cite{Villar.etal:2017} and references therein. 
The bolometric luminosity of the event 1.5 days after merger was $\sim 3.2 \times 10^{41}{\rm erg~s^{-1}}$, while it decayed approximately following a $t^{-1.3}$ power-law during the first week so that around 7 days $L_{\gamma} \approx 6 \times 10^{40} {\rm erg~s^{-1}}$.
The spectrum at 1.5 days was very close to a black-body of $5 \times 10^3K$, while around 7 days it was broadly compatible with a $2.2 \times 10^3K$ black-body spectrum with a forest of absorption features, e.g. \cite{Pian.etal:2017,Smartt.etal:2017,Tanvir.etal:2017}.
The identification of elements in the spectrum is very challenging due to the high density of lines and to their broadening due to the high expansion velocities.
An analysis of the spectra recorded during the first days has revealed features compatible with the presence of Sr, an element of the first $r$-process peak \cite{Watson.etal:2019}.

\begin{figure}[t!]
    \centering
    \includegraphics[width=\linewidth]{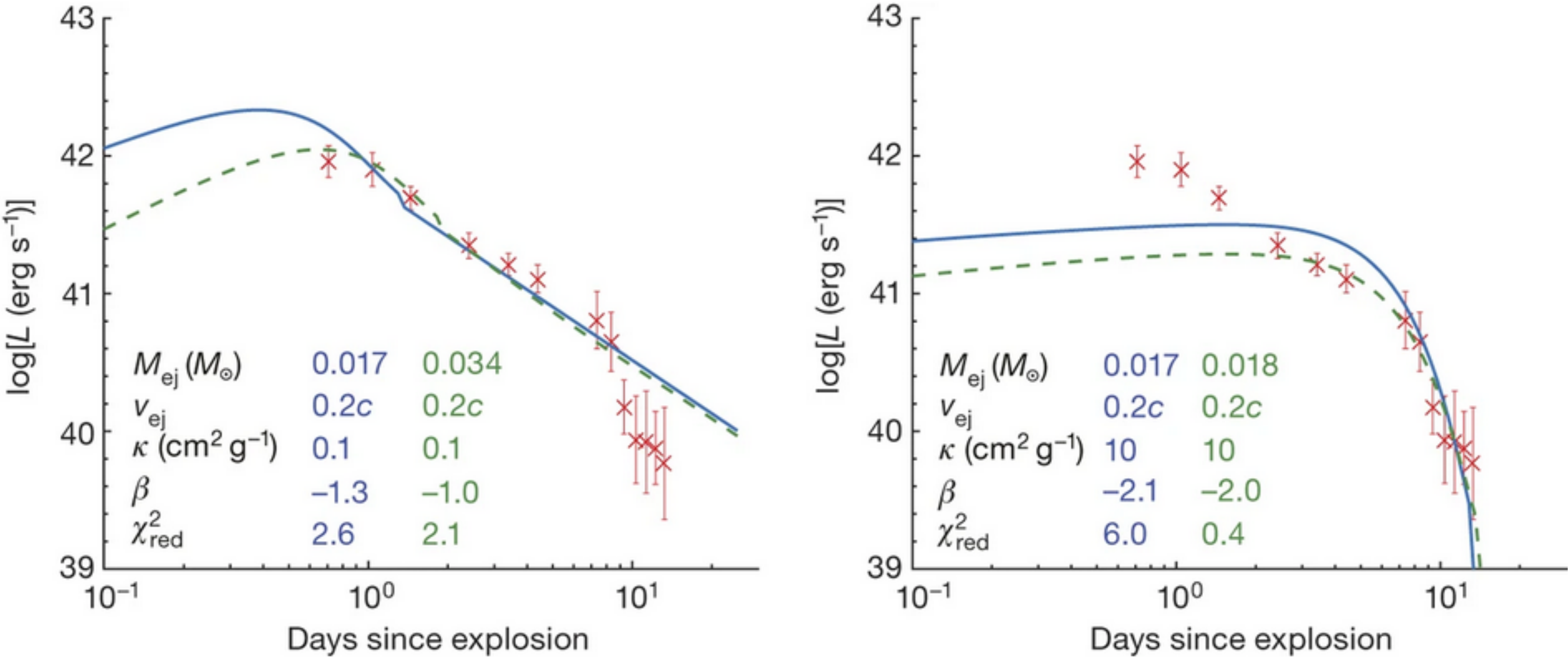}
    \caption{Modeling of the bolometric light curve of AT2017gfo, the kilonova associated with the GW170817 BNS merger event. The slope of the light curved, fitted with four different models characterized by different ejecta mass and opacity, is consistent with a power-law decay, $t^{-\beta}$ such that $\beta \approx 1.0-2.0$ ($\alpha$ in the text), compatible with the decay of $r$-process material in the ejecta of compact binary mergers. Figures taken from from \protect\cite{Smartt.etal:2017}.}
    \label{fig: kilonova model}
\end{figure}

Theoretical modelling of AT2017gfo requires the presence of more than one component of the ejecta. The different components are characterized by different masses, velocities and opacities, and possibly a non-trivial dependence from the geometry of the ejection. This is indeed necessary to explain the observed blue and red peaks \cite{Villar.etal:2017,Perego.etal:2017}. 
For example, one can consider the presence of two distinct components to explain the color evolution of AT2917gfo.
Using Eq.~(\ref{eq: peak time estimate})-(\ref{eq: BB temperature estimate}), assuming a $\alpha \sim 1.3$ and a thermalization efficiency of 0.8 and 0.4 for the peaks at 1.5 and 7 days, respectively, the above peak properties imply that the blue peak was characterized by $m_{\rm blue} \approx 0.019$ \msun, $\kappa_{\gamma,{\rm blue}} \approx 1~{\rm cm^{2}~g^{-1}}$ and $v_{\rm ej,blue} \approx 0.20 c$, while for the red peak $m_{\rm red} \approx 0.058$ \msun, $\kappa_{\gamma,{\rm red}} \approx 4.2~{\rm cm^{2}~g^{-1}}$ and $v_{\rm ej,red} \approx 0.09 c$.
The estimated total amount of ejecta is thus of the order of several percents of a solar mass. Results from one of these models are presented in Figure~\ref{fig: kilonova model}.
According to more detailed models, e.g \cite{Kasen.etal:2017,Tanaka.etal:2017,Wollaeger.etal:2018}, the amount of ejecta in this event was $\sim 0.02-0.05$~\msun. Nuclear physics input (e.g. the nuclear mass model) can introduce 
an additional uncertainty factor of a few, possibly reducing the total ejecta mass.
The inferred opacities suggest a negligible amount of lanthanides in the blue component and a lanthanide mass fraction between $10^{-3}$ and $10^{-2}$ in the red one. 

The emerging picture  is certainly very compatible with results of compact binary merger simulations and nucleosynthesis calculations.
In particular, the decline rate of the bolometric light curve and the inferred opacities are compatible with what expected from the collective decay of freshly synthetized $r$-process elements (see for example Figure~\ref{fig: kilonova model}).

\subsection{\textit{Compact binary mergers and the chemical evolution}}

\begin{figure}[t!]
    \centering
    \includegraphics[width=0.49 \linewidth]{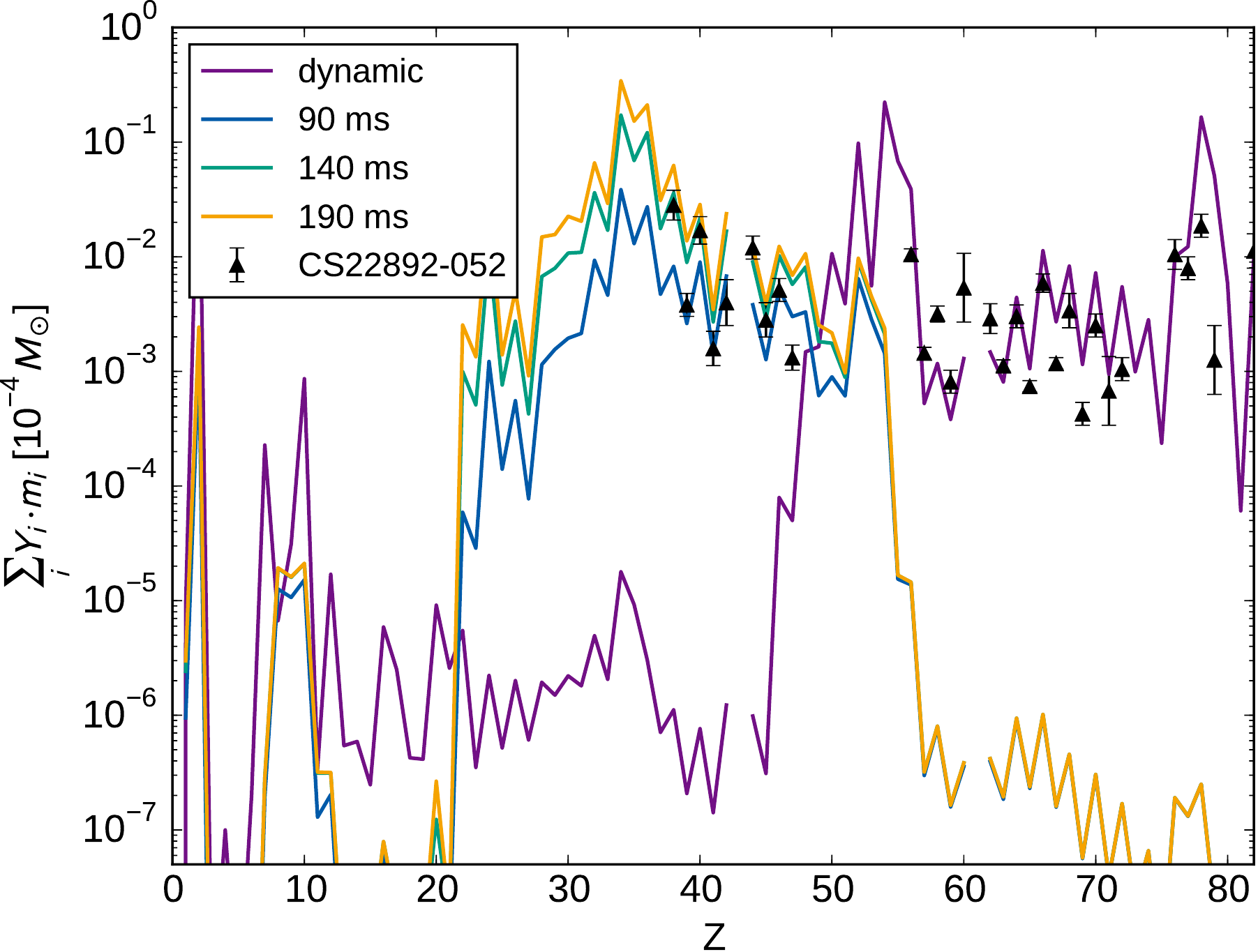}
    \includegraphics[width=0.49 \linewidth]{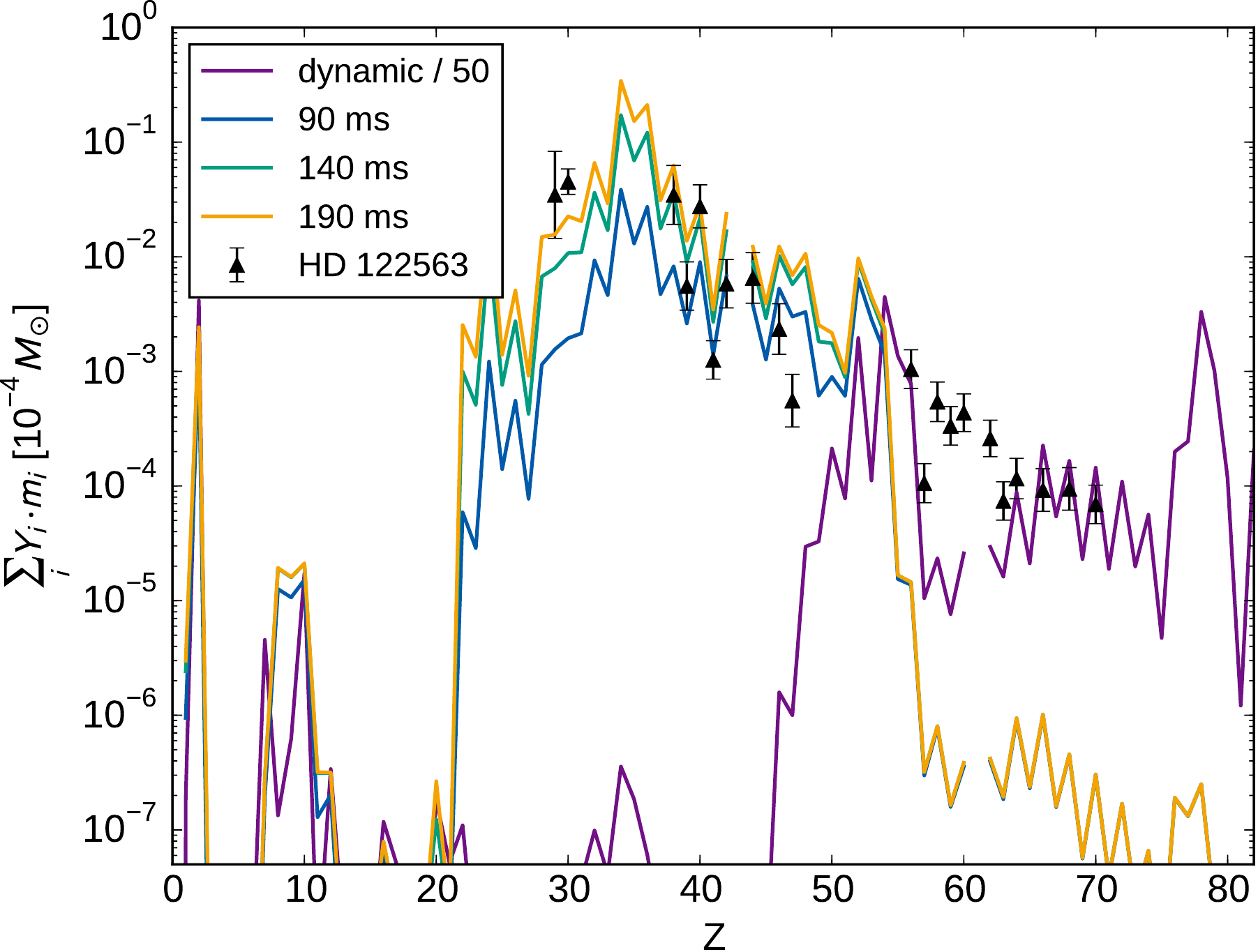}    
    \caption{Elemental abundance comparison between BNS merger models and metal poor star observations. The violet curve represents nucleosynthesis yields for tidally dominated dynamical ejecta, while the other curves yields from neutrino-driven wind ejecta for different massive NS lifetimes. Theoretical abundances are compared with two classes of metal poor stars. In the right panel dynamical ejecta have been diluted by a factor of 50 with respect to the neutrino-driven wind one. Figures taken from \protect\cite{Martin.etal:2015}.}
    \label{fig: ejecta VS stars}
\end{figure}

\begin{figure}[t!]
    \centering
    \includegraphics[width=\linewidth]{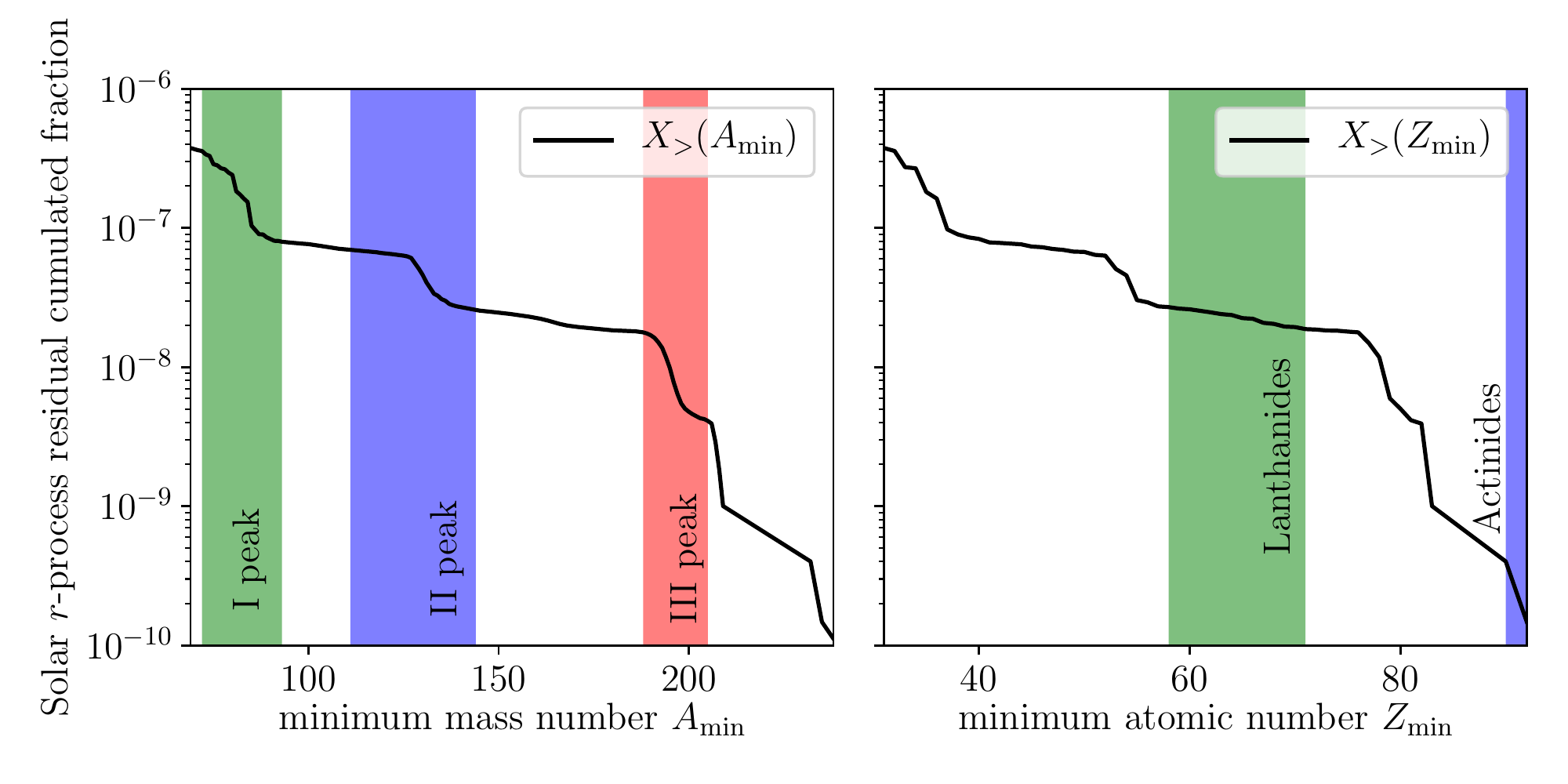}
    \caption{Cumulative mass fraction obtained by the solar residual $r$-process abundances from \protect\cite{Prantzos.etal:2020} both as a function of the mass numbers (left panel) and the atomic numbers (right panel).}
    \label{fig: cumulative fraction sun}
\end{figure}

Due to the production and ejection of heavy nuclei, compact binary mergers are a possible key player in shaping the evolution of chemical abundances in the Universe. 
A first relevant question is whether they are primary astrophysical sites for the production of $r$-process elements.
Two main ingredients are necessary to answer this question: detailed abundance predictions and reliable merger rates. The former are obtained by combining the amount of ejecta predicted by merger simulations with the distributions of the nucleosynthesis yields corresponding to the properties of the ejecta.
The latter are related to the binary formation channels and evolution. In particular, they depend on the probability that a compact binary forms and on the time that it takes for it to reach coalescence due to GW emission.
To provide a simple answer, we compare the amount of $r$-process elements in the Milky-Way (MW, $M_{r,{\rm MW}}$) to the whole amount of $r$-process material ejected by compact binary mergers during the galactic evolution ($M_{r,{\rm CBM}}$). 
Their ratio, $f_{\rm CBM} \equiv M_{r,{\rm CBM}}/M_{r,{\rm MW}}$, quantifies the relevance of compact binary mergers in accounting for the observed $r$-process elements.
The enumerator can be evaluated as the product of the average merger rate in the Galaxy, $R_{\rm CBM}$, times the average mass of $r$-process elements expelled by a single event, $m_{r,{\rm CBM}}$, times the age of the Galaxy, $t_{\rm MW} \approx 13.5~{\rm Gyrs}$.
The mass per event can be estimated from the lower bound obtained from GW170817 ejecta, i.e. $m_{r,{\rm CBM}} \sim 0.02M_{\odot}$.
The merger rate is still very uncertain, but it can be measured in different ways. 
From the GW events detected so far in the first two observing runs, the Ligo-Virgo collaborations have provided a rate of $110-3840{\rm Gpc^{-3}~yrs^{-1}}$ for BNS mergers, and a robust upper bound of $660~{\rm Gpc^{-3}~yrs^{-1}}$ for BHNS mergers \cite{Abbott:2019}. 
Assuming that a significant fraction of mergers produce also a short gamma-ray burst (SGRB), the merger rate can be inferred also from the SGRB rate. The latter is bracket by $R_{\rm SGRB} = 0.6-6 ~{\rm Gpc^{-3}{\rm yrs}^{-1}}$ \cite{Wanderman.Piran:2015,Ghirlanda.etal:2016}, and since GRBs are collimated emissions with a beaming correction factor $f_{\rm beam} \sim 100$ ($f_{\rm beam} = 1-\cos{\theta_{\rm jet}}$ and typical opening angles are $\theta_{\rm jet} \approx 5- 10^{\circ}$, \cite{Fong.etal:2015}), the resulting lower bound on $R_{\rm CBM}$ rate is $\sim 80-600~{\rm Gpc^{-3}{\rm yrs}^{-1}}$.
More theoretical bounds can be obtained from population synthesis studies and from extrapolations of the observed populations of galactic BNS system (see for example \cite{abadie:2010} and references therein). These rates broadly agree with the measured rates, even if with larger uncertainties.
Assuming a galaxy density of $\sim 0.01~{\rm galaxy}~{\rm Mpc^{-3}}$, a conservative merger rate inferred from theoretical models and observations is $R_{\rm CBM} \approx 30~{\rm Myr^{-1}}{\rm galaxy^{-1}}$.
For the total mass of $r$-process elements in the MW, we consider the Sun as representative of the averaged enrichment of the Galaxy, a common assumption in Galactic chemical evolution models. We also assume that all this mass is produced by the one single type of events and with a relative constant yield and rate for the entire evolution of our Galaxy.
These hypothesis are also grounded in the observations of the abundances of $r$-process elements in metal poor stars.
Indeed the spectroscopic analysis of the light emitted by a star reveals the chemical composition of the ISM from which the star has formed. 
Low mass stars can live longer than the Galaxy, thus some of them are among the oldest objects in the MW. These old stars have an extremely low metallicity since they formed very early in the Galactic history when only a few SNe had exploded and polluted the ISM. They can be considered fossils of the early chemical enrichment of our Galaxy and the chemical abundances measured in their spectra can be studied to infer the characteristics of the first stellar nucleosynthesis events. 
Noteworthy, in the spectra of these old stars the lines of neutron capture elements are identified and their abundances measured. Among these fossil stars, there is a group that appears particularly rich in $r$-process elements (so called $r$-process rich stars, see e.g. \cite{Sneden.etal:2008}). For this group, the derived abundance pattern reproduces very closely the solar residual $r$-process pattern, at least beyond the second peak, while the first peak presents a significant dispersion. Since the low metallicity implies that these stars have been polluted by very few (if not one single) $r$-process sources, the common patterns observed in the solar $r$-process residual, in this group of metal poor stars and in compact binary merger nucleosynthesis calculations reveal the presence of robust features that characterize the $r$-process nucleosynthesis in very general terms.
Nevertheless, the emerging picture is more complicated than that: other metal poor stars present a substantial amount of first peak elements and a lower enrichment in the heavy ones \cite{Honda.etal:2004}. This is still compatible with the large variability expected in the the yields of compact mergers (see Figure~\ref{fig: ejecta VS stars}), but it leaves also space for other production sites. 
To account for these uncertainties, starting from the solar $r$-process abundances, we compute the cumulative mass fraction of the $r$-process materials above a certain mass number $A_{\rm min}$, $X_>(A > A_{\rm min})$, see Figure~\ref{fig: cumulative fraction sun}. To estimate the typical total mass fraction of $r$-process elements in the MW we assume $X_{r-{\rm proc}} \approx X_>(A > A_{\rm min})$ and we consider $A_{\rm min} = 68, 89, 124$ such that $X_{r-{\rm proc}}(A > A_{\rm min}) \approx 40,9,6 \times 10^{-8}$, respectively.
Since the mass in stars and ISM of the MW is $M_{\rm MW} \approx 6 \times 10^{10}$\msun, $f_{\rm CMB}$ can be finally estimated as:
\begin{equation}
f_{\rm CBM} \approx 1.35 \left( \frac{R_{\rm CBM}}{30 ~{\rm Myr^{-1}}} \right)
\left( \frac{m_{r,{\rm CBM}}}{0.02 M_{\odot}} \right)
\left( \frac{t_{\rm MW}}{13.5 ~{\rm Gyr}} \right)
\left( \frac{X_{r-{\rm proc}}}{10^{-7}} \right) \, ,
\label{eq: CBM mass estimate}
\end{equation}
and $f_{\rm CBM} \approx 0.34,1.50,2.25$ for $A_{\rm min} = 68, 89, 124$, respectively.

These estimates are rather crude in many ways. 
For example, they assume that the present day merger rate is representative of the average merger rate, while we know that it has significantly changed during the Galaxy history. Chemical evolution models can be used to better consider the merger rate and set more stringent constrain on the impact of the enrichment of compact mergers to the total balance of $r$-process in the Galaxy, see \cite{Cowan.etal:2019}. Moreover, it is not obvious that the reference amount of ejecta is representative for the merger population. However, it is worth noticing that an IR excess, observed in the afterglow light curves of a few SGRB, is broadly compatible with the emission expected from a kilonova and it usually points to the presence of large ejecta masses ($10^{-2}-10^{-1}$ \msun), e.g. \cite{Tanvir.etal:2013}. In addition, assuming that a significant fraction of the accretion disk is expelled in disk winds, our reference value is also well compatible with the results reported in Figure~\ref{fig: disk VS dyn ejecta}.
Some astrophysical processes are also neglected for simplicity. For example, when compact objects emerge from CCSNe, they receive a kick at birth. This kick, in addition of being a threat for the survival of the binary, can put the binary in a wide orbit inside the galactic potential, such that the merger could happen with a significant offset from the stellar and gas distributions inside the Galaxy (see \cite{Hotokezaka.etal:2018} for a more detailed discussion). 
Despite the large uncertainties, $f_{\rm CBM} \sim 1$ and this testifies that compact binary mergers are primary astrophysical site where $r$-process elements are produced. This is especially true for elements above the second $r$-process peak, while the explanation of the first peak could be a clear confirmation that different mergers can produce different yields or could require additional production sites.

\begin{figure}[t!]
    \centering
    \includegraphics[width=0.41 \linewidth]{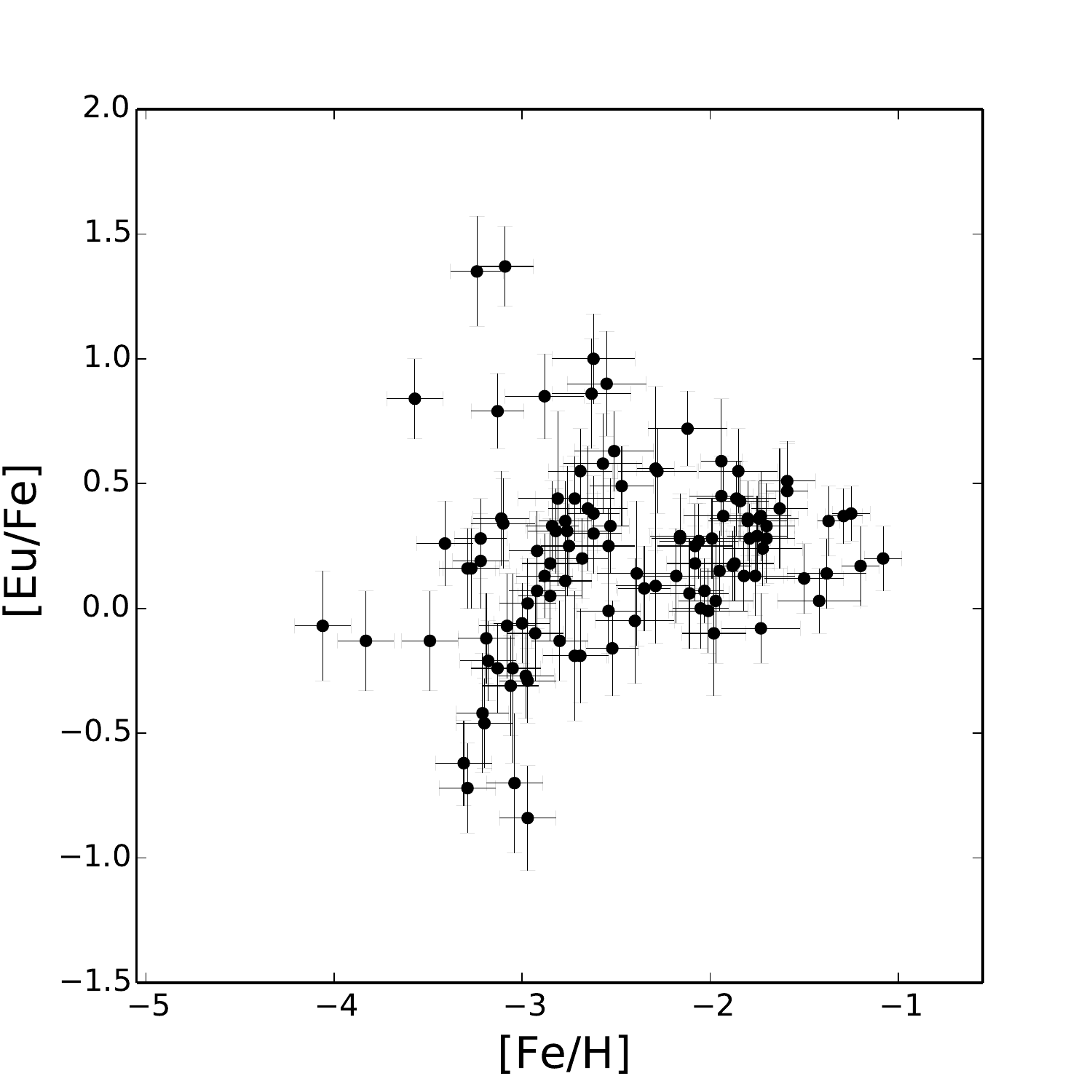}
    \includegraphics[width=0.58\linewidth]{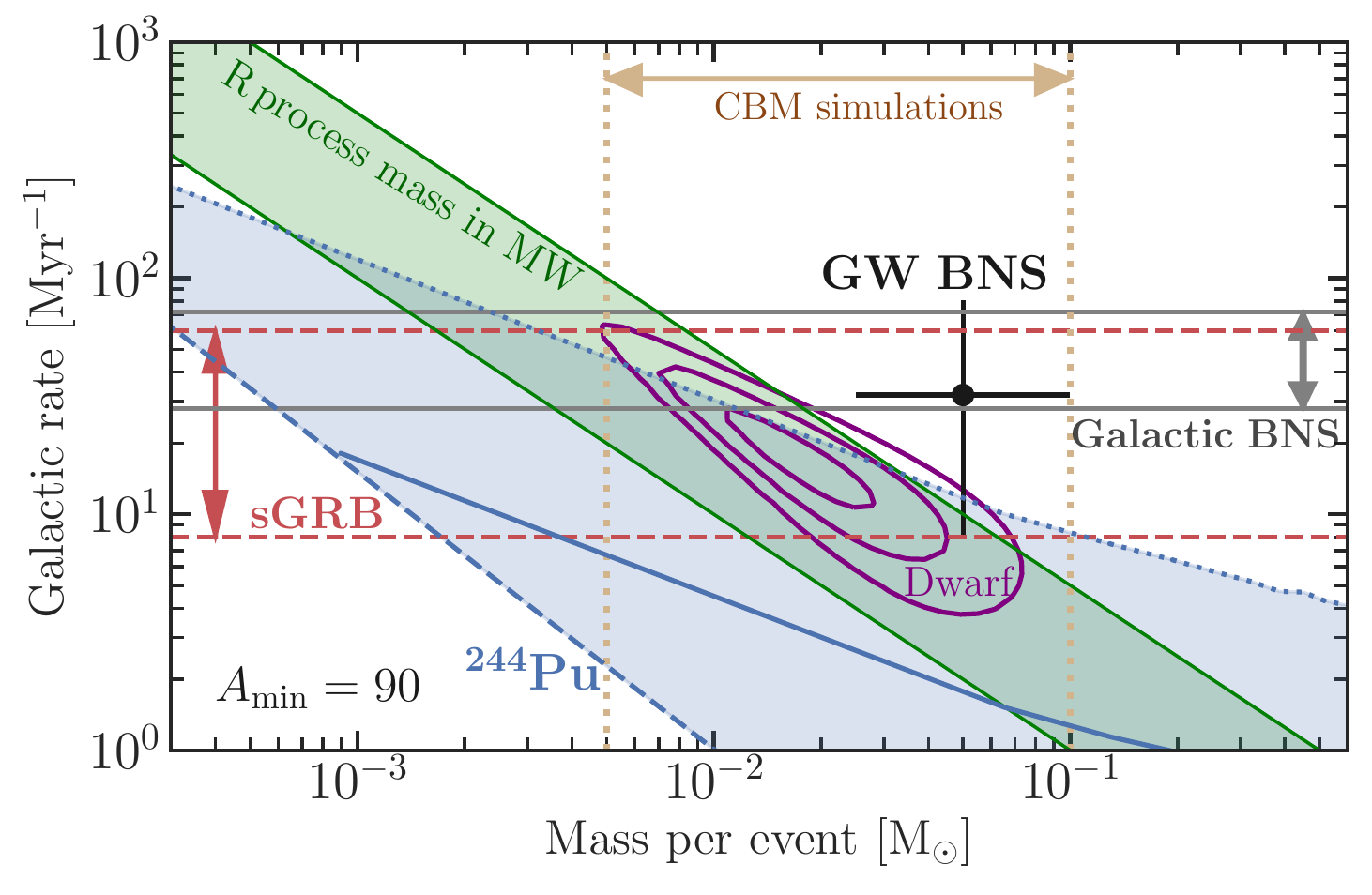}
    \caption{Left: Scatter plot of the relative abundance ratio of iron over hydrogen versus europium over iron, for a large sample of metal poor stars. The quantity ${\rm [A/B]}$ is relative with respect to the solar one, i.e. it is defined as ${\rm [A/B]} \equiv \log_{10}\left( Y_A/Y_B \right) - \log_{10}\left( Y_A/Y_B \right)_{\odot} $.  Data have been taken from \protect\cite{Roederer.etal:2014}. Right: Summary of the observational evidences for the production of $r$-process elements expressed on the $r$-process mass ($m_{\rm r}$) versus galactic event rate ($R_{\rm MW}$) plane. 
    For concreteness, $r$-process abundances are assumed to be equal to the solar residual for $A_{\rm min} = 90$ while the BNS rate is inferred from galactic BNS observations \protect\cite{Pol.etal:2019}. (Courtesy of K. Hotokezaka; see also \protect\cite{Hotokezaka.etal:2018}).}
    \label{fig: observational constraints}
\end{figure}

A related question is whether other astrophysical sites are able to synthesise heavy elements through the $r$-process.
Historically, proto-neutron star winds emerging after a successful CCSN explosion have long been thought to be a possible $r$-process nucleosynthesis site \cite{Qian.Woosley:1996}. The argument applied above to relate the rates and the ejecta from compact binary mergers to the observed amount of $r$-process elements in the MW could be easily adapted to regular CCSNe and it is not able to distinguish alone between these two scenarios. However CCSNe are significantly more frequent than compact binary mergers and their rate is well constrained, $R_{\rm CCSN} \approx 2.8 \times 10^4{\rm Myr^{-1}}$ (this rate can be related to $R_{\rm CBM}$ by considering that approximately half of the massive stars are in binaries, two CCSNe are required to form a compact binary and only $\sim 1\%$ of the stellar binaries survives the two CCSN explosions). From Eq.(\ref{eq: CBM mass estimate}) it is evident that CCSNe could explain the bulk of the galactic $r$-process nucleosynthesis if every SN would eject $\sim 10^{-5}$ \msun \, of $r$-process elements. Thus, the competition here is between rare events that expel large amounts of $r$-process material (e.g. compact binary mergers) or frequent events with much smaller amounts (e.g. regular CCSN).
Over the past years, several observational evidences (in addition to GW170817) have accumulated pointing to the fact that the $r$-process elements come from rare events that produce a significantly large amount of $r$-process elements. In the following we will briefly review them:\\
\textbf{Eu abundance in galactic metal poor star}. The analysis of the abundance of $r$-process elements, and especially of Eu, as a function of metallicity (and in particular of iron and of $\alpha$ elements, which are good tracers of the early evolution of the ISM composition due to the explosions of the very first massive stars) has revealed that the \text{\rm average} abundance of Eu correlates with the one of $\alpha$ elements produced in CCSNe through the entire metallicity evolution (it correlates also with the one of iron, but only before type Ia SN start to explode, i.e. for low metallicity).
However, differently from the $\alpha$ elements case, the distribution of the \textit{single} observations changes considerably: while at present metallicity the cumulative effect of several nucleosynthesis episodes and the efficient mixing of gas inside the Galaxy has homogenized the ISM composition and reduced the observed spread, at early times, when the amount of iron was $10^2-10^3$ times smaller than the present one, the ratio of the Eu over iron abundances shows a two orders of magnitude scatter, ranging from metal poor stars where Eu is underrepresented to cases where this ratio is more than 10 times the one observed in the present solar system (see the left panel of Figure~\ref{fig: observational constraints} and \cite{Sneden.etal:2008}). 
This large scatter suggests that the possible $r$-process elements pollution comes from a single rare event that ejects large amount of $r$-process material in ISM clouds that have seen only a few SNe.
Considering that ISM mixing over the entire Galaxy requires a timescale much larger than the star formation timescale, this naturally introduces the inhomogeneous character to the early chemical evolution necessary to explain the observations. Sophisticated Galactic chemical evolution models have been specifically developed to mimic this observational spread (see \cite{Cowan.etal:2019} and references therein). It shall be underlined that this is a peculiar characteristics of the $r$-process elements,  not present in other chemical elements, such as $\alpha-$elements produced together with iron by the more frequent SNe. \\
\textbf{$r$-process abundance in ultra-faint dwarf galaxies}. In addition to galactic observation, Eu can be observed also in classical dwarf and ultra-faint dwarf (UFD) galaxies, satellites of the MW and formed by very old stars (interestingly, these observations are possibly related with the ones of galactic metal poor stars, because the latter are often located in the so called galactic halo, that is thought to be formed by accretion episodes of dwarf galaxies on the MW). In particular, in the case of UFD galaxies, while Fe is observed in all cases, Eu has been detected so far only in a couple of cases (Reticulum II and Tucana III), corresponding to $\sim$30\% of the available sample \cite{Ji.etal:2016,Hansen.etal:2017}. Since UFD galaxies are formed by $10^3-10^5$ stars, such a low detection rate points again to a single, rare event (no more than one per $10^{4}-10^{5}$ stars).
Moreover, it is interesting to compare the ratio between the amount of iron and of europium observed in Reticulum II ($M_{\rm Fe,Ret~II} \approx 0.7$ \msun and $M_{\rm Eu,Ret~II} \approx 10^{-5}$\msun), with the theoretical value predicted by assuming that Fe is produced by CCSNe (typically, $m_{\rm Fe,CCSN} \approx 0.07$\msun) and Eu by merger events (for which we conservatively assume again $m_{r,{\rm CBM}}\sim 0.02 M_{\odot}$) with an abundance equal to the solar one ($X_{\rm Eu} \approx 4.1 \times 10^{-10} $). Since $\left( M_{\rm Fe}/M_{\rm Eu}\right)_{\rm theory} \sim m_{\rm Fe,CCSN} R_{\rm CCSN} / m_{\rm Eu,CBM} R_{\rm CBM}  $ and
$m_{\rm Eu,CBM} \sim m_{r,{\rm CBM}}~X_{\rm Eu}/X_{r-{\rm proc}}$, we finally obtain:
\begin{eqnarray}
    \frac{\left( M_{\rm Fe}/M_{\rm Eu} \right)_{\rm theory}}{\left( M_{\rm Fe}/M_{\rm Eu} \right)_{\rm Ret~II} } & \approx & 5.69
    \left( \frac{\left( M_{\rm Fe}/M_{\rm Eu} \right)_{\rm Ret~II}}{7 \times 10^4} \right)^{-1}
    \left( \frac{R_{\rm CCSN}/R_{\rm CBM}}{4.7 \times 10^2} \right) \nonumber \\
    & & \left( \frac{m_{\rm Fe,CCSN}/m_{r,{\rm CBM}}}{7/2} \right)
    \left( \frac{X_{\rm Eu}}{4.1 \times 10^{-10}} \right)^{-1}
    \left( \frac{X_{>}}{10^{-7}} \right) \, .
\end{eqnarray}
Thus, despite the crude approximations, the large uncertainties and the possible peculiar character of Reticulum II, the relative amount of Fe over Eu is broadly compatible with what expected from compact binary enrichment.\\
\textbf{Radioactive elements in Earth sediments and in meteorites}. The analysis of the abundances of long-lived $r$-process isotopes ($t_{1/2} > 10^7 {\rm yr}$) and of their daughter nuclei in meteorites (formed when the solar system was formed) and in sediments (e.g. deep see floor) reveals information on the local isotopes production at specific times (see \cite{Hotokezaka.etal:2018} and references therein). Due to the paucity of the $r$-process nuclides ${~}^{129}$I and ${~}^{247}$Cm, whose lifetimes are $\lesssim 20~{\rm Myr}$, and to the relative large abundance ratio (0.008) between ${~}^{244}$Pt ($t_{1/2} \sim 80~{\rm Myr}$) and its stable daughter nucleus ${~}^{238}$U in meteorites, it is possible to set the delay time between the solar system formation and its last $r$-process pollution to $\sim 100-120~{\rm Myr}$. In addition, the analysis of the deposition rates in deep see sediments of radioactive nuclides like ${}^{60}$Fe (very short-lived nuclide usually produced in SNe) and the $r$-only ${~}^{244}$Pt over the past 25 Myr shows a large fluctuation, with the present rate $\sim 10^{-2}$ times smaller than the rate 25 Myr ago. This indicates that the enrichment of $r$-process elements is uncorrelated with the CCSNe enrichment and the former is due to rarer events that produce large yields \cite{Hotokezaka.etal:2015}. 

A summary of the above observational constraints is provided in the right panel of Figure~(\ref{fig: observational constraints}). All of them work against regular CCSNe and in favor of compact binary merger as major site for the $r$-process production.
A final interesting question is whether compact binary mergers are the \textit{only} sources of $r$-process elements. The large abundances of the first peak elements in the solar $r$-process pattern and in some metal poor stars, with respect to the elements beyond it, could be an indication of the occurrence of mergers in which the production of heavy $r$-process elements is disfavored (for example, due to a strong neutrino irradiation). However, it leaves also space for other sources able to provide weak $r$-process nucleosynthesis (e.g. electron capture SN, \cite{Wanajo.etal:2011}). Moreover, it is non-trivial for compact binary mergers to explain the abundances at very low metallicity. Indeed, a compact binary merger requires the successful explosion of at least one CCSN to form it (two for a BNS system) and, in addition to the stellar evolution timescale, it is necessary to wait for the GW-driven inspiral timescale. The latter can be estimated from the semi-major axis, $a$, the eccentricity, $e$, the total and reduced mass, $M$ and $\mu$, of the binary at formation as:
\begin{equation}
    t_{\rm GW} \sim 0.66~{\rm Gyr} 
    \left( \frac{a}{ 0.01 {\rm AU}} \right)^{4}
    \left( \frac{M}{2.7~M_{\odot}} \right)^{-2}
    \left( \frac{\mu}{0.68 M_{\odot}} \right)^{-1}
    \left( 1-e^2 \right)^{7/2} \, .
\end{equation}
This is a possibly rather long timescale to be reconciled with fast mergers happening at extremely low metallicity. However, the strong dependence of $t_{\rm GW}$ on $a$ and $e$ does not exclude them, but it requires a population of tight, possibly eccentric, compact binaries. Possible alternative sites are represented by rare classes of SNe characterized, for example, by very intense magnetic fields and fast rotating cores (e.g. \cite{Winteler.etal:2012}). It is important to stress that the relation between time and metallicity is not obvious: yields expelled by stars, SNe and compact binary mergers need to mix with the ISM before entering in the composition of the next stellar generations. Moreover, due to kicks and fast traveling ejecta, the places where SNs explode, compact binaries merge and new stars form could be distinct. All these effects are amplified in low metallicity conditions, where only a few enrichment episodes have happened.

\section{Summary and outlook}
Compact binary mergers involving at least one neutron star represent ideal environments where the production of heavy elements through the $r$-process nucleosynthesis takes place. The prediction of the precise composition of the ejecta and a clear understanding of its origin depend both on the modeling of the astrophysical sites and on detailed nuclear physics knowledge. A lot of progress has been achieved in the past few years and the predicted abundances are able to explain many independent observations. Compact binary mergers are also possibly very relevant in explaining the nuclear evolution of the Universe, in terms of the abundances that we observe at different epochs and in different astrophysical environments.

Still, many open questions remain. 
On the one hand, more realistic and sophisticated compact binary merger models are required to predict the properties and the amount of ejecta with better accuracy and properly taking into account all the relevant physics.
On the other hand, a more robust knowledge of the properties of exotic neutron-rich nuclei is key to reduce present nuclear uncertainties. In this respect, existing and upcoming world-wide nuclear facilities (including FAIR, FRIB, HIAF, RAON, RIKEN and SPIRAL) will finally produce some of the neutron-rich nuclei relevant for the $r$-process and measure their properties. Multimessenger observations, as well as the study of the composition of matter in different astrophysical and terrestrial contexts, will also sharpen our understanding and test our models, helping reducing our ignorance and forcing us to look at the problem from many viewing angles. Despite a conclusive answer about the presence and the role of other possible $r$-process nucleosynthesis sites still needs deeper investigations and clear evidences, it is nowadays certain that compact binary mergers are one of the major sources of $r$-process elements in the Universe.


%
%
%
\bibliographystyle{plain}
\bibliography{bibref}

\begin{thebibliography}{10}

\bibitem{abadie:2010}
J.~{Abadie} et~al.
\newblock {TOPICAL REVIEW: Predictions for the rates of compact binary
  coalescences observable by ground-based gravitational-wave detectors}.
\newblock {\em Classical and Quantum Gravity}, 27(17):173001, September 2010.

\bibitem{Abbott:2017a}
B.~P. {Abbott} et~al.
\newblock {GW170817: Observation of Gravitational Waves from a Binary Neutron
  Star Inspiral}.
\newblock {\em Physical Review Letter}, 119(16):161101, October 2017.

\bibitem{Abbott:2017b}
B.~P. {Abbott} et~al.
\newblock {Multi-messenger Observations of a Binary Neutron Star Merger}.
\newblock {\em The Astrophysical Journal Letters}, 848(2):L12, October 2017.

\bibitem{Abbott.etal:2018}
B.~P. {Abbott} et~al.
\newblock {GW170817: Measurements of Neutron Star Radii and Equation of State}.
\newblock {\em Physical Review Letters}, 121(16):161101, October 2018.

\bibitem{Abbott:2019}
B.~P. {Abbott} et~al.
\newblock {GWTC-1: A Gravitational-Wave Transient Catalog of Compact Binary
  Mergers Observed by LIGO and Virgo during the First and Second Observing
  Runs}.
\newblock {\em Physical Review X}, 9(3):031040, July 2019.

\bibitem{Barnes.etal:2016}
Jennifer {Barnes}, Daniel {Kasen}, Meng-Ru {Wu}, and Gabriel
  {Mart{\'\i}nez-Pinedo}.
\newblock {Radioactivity and Thermalization in the Ejecta of Compact Object
  Mergers and Their Impact on Kilonova Light Curves}.
\newblock {\em The Astrophysical Journal}, 829(2):110, October 2016.

\bibitem{Bernuzzi.etal:2020}
Sebastiano {Bernuzzi}, Matteo {Breschi}, Boris {Daszuta}, Andrea {Endrizzi},
  Domenico {Logoteta}, Vsevolod {Nedora}, Albino {Perego}, David {Radice},
  Federico {Schianchi}, Francesco {Zappa}, Ignazio {Bombaci}, and Nestor
  {Ortiz}.
\newblock {Accretion-induced prompt black hole formation in asymmetric neutron
  star mergers, dynamical ejecta, and kilonova signals}.
\newblock {\em Monthly Notices of the Royal Astronomical Society},
  497(2):1488--1507, June 2020.

\bibitem{BBFH:1957}
E.~Margaret {Burbidge}, G.~R. {Burbidge}, William~A. {Fowler}, and F.~{Hoyle}.
\newblock {Synthesis of the Elements in Stars}.
\newblock {\em Reviews of Modern Physics}, 29(4):547--650, January 1957.

\bibitem{Cameron:1957}
A.~G.~W. {Cameron}.
\newblock {Nuclear Reactions in Stars and Nucleogenesis}.
\newblock {\em Publications of the Astronomy Society of the Pacific},
  69(408):201, June 1957.

\bibitem{Clayton:1983}
Donald~D. {Clayton}.
\newblock {\em {Principles of stellar evolution and nucleosynthesis}}.
\newblock University of Chicago Press, Chicago.

\bibitem{Cowan.etal:2019}
John~J. {Cowan}, Christopher {Sneden}, James~E. {Lawler}, Ani {Aprahamian},
  Michael {Wiescher}, Karlheinz {Langanke}, Gabriel {Mart{\'\i}nez-Pinedo}, and
  Friedrich-Karl {Thielemann}.
\newblock {Origin of the Heaviest Elements: the Rapid Neutron-Capture Process}.
\newblock {\em arXiv e-prints}, page arXiv:1901.01410, January 2019.

\bibitem{Eichler:1989}
David {Eichler}, Mario {Livio}, Tsvi {Piran}, and David~N. {Schramm}.
\newblock {Nucleosynthesis, neutrino bursts and {\ensuremath{\gamma}}-rays from
  coalescing neutron stars}.
\newblock {\em Nature}, 340(6229):126--128, July 1989.

\bibitem{Eichler.etal:2015}
M.~{Eichler}, A.~{Arcones}, A.~{Kelic}, O.~{Korobkin}, K.~{Langanke},
  T.~{Marketin}, G.~{Martinez-Pinedo}, I.~{Panov}, T.~{Rauscher}, S.~{Rosswog},
  C.~{Winteler}, N.~T. {Zinner}, and F.~K. {Thielemann}.
\newblock {The Role of Fission in Neutron Star Mergers and Its Impact on the
  r-Process Peaks}.
\newblock {\em The Astrophysical Journal}, 808(1):30, July 2015.

\bibitem{Fernandez.Metzger:2016}
Rodrigo {Fern{\'a}ndez} and Brian~D. {Metzger}.
\newblock {Electromagnetic Signatures of Neutron Star Mergers in the Advanced
  LIGO Era}.
\newblock {\em Annual Review of Nuclear and Particle Science}, 66(1):23--45,
  October 2016.

\bibitem{Fong.etal:2015}
W.~{Fong}, E.~{Berger}, R.~{Margutti}, and B.~A. {Zauderer}.
\newblock {A Decade of Short-duration Gamma-Ray Burst Broadband Afterglows:
  Energetics, Circumburst Densities, and Jet Opening Angles}.
\newblock {\em The Astrophysical Journal}, 815(2):102, December 2015.

\bibitem{Freiburghaus.etal:1999a}
C.~{Freiburghaus}, J.~F. {Rembges}, T.~{Rauscher}, E.~{Kolbe}, F.~K.
  {Thielemann}, K.~L. {Kratz}, B.~{Pfeiffer}, and J.~J. {Cowan}.
\newblock {The Astrophysical r-Process: A Comparison of Calculations following
  Adiabatic Expansion with Classical Calculations Based on Neutron Densities
  and Temperatures}.
\newblock {\em The Astrophysical Journal}, 516(1):381--398, May 1999.

\bibitem{Freiburghaus.etal:1999}
C.~{Freiburghaus}, S.~{Rosswog}, and F.~K. {Thielemann}.
\newblock {R-Process in Neutron Star Mergers}.
\newblock {\em The Astrophysical Journal Letters}, 525(2):L121--L124, November
  1999.

\bibitem{Ghirlanda.etal:2016}
G.~{Ghirlanda}, O.~S. {Salafia}, A.~{Pescalli}, G.~{Ghisellini},
  R.~{Salvaterra}, E.~{Chassande-Mottin}, M.~{Colpi}, F.~{Nappo},
  P.~{D'Avanzo}, A.~{Melandri}, M.~G. {Bernardini}, M.~{Branchesi},
  S.~{Campana}, R.~{Ciolfi}, S.~{Covino}, D.~{G{\"o}tz}, S.~D. {Vergani},
  M.~{Zennaro}, and G.~{Tagliaferri}.
\newblock {Short gamma-ray bursts at the dawn of the gravitational wave era}.
\newblock {\em Astronomy and Astrophysics}, 594:A84, October 2016.

\bibitem{Goriely.etal:2013}
S.~{Goriely}, J.~L. {Sida}, J.~F. {Lema{\^\i}tre}, S.~{Panebianco},
  N.~{Dubray}, S.~{Hilaire}, A.~{Bauswein}, and H.~T. {Janka}.
\newblock {New Fission Fragment Distributions and r-Process Origin of the
  Rare-Earth Elements}.
\newblock {\em Physical Review Letters}, 111(24):242502, December 2013.

\bibitem{Hansen.etal:2017}
T.~T. {Hansen} et~al.
\newblock {An r-process Enhanced Star in the Dwarf Galaxy Tucana III}.
\newblock {\em The Astrophysical Journal}, 838(1):44, March 2017.

\bibitem{Hoffman.etal:1997}
R.~D. {Hoffman}, S.~E. {Woosley}, and Y.~Z. {Qian}.
\newblock {Nucleosynthesis in Neutrino-driven Winds. II. Implications for Heavy
  Element Synthesis}.
\newblock {\em The Astrophysical Journal}, 482(2):951--962, June 1997.

\bibitem{Honda.etal:2004}
Satoshi {Honda}, Wako {Aoki}, Toshitaka {Kajino}, Hiroyasu {Ando}, Timothy~C.
  {Beers}, Hideyuki {Izumiura}, Kozo {Sadakane}, and Masahide {Takada-Hidai}.
\newblock {Spectroscopic Studies of Extremely Metal-Poor Stars with the Subaru
  High Dispersion Spectrograph. II. The r-Process Elements, Including Thorium}.
\newblock {\em The Astrophysical Journal}, 607(1):474--498, May 2004.

\bibitem{Hotokezaka.etal:2016}
K.~{Hotokezaka}, S.~{Wanajo}, M.~{Tanaka}, A.~{Bamba}, Y.~{Terada}, and
  T.~{Piran}.
\newblock {Radioactive decay products in neutron star merger ejecta: heating
  efficiency and {\ensuremath{\gamma}}-ray emission}.
\newblock {\em Monthly Notices of the Royal Astronomical Society},
  459(1):35--43, June 2016.

\bibitem{Hotokezaka.etal:2018}
Kenta {Hotokezaka}, Paz {Beniamini}, and Tsvi {Piran}.
\newblock {Neutron star mergers as sites of r-process nucleosynthesis and short
  gamma-ray bursts}.
\newblock {\em International Journal of Modern Physics D}, 27(13):1842005,
  January 2018.

\bibitem{Hotokezaka.etal:2015}
Kenta {Hotokezaka}, Tsvi {Piran}, and Michael {Paul}.
\newblock {Short-lived $^{244}$Pu points to compact binary mergers as sites for
  heavy r-process nucleosynthesis}.
\newblock {\em Nature Physics}, 11(12):1042, December 2015.

\bibitem{Hulse.Taylor1975}
R.~A. {Hulse} and J.~H. {Taylor}.
\newblock {Discovery of a pulsar in a binary system.}
\newblock {\em The Astrophysical Journal Letters}, 195:L51--L53, January 1975.

\bibitem{Iliadis:2007}
Christian {Iliadis}.
\newblock {\em {Nuclear Physics of Stars}}.
\newblock 2007.

\bibitem{Ji.etal:2016}
Alexander~P. {Ji}, Anna {Frebel}, Anirudh {Chiti}, and Joshua~D. {Simon}.
\newblock {R-process enrichment from a single event in an ancient dwarf
  galaxy}.
\newblock {\em Nature}, 531(7596):610--613, March 2016.

\bibitem{Just.etal:2015}
O.~{Just}, A.~{Bauswein}, R.~{Ardevol Pulpillo}, S.~{Goriely}, and H.~T.
  {Janka}.
\newblock {Comprehensive nucleosynthesis analysis for ejecta of compact binary
  mergers}.
\newblock {\em The Monthly Notices of the Royal Astronomical Society},
  448(1):541--567, March 2015.

\bibitem{Kasen.etal:2013}
Daniel {Kasen}, N.~R. {Badnell}, and Jennifer {Barnes}.
\newblock {Opacities and Spectra of the r-process Ejecta from Neutron Star
  Mergers}.
\newblock {\em The Astrophysical Journal}, 774(1):25, September 2013.

\bibitem{Kasen.etal:2017}
Daniel {Kasen}, Brian {Metzger}, Jennifer {Barnes}, Eliot {Quataert}, and
  Enrico {Ramirez-Ruiz}.
\newblock {Origin of the heavy elements in binary neutron-star mergers from a
  gravitational-wave event}.
\newblock {\em Nature}, 551(7678):80--84, November 2017.

\bibitem{Korobkin.etal:2012}
O.~{Korobkin}, S.~{Rosswog}, A.~{Arcones}, and C.~{Winteler}.
\newblock {On the astrophysical robustness of the neutron star merger
  r-process}.
\newblock {\em Monthly Notices of the Royal Astronomical Society},
  426(3):1940--1949, November 2012.

\bibitem{Kratz.etal:1993}
Karl-Ludwig {Kratz}, Jean-Philippe {Bitouzet}, Friedrich-Karl {Thielemann},
  Peter {Moeller}, and Bernd {Pfeiffer}.
\newblock {Isotopic r-Process Abundances and Nuclear Structure Far from
  Stability: Implications for the r-Process Mechanism}.
\newblock {\em The Astrophysical Journal}, 403:216, January 1993.

\bibitem{Lattimer:1974a}
J.~M. {Lattimer} and D.~N. {Schramm}.
\newblock {Black-hole-neutron-star collisions}.
\newblock {\em The Astrophysical Journal Letters}, 192:L145--L147, September
  1974.

\bibitem{Lattimer:1976}
J.~M. {Lattimer} and D.~N. {Schramm}.
\newblock {The tidal disruption of neutron stars by black holes in close
  binaries.}
\newblock {\em The Astrophysical Journal}, 210:549--567, December 1976.

\bibitem{Li.Paczynski:1998}
Li-Xin {Li} and Bohdan {Paczy{\'n}ski}.
\newblock {Transient Events from Neutron Star Mergers}.
\newblock {\em The Astrophysical Journal Letters}, 507(1):L59--L62, November
  1998.

\bibitem{Lippuner.etal:2017}
Jonas {Lippuner}, Rodrigo {Fern{\'a}ndez}, Luke~F. {Roberts}, Francois
  {Foucart}, Daniel {Kasen}, Brian~D. {Metzger}, and Christian~D. {Ott}.
\newblock {Signatures of hypermassive neutron star lifetimes on r-process
  nucleosynthesis in the disc ejecta from neutron star mergers}.
\newblock {\em Monthly Notices of the Royal Astronomical Society},
  472(1):904--918, November 2017.

\bibitem{Lippuner.Roberts:2015}
Jonas {Lippuner} and Luke~F. {Roberts}.
\newblock {r-process Lanthanide Production and Heating Rates in Kilonovae}.
\newblock {\em The Astrophysical Journal}, 815(2):82, December 2015.

\bibitem{Lippuner.Roberts:2017}
Jonas {Lippuner} and Luke~F. {Roberts}.
\newblock {SkyNet: A Modular Nuclear Reaction Network Library}.
\newblock {\em The Astrophysical Journal Supplement Series}, 233(2):18,
  December 2017.

\bibitem{Martin.etal:2015}
D.~{Martin}, A.~{Perego}, A.~{Arcones}, F.~K. {Thielemann}, O.~{Korobkin}, and
  S.~{Rosswog}.
\newblock {Neutrino-driven Winds in the Aftermath of a Neutron Star Merger:
  Nucleosynthesis and Electromagnetic Transients}.
\newblock {\em The Astrophysical Journal}, 813(1):2, November 2015.

\bibitem{Martinez-Pinedo:2008}
G.~{Mart{\'\i}nez-Pinedo}.
\newblock {Selected topics in nuclear astrophysics}.
\newblock {\em European Physical Journal Special Topics}, 156(1):123--149,
  April 2008.

\bibitem{Martinez-Pinedo.etal:2012}
G.~{Mart{\'\i}nez-Pinedo}, T.~{Fischer}, A.~{Lohs}, and L.~{Huther}.
\newblock {Charged-Current Weak Interaction Processes in Hot and Dense Matter
  and its Impact on the Spectra of Neutrinos Emitted from Protoneutron Star
  Cooling}.
\newblock {\em Physical Review Letters}, 109(25):251104, December 2012.

\bibitem{Mendoza.etal:2015}
Joel de~Jes{\'u}s {Mendoza-Temis}, Meng-Ru {Wu}, Karlheinz {Langanke}, Gabriel
  {Mart{\'\i}nez-Pinedo}, Andreas {Bauswein}, and Hans-Thomas {Janka}.
\newblock {Nuclear robustness of the r process in neutron-star mergers}.
\newblock {\em Physical review C}, 92(5):055805, November 2015.

\bibitem{Metzger.etal:2010}
B.~D. {Metzger}, G.~{Mart{\'\i}nez-Pinedo}, S.~{Darbha}, E.~{Quataert},
  A.~{Arcones}, D.~{Kasen}, R.~{Thomas}, P.~{Nugent}, I.~V. {Panov}, and N.~T.
  {Zinner}.
\newblock {Electromagnetic counterparts of compact object mergers powered by
  the radioactive decay of r-process nuclei}.
\newblock {\em Monthly Notices of the Royal Astronomical Society},
  406(4):2650--2662, August 2010.

\bibitem{Metzger:2019}
Brian~D. {Metzger}.
\newblock {Kilonovae}.
\newblock {\em Living Reviews in Relativity}, 23(1):1, December 2019.

\bibitem{Oertel.etal:2017}
M.~{Oertel}, M.~{Hempel}, T.~{Kl{\"a}hn}, and S.~{Typel}.
\newblock {Equations of state for supernovae and compact stars}.
\newblock {\em Reviews of Modern Physics}, 89(1):015007, January 2017.

\bibitem{Perego.etal:2014}
A.~{Perego}, S.~{Rosswog}, R.~M. {Cabez{\'o}n}, O.~{Korobkin},
  R.~{K{\"a}ppeli}, A.~{Arcones}, and M.~{Liebend{\"o}rfer}.
\newblock {Neutrino-driven winds from neutron star merger remnants}.
\newblock {\em Monthly notices of the Royal Astronomical Society},
  443(4):3134--3156, October 2014.

\bibitem{Perego.etal:2017}
Albino {Perego}, David {Radice}, and Sebastiano {Bernuzzi}.
\newblock {AT 2017gfo: An Anisotropic and Three-component Kilonova Counterpart
  of GW170817}.
\newblock {\em The Astrophysical Journal Letters}, 850(2):L37, December 2017.

\bibitem{Pian.etal:2017}
E.~{Pian} et~al.
\newblock {Spectroscopic identification of r-process nucleosynthesis in a
  double neutron-star merger}.
\newblock {\em Nature}, 551(7678):67--70, November 2017.

\bibitem{Pol.etal:2019}
Nihan {Pol}, Maura {McLaughlin}, and Duncan~R. {Lorimer}.
\newblock {Future Prospects for Ground-based Gravitational-wave Detectors: The
  Galactic Double Neutron Star Merger Rate Revisited}.
\newblock {\em The Astrophysical Journal}, 870(2):71, January 2019.

\bibitem{Prantzos.etal:2020}
N.~{Prantzos}, C.~{Abia}, S.~{Cristallo}, M.~{Limongi}, and A.~{Chieffi}.
\newblock {Chemical evolution with rotating massive star yields II. A new
  assessment of the solar s- and r-process components}.
\newblock {\em Monthly Notices of the Royal Astronomical Society},
  491(2):1832--1850, January 2020.

\bibitem{Qian.Woosley:1996}
Y.~Z. {Qian} and S.~E. {Woosley}.
\newblock {Nucleosynthesis in Neutrino-driven Winds. I. The Physical
  Conditions}.
\newblock {\em The Astrophysics Journal}, 471:331, November 1996.

\bibitem{Radice.etal:2020}
David {Radice}, Sebastiano {Bernuzzi}, and Albino {Perego}.
\newblock {The Dynamics of Binary Neutron Star Mergers and GW170817}.
\newblock {\em Annual Review of Nuclear and Particle Science}, 70(1):annurev,
  October 2020.

\bibitem{Radice.etal:2018}
David {Radice}, Albino {Perego}, Kenta {Hotokezaka}, Steven~A. {Fromm},
  Sebastiano {Bernuzzi}, and Luke~F. {Roberts}.
\newblock {Binary Neutron Star Mergers: Mass Ejection, Electromagnetic
  Counterparts, and Nucleosynthesis}.
\newblock {\em The Astrophysics Journal}, 869(2):130, December 2018.

\bibitem{Roberts.etal:2017}
Luke~F. {Roberts}, Jonas {Lippuner}, Matthew~D. {Duez}, Joshua~A. {Faber},
  Francois {Foucart}, Jr. {Lombardi}, James~C., Sandra {Ning}, Christian~D.
  {Ott}, and Marcelo {Ponce}.
\newblock {The influence of neutrinos on r-process nucleosynthesis in the
  ejecta of black hole-neutron star mergers}.
\newblock {\em Monthly Notices of the Royal Astronomical Society},
  464(4):3907--3919, February 2017.

\bibitem{Roederer.etal:2014}
Ian~U. {Roederer}, George~W. {Preston}, Ian~B. {Thompson}, Stephen~A.
  {Shectman}, Christopher {Sneden}, Gregory~S. {Burley}, and Daniel~D.
  {Kelson}.
\newblock {A Search for Stars of Very Low Metal Abundance. VI. Detailed
  Abundances of 313 Metal-poor Stars}.
\newblock {\em The Astronomical Journal}, 147(6):136, June 2014.

\bibitem{Rosswog.etal:2017}
S.~{Rosswog}, U.~{Feindt}, O.~{Korobkin}, M.~R. {Wu}, J.~{Sollerman},
  A.~{Goobar}, and G.~{Martinez-Pinedo}.
\newblock {Detectability of compact binary merger macronovae}.
\newblock {\em Classical and Quantum Gravity}, 34(10):104001, May 2017.

\bibitem{Rosswog.etal:2014}
S.~{Rosswog}, O.~{Korobkin}, A.~{Arcones}, F.~K. {Thielemann}, and T.~{Piran}.
\newblock {The long-term evolution of neutron star merger remnants - I. The
  impact of r-process nucleosynthesis}.
\newblock {\em Monthly Notices of the Royal Astronomical Society},
  439(1):744--756, March 2014.

\bibitem{Shibata.Hotokezaka:2019}
Masaru {Shibata} and Kenta {Hotokezaka}.
\newblock {Merger and Mass Ejection of Neutron Star Binaries}.
\newblock {\em Annual Review of Nuclear and Particle Science}, 69:41--64,
  October 2019.

\bibitem{Smartt.etal:2017}
S.~J. {Smartt} et~al.
\newblock {A kilonova as the electromagnetic counterpart to a
  gravitational-wave source}.
\newblock {\em Nature}, 551(7678):75--79, November 2017.

\bibitem{Sneden.etal:2008}
C.~{Sneden}, J.~J. {Cowan}, and R.~{Gallino}.
\newblock {Neutron-capture elements in the early galaxy.}
\newblock {\em Annual Review of Astronomy and Astrophysics}, 46:241--288,
  September 2008.

\bibitem{Symbalisty:1982}
E.~{Symbalisty} and D.~N. {Schramm}.
\newblock {Neutron Star Collisions and the r-Process}.
\newblock {\em The Astrophysical Journal Letters}, 22:143, January 1982.

\bibitem{Tanaka.etal:2017}
Masaomi {Tanaka} et~al.
\newblock {Kilonova from post-merger ejecta as an optical and near-Infrared
  counterpart of GW170817}.
\newblock {\em Publications of the Astronomical Society of Japan}, 69(6):102,
  December 2017.

\bibitem{Tanaka.etal:2013}
Masaomi {Tanaka} and Kenta {Hotokezaka}.
\newblock {Radiative Transfer Simulations of Neutron Star Merger Ejecta}.
\newblock {\em The Astrophysical Journal}, 775(2):113, October 2013.

\bibitem{Tanvir.etal:2017}
N.~R. {Tanvir} et~al.
\newblock {The Emergence of a Lanthanide-rich Kilonova Following the Merger of
  Two Neutron Stars}.
\newblock {\em The Astrophysical Journal Letters}, 848(2):L27, October 2017.

\bibitem{Tanvir.etal:2013}
N.~R. {Tanvir}, A.~J. {Levan}, A.~S. {Fruchter}, J.~{Hjorth}, R.~A. {Hounsell},
  K.~{Wiersema}, and R.~L. {Tunnicliffe}.
\newblock {A `kilonova' associated with the short-duration
  {\ensuremath{\gamma}}-ray burst GRB 130603B}.
\newblock {\em Nature}, 500(7464):547--549, August 2013.

\bibitem{Thielemann.etal:2017}
F.~K. {Thielemann}, M.~{Eichler}, I.~V. {Panov}, and B.~{Wehmeyer}.
\newblock {Neutron Star Mergers and Nucleosynthesis of Heavy Elements}.
\newblock {\em Annual Review of Nuclear and Particle Science}, 67:253--274,
  October 2017.

\bibitem{Villar.etal:2017}
V.~A. {Villar}, J.~{Guillochon}, E.~{Berger}, B.~D. {Metzger}, P.~S.
  {Cowperthwaite}, M.~{Nicholl}, K.~D. {Alexand er}, P.~K. {Blanchard},
  R.~{Chornock}, T.~{Eftekhari}, W.~{Fong}, R.~{Margutti}, and P.~K.~G.
  {Williams}.
\newblock {The Combined Ultraviolet, Optical, and Near-infrared Light Curves of
  the Kilonova Associated with the Binary Neutron Star Merger GW170817: Unified
  Data Set, Analytic Models, and Physical Implications}.
\newblock {\em The Astrophysical Journal Letters}, 851(1):L21, December 2017.

\bibitem{Wanajo.etal:2011}
Shinya {Wanajo}, Hans-Thomas {Janka}, and Bernhard {M{\"u}ller}.
\newblock {Electron-capture Supernovae as The Origin of Elements Beyond Iron}.
\newblock {\em The Astrophysical Journal Letters}, 726(2):L15, January 2011.

\bibitem{Wetal:2014}
Shinya {Wanajo}, Yuichiro {Sekiguchi}, Nobuya {Nishimura}, Kenta {Kiuchi},
  Koutarou {Kyutoku}, and Masaru {Shibata}.
\newblock {Production of All the r-process Nuclides in the Dynamical Ejecta of
  Neutron Star Mergers}.
\newblock {\em The Astrophysical Journal Letters}, 789(2):L39, July 2014.

\bibitem{Wanderman.Piran:2015}
David {Wanderman} and Tsvi {Piran}.
\newblock {The rate, luminosity function and time delay of non-Collapsar short
  GRBs}.
\newblock {\em Monthly Notices of the Royal Astronomical Society},
  448(4):3026--3037, April 2015.

\bibitem{Watson.etal:2019}
Darach {Watson}, Camilla~J. {Hansen}, Jonatan {Selsing}, Andreas {Koch},
  Daniele~B. {Malesani}, Anja~C. {Andersen}, Johan P.~U. {Fynbo}, Almudena
  {Arcones}, Andreas {Bauswein}, Stefano {Covino}, Aniello {Grado}, Kasper~E.
  {Heintz}, Leslie {Hunt}, Chryssa {Kouveliotou}, Giorgos {Leloudas}, Andrew~J.
  {Levan}, Paolo {Mazzali}, and Elena {Pian}.
\newblock {Identification of strontium in the merger of two neutron stars}.
\newblock {\em Nature}, 574(7779):497--500, October 2019.

\bibitem{Winteler.etal:2012}
C.~{Winteler}, R.~{K{\"a}ppeli}, A.~{Perego}, A.~{Arcones}, N.~{Vasset},
  N.~{Nishimura}, M.~{Liebend{\"o}rfer}, and F.~K. {Thielemann}.
\newblock {Magnetorotationally Driven Supernovae as the Origin of Early Galaxy
  r-process Elements?}
\newblock {\em The Astrophysical Journal Letters}, 750(1):L22, May 2012.

\bibitem{Wollaeger.etal:2018}
Ryan~T. {Wollaeger}, Oleg {Korobkin}, Christopher~J. {Fontes}, Stephan~K.
  {Rosswog}, Wesley~P. {Even}, Christopher~L. {Fryer}, Jesper {Sollerman},
  Aimee~L. {Hungerford}, Daniel~R. {van Rossum}, and Allan~B. {Wollaber}.
\newblock {Impact of ejecta morphology and composition on the electromagnetic
  signatures of neutron star mergers}.
\newblock {\em Monthly Notices of the Royal Astronomical Society},
  478(3):3298--3334, August 2018.

\bibitem{Wu.etal:2019}
Meng-Ru {Wu}, J.~{Barnes}, G.~{Mart{\'\i}nez-Pinedo}, and B.~D. {Metzger}.
\newblock {Fingerprints of Heavy-Element Nucleosynthesis in the Late-Time
  Lightcurves of Kilonovae}.
\newblock {\em Physical Review Letters}, 122(6):062701, February 2019.

\bibitem{Wu.etal:2016}
Meng-Ru {Wu}, Rodrigo {Fern{\'a}ndez}, Gabriel {Mart{\'\i}nez-Pinedo}, and
  Brian~D. {Metzger}.
\newblock {Production of the entire range of r-process nuclides by black hole
  accretion disc outflows from neutron star mergers}.
\newblock {\em Monthly Notices of the Royal Astronomical Society},
  463(3):2323--2334, December 2016.

\bibitem{Zhu.etal:2018}
Y.~{Zhu}, R.~T. {Wollaeger}, N.~{Vassh}, R.~{Surman}, T.~M. {Sprouse}, M.~R.
  {Mumpower}, P.~{M{\"o}ller}, G.~C. {McLaughlin}, O.~{Korobkin}, T.~{Kawano},
  P.~J. {Jaffke}, E.~M. {Holmbeck}, C.~L. {Fryer}, W.~P. {Even}, A.~J.
  {Couture}, and J.~{Barnes}.
\newblock {Californium-254 and Kilonova Light Curves}.
\newblock {\em The Astrophysical Journal Letters}, 863(2):L23, August 2018.

\end{thebibliography}


\end{document}